# 2D Nb-doped $MoS_2$: Tuning the Exciton Transitions and Application to p-type FETs

Baokun Song,[†,#] Honggang Gu,[†,#,*] Mingsheng Fang,[†] Zhengfeng Guo,[†] Yen-Teng Ho,[‡,*] Xiuguo Chen,[†] Hao Jiang,[†] Shiyuan Liu[†,*]

[†]State Key Laboratory of Digital Manufacturing Equipment and Technology, Huazhong University of Science and Technology, Wuhan 430074, China.

[‡]International College of Semiconductor Technology, National Chiao Tung University, Hsinchu 30010, Taiwan.

[#] These authors contributed equally to this work.

**KEYWORDS:** 2D Nb-doped $MoS_2$, exciton transitions, dielectric functions, band renormalization, p-type FETs, spectroscopic ellipsometry

## ABSTRACT

Two-dimensional (2D) $MoS_2$ has been intensively investigated for its use in the fields of microelectronics, nanoelectronics, and optoelectronics. However, intrinsic 2D $MoS_2$ is usually used as the n-type semiconductor due to the unintentional sulphur vacancies and surface gas adsorption. The synthesis and characterization of 2D $MoS_2$ semiconductor of p-type are crucial for the development of relevant p-n junction devices, as well as the practical applications of 2D $MoS_2$ in the next-generation CMOS integrated circuit. Here, we synthesize high-quality, wafer-scale, 2D p-type $MoS_2$ ($Mo_{1-x}Nb_xS_2$) with various niobium (Nb) mole fractions from 0 to 7.6% by a creative two-step method. The dielectric functions of 2D $Mo_{1-x}Nb_xS_2$ are accurately determined by spectroscopic ellipsometry. We find that the increasing fraction of Nb dopant in 2D $MoS_2$ can modulate and promote the combination of A and B exciton peaks of 2D $MoS_2$. The direct causes of this impurity-tunable combination are interpreted as the joint influence of decreasing peak A and broadening peak B. We explain the broadening peak B as the multiple transitions from the impurity-induced valance bands to the conductive band minimum at K point of Brillouin zone by comparing and analyzing the simulated electronic structure of intrinsic and 2D Nb-doped $MoS_2$. A p-type FET based on the 2D Nb-doped $MoS_2$ was fabricated for characterization, and its working performance is expected to be adjustable as a function of concentration of Nb dopant according to our theoretical research. Our study is informative for comprehending optical and electronic properties of extrinsic 2D transitional metal dichalcogenides, which is important and imperative for the development and optimization of corresponding photonics and optoelectronics devices.

## INTRODUCTION

Two-dimensional (2D) materials, such as graphene, transitional metal dichalcogenides (TMDCs), and phosphorene, have received considerable attention due to the novel optoelectronic, thermal, and mechanical properties.[1–7] Further, the well-designed modulating engineering, including doping (defect) engineering,[8,9] strain engineering,[10] and band engineering,[11] endows 2D materials with exotic properties and extends their applications. Recently, 2D $MoS_2$ and customized heterostructures hybridized with other 2D materials have been widely applied in novel microelectronics, nanoelectronics, and optoelectronics devices,[12–17] benefiting from the layer-controllable electronic structures, the moderate carrier mobility, the excellent thermostability, and multiple feasible doping modes.[5] Dopants can renormalize the electronic structure of 2D $MoS_2$ to some extent, therefore, modify its basic properties.[18,19] For instance, by introducing chlorine dopants, the contact resistivity ($R_c$) of field effect transistors (FETs) based on the 2D $MoS_2$ can be strikingly reduced.[20] When the rhenium atoms are selected as dopants, the structural and photonic responses of monolayer (1L) $MoS_2$ can be improved effectively.[21] Mouri *et al*. achieves the tunable photoluminescence (PL) of 1L $MoS_2$ by adding some organic impurities.[22] Furthermore, some researchers predict the possible doping paradigms of $MoS_2$ and study the properties of these extrinsic $MoS_2$ theoretically.[23–26] Doping engineering could extremely extend the application fields of 2D $MoS_2$ via renormalizing electronic structures, adjusting conduction types, and optimizing physicochemical properties.[27] Compared with the non-intrinsic 2D $MoS_2$ obtained by electrostatic doping and contact engineering, 2D doped $MoS_2$ directly synthesized by chemical doping processes should have excellent thermal and chemical stability. Optoelectronic devices based on the chemically doped 2D $MoS_2$ should be with better environmental adaptability, more stable working performance, and better durability. These devices are fabricated with less complicated COMS processes, which greatly reduces the production costs, and it is expected to achieve large-scale applications in the next-generation digital integrated circuit industry. Additionally, the basic optical and dielectric parameters of 2D $MoS_2$ exhibit remarkable gate-tunability, which is conducive to the development of high-performance modulators, sensors, photodetectors, and so on.[28–30]

In general, 2D $MoS_2$ is a natural n-type conductive semiconductor due to the omnipresent electron-donating sulfur vacancies,[31] thereby the intrinsic 2D $MoS_2$ is essentially suitable for preparing n-type devices, like n-type FETs.[32] In order to introduce 2D $MoS_2$ into the low-power, high-performance complementary logic applications, both n- and p-type FETs should be fabricated.[33,34] Thus, the high-quality substitutional p-type doping 2D $MoS_2$ is highly desired, which is promised to be the preferred platform for designing p-type optoelectronic devices, owning to the facile and robust integration and applications in advanced nanophotonics and nanoelectronics.[27] Among various doping strategies of 2D $MoS_2$, the substitutional Nb doping is considered to be one of the most promising doping modes to achieve the p-type $MoS_2$.[23,35] It is predicted to have a negative formation energy since the Nb atom has one less d-electron comparing with Mo,[19,23] which is important from the perspective of thermodynamics. Theoretically, an additional hole will be released into the material system when a Nb atom replaces a Mo atom. Recently, some researchers synthesize Nb-doped $MoS_2$ by the chemical vapor transport, physical vapor transport, and chemical vapor deposition,[18,31,36-38] and some devices have also been developed based on 2D Nb-doped $MoS_2$.[39,40] The performances of these devices strongly depend on the basic optical and dielectric properties of 2D Nb-doped $MoS_2$.[41] Nevertheless, on account of the lack of sufficiently high-quality, large-area 2D Nb-doped $MoS_2$ films, the inherent optical and dielectric properties and relevant comprehensive analysis, especially the attributive optical and dielectric parameters, of 2D Nb-doped $MoS_2$ are rarely reported. Performing a detailed study on the basic optical and dielectric parameters of 2D Nb-doped $MoS_2$ could provide a quantitative guidance for the development of corresponding electronic and optoelectronics devices. Additionally, revealing the impurity-tunable properties of 2D Nb-doped $MoS_2$ can advance the understanding of electronic structures in the extrinsic 2D TMDCs.

In this work, high-quality, wafer-scale, few-layer (~3L) $Mo_{1-x}Nb_xS_2$ are fabricated by an innovative two-step method. Spectroscopic ellipsometry (SE) was used to study the dielectric functions $\varepsilon(E)$ of extrinsic 2D $MoS_2$, we find that the exciton transition peaks A and B gradually combine into one broad peak A+B with the increasing Nb mole fraction. We ascribe this intriguing combination phenomenon to the decrease of peak A and the broadening of peak B. According to the simulated electronic structure of 2D intrinsic and Nb-doped $MoS_2$ via first principles, we infer that the broadening exciton peak B is related to multiple transitions occurring between the impurity-induced extra valance bands (VBs) and the conductive band (CB) minimum. Interestingly, we find that the center energies of exciton transition peaks A and B ($E_0$(A) and $E_0$(B)) are moving in opposite directions with the increasing Nb mole fraction. It is interpreted as the competition between the impurity-induced band shrinkage (especially the uplifted VBs) and the decreasing exciton binding energy ($E_b$). We also prepare a p-type FET based on 2D $Mo_{1-x}Nb_xS_2$, whose transfer characteristic suggests that the Schottky barrier height (SBH) can be effectively reduced, and even ohmic contact can be achieved by adjusting the Nb mole fraction.

## RESULTS AND DISCUSSION

High-quality and wafer-scale 2D $Mo_{1-x}Nb_xS_2$ are directly synthesized onto *c*-sapphire substrates via a two-step method. Detailed procedures and process parameters about the fabrication method are available in the Methods section and Table S1. The optical photograph (Figure 1a) and the micrograph (Figure 1b) with a consistent optical contrast across the wafer suggest that the 2D $Mo_{1-x}Nb_xS_2$ film is with well thickness uniformity. The high-resolution transmission electron microscope (HRTEM) images show that all 2D $Mo_{1-x}Nb_xS_2$ films adopted in this work are about 1.90 nm (~3L) (Figure 1c,d, Figure S1, and Table S2). The layer number of 2D $MoS_2$ is evaluated based on the ratio between the thickness and the nominal interlayer spacing 6.15 Å of bulk $MoS_2$.[42] The clear layered structures in the HRTEM images demonstrate that the 2D $Mo_{1-x}Nb_xS_2$ films are with remarkable crystal quality. The doping levels of these samples are checked by the X-ray photoelectron spectroscopy (XPS) (Figure 1e–g, Figure S2). The XPS peaks at 204 eV and 207 eV (Nb $3d_{5/2}$ and Nb $3d_{3/2}$) corresponding to the Nb-S bonds can be deconvoluted with only one Voigt function, meaning that only one niobium-containing chemical species exists at the surface of samples. A similar XPS analysis can also be applied to estimate the chemical components of Mo and S elements (Figure 1e,f). The Nb mole fraction of 2D $Mo_{1-x}Nb_xS_2$ specimen is quantitatively assessed by normalizing the integrated area of Nb XPS peaks to the area of Mo XPS peak. In this work, five representative 2D $Mo_{1-x}Nb_xS_2$ specimens with Nb mole fractions of 0, 2.4%, 4.3%, 5.1%, 7.6% are selected as the research objects. For the sake of brevity, hereafter, these five 2D $Mo_{1-x}Nb_xS_2$ samples with different doping concentrations are referred to as #1–#5 in sequence. The Raman spectra of #1–#5 samples are measured and plotted in Figure 1h, and two typical phonon modes $A_{1g}$ and $E^1_{2g}$ are observed. Referring to the previous reports, the layer number of TMDCs can be approximatively estimated according to the frequency difference between the out-of-plane phonon mode $A_{1g}$ and the in-plane phonon mode $E^1_{2g}$.[43,44] As illustrated in Figure 1h, the differences $A_{1g} - E^1_{2g}$ of #1–#5 samples are stable around 23.5 $cm^{-1}$, which are highly in accordance with that of mechanical exfoliated single-crystal 3L $MoS_2$.[45] The sharp Raman peaks also indicate that the 2D $Mo_{1-x}Nb_xS_2$ films used in this experiment are with high purity and excellent crystal quality. We have also measured the Raman spectra of the center and edge areas of 2D $Mo_{1-x}Nb_xS_2$ wafer (see Figure S3). Both intensity and frequency positions of phonon modes $E_{2g}^1$ and $A_{1g}$ in these two measurement regions are in good agreement, indicating that the 2D $Mo_{1-x}Nb_xS_2$ wafer is with remarkable quality consistency. It is worth pointing out that 2D $Mo_{1-x}Nb_xS_2$ films here are mainly polycrystalline with localized monocrystalline regions.

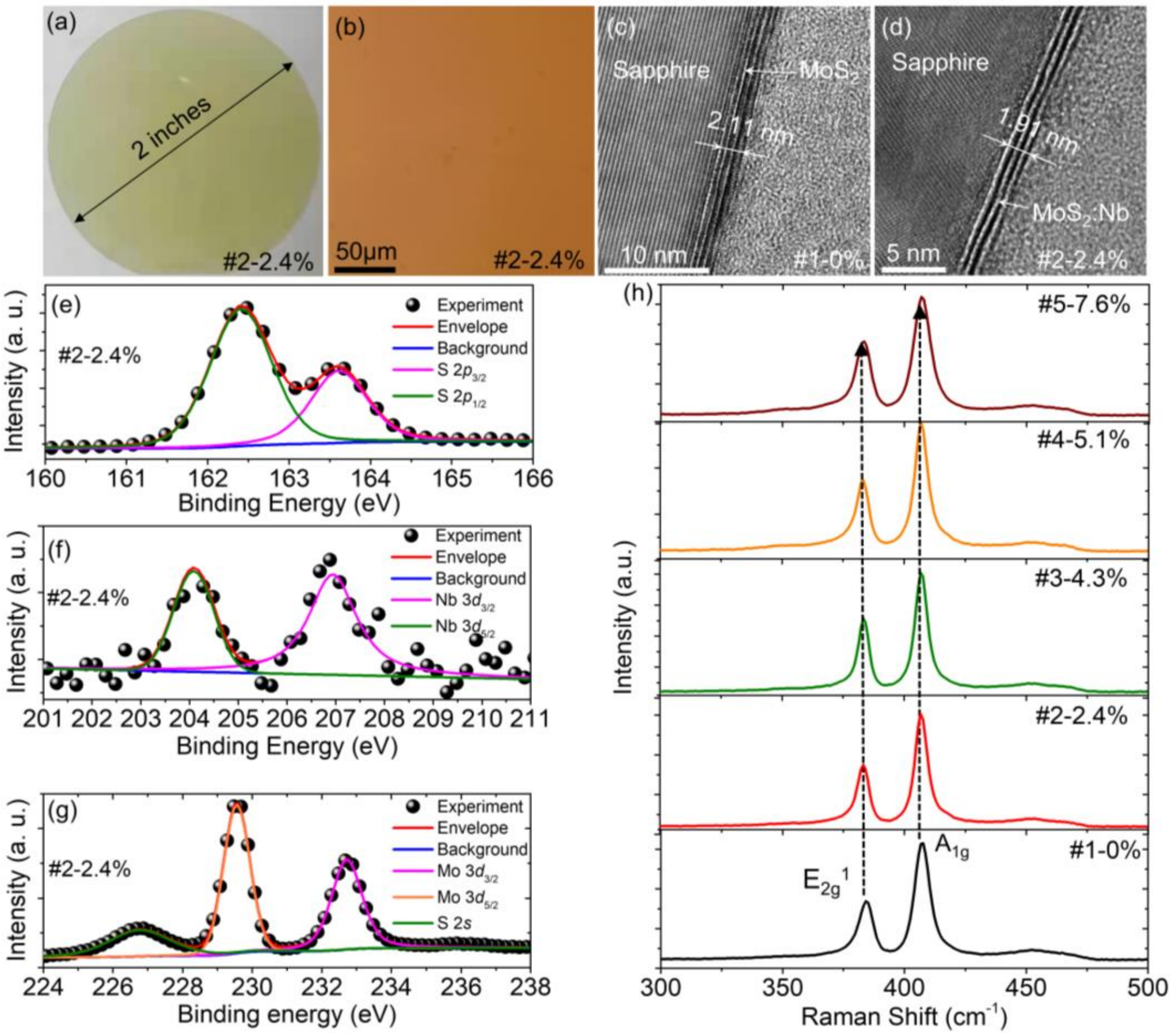

**Figure 1.** (a) Optical photograph and (b) micrograph of #2 2D $Mo_{1-x}Nb_xS_2$ sample. (c,d) HRTEM images of #1 and #2 2D $Mo_{1-x}Nb_xS_2$ samples. (e–g) Measured XPS spectra and corresponding peak-differentiation-imitating analysis results of #2 2D $Mo_{1-x}Nb_xS_2$ sample. (f) Typical Raman spectra of #1–#5 2D $Mo_{1-x}Nb_xS_2$ samples, the wavelength of exciting laser is 532 nm.

The PL spectra and absorbance of 2D $Mo_{1-x}Nb_xS_2$ are also measured to assist in understanding the impurity-dependent optical properties. Two obvious PL peaks, A and B, can be observed in the PL spectra of 2D $Mo_{1-x}Nb_xS_2$ (Figure 2a), which corresponds to the exciton recombination radiations occurring at K point of the Brillouin zone (BZ).[46] With the increase of Nb dopant, peak A in PL spectra show a slight redshift, while the high-energy PL peak B exhibits an obvious blueshift. The opposite shifts of peaks A and B are mainly associated with the impurity-induced band shrinkage and the prominent decreasing exciton binding energy $E_b$ of 2D $Mo_{1-x}Nb_xS_2$, respectively. Further, Nb doping is expected to reconfigure the electronic structures of $MoS_2$, generate several new impurity-induced valance bands, and compress the energy gap at K point of BZ, thereby results in the redshift of PL peak A.[19] Furthermore, in the low-dimensional materials with intense dielectric confinement (image charge effect),[47] the impurity-induced band renormalization should be more significant. The dense hole gas provided by the Nb atoms can effectively shield and decrease the $E_b$, and offset the opposite effects caused by the band shrinkage. Thus, the transition energy ($E_t$) of exciton in 2D $MoS_2$ will increase with the mole fraction of dopant Nb, which will eventually lead to the blueshift of PL peak B. Another relatively weak PL peak around 1.88 eV shown in Figure 2a, can be designated as the exciton recombination radiations from the CB minimum to the non-degenerate subbands near the first VB at K point of BZ, where the quantized subbands are relevant to the SOC-induced (SOC is the abbreviation of spin-orbit-coupling) band splitting.[48] Due to the impurity-induced band renormalization, the weaker PL peak also exhibits a slight redshift. As shown in Figure 2b, the absorption peaks A and B exhibit similar impurity-dependent shift patterns to the PL peaks, and we infer that the corresponding modulation mechanisms should also be the same. It is notable to mention that the flat absorption feature in some heavily-doped 2D $Mo_{1-x}Nb_xS_2$ specimens (#3–#5) prevents us from identifying the exact position of absorption peak A.

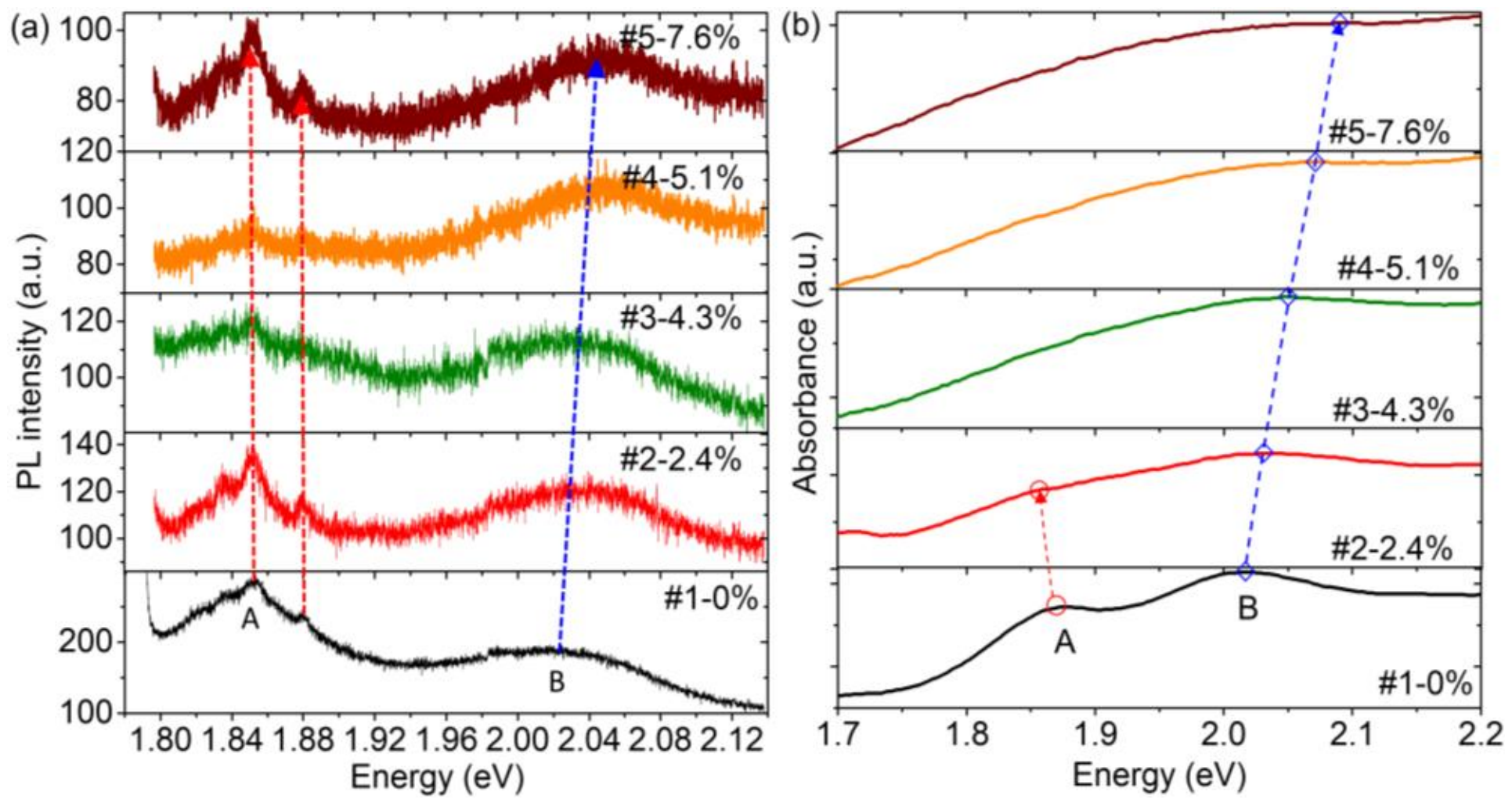

**Figure 2**. (a) PL spectra and (b) absorbance of #1–#5 2D $Mo_{1-x}Nb_xS_2$ samples.

The dielectric function spectra $\varepsilon(E)$ ($\varepsilon = \varepsilon_r - i\varepsilon_i$) of #1–#5 2D $Mo_{1-x}Nb_xS_2$ are accurately determined by SE over an ultra-broad energy range of 0.73–6.42 eV (i.e., the spectral range of 193–1690 nm). In the ellipsometric analysis, a vertical stacking optical model (Figure S4) and a parameterized dielectric function combining several classical oscillators are established to describe the geometric structure and dielectric dispersion of 2D $Mo_{1-x}Nb_xS_2$. The theoretical ellipsometric spectra of these samples are calculated by the transfer matrix method.[49] Finally, the dielectric functions and thicknesses of 2D $Mo_{1-x}Nb_xS_2$ samples are simultaneously determined by fitting the measured ellipsometric spectra with theoretical ones. The dielectric function spectra reported here (Figure S6) quantitatively describe the dielectric responses of polycrystalline 2D $Mo_{1-x}Nb_xS_2$ films. Further ellipsometry principles and elaborated analysis procedures are available in the Supporting Information and some previous publications.[50–58]

Ellipsometric analysis results of #1–#5 2D $Mo_{1-x}Nb_xS_2$ samples are shown in Figures S5–S7, Table S2, and Table S5. The thicknesses of 2D $Mo_{1-x}Nb_xS_2$ films determined by SE are all about 1.9 nm, which are in good agreement with the HRTEM results. The micro-interface layers, including the surface roughness layer of 2D $Mo_{1-x}Nb_xS_2$ and the extremely thin interlayer between the 2D $Mo_{1-x}Nb_xS_2$ and *c*-sapphire substrates, may bring about a slight overestimation of the thickness determined by HRTEM. The highly uniform thicknesses of 2D $Mo_{1-x}Nb_xS_2$ are helpful for investigating the doping concentration-dependent optical and dielectric properties, largely avoiding the interference from the quantum scale effect.[52] Therefore, the overall trends and magnitudes of dielectric functions and complex refractive indices $N(E)$ ($N = n - i\kappa$) (see Figure S6) from #1–#5 samples have no significant difference over the measurement spectral region. While the dielectric function spectra around the exciton transition peaks A and B exhibit interesting impurity-dependent subtle changes. It's worth emphasizing here that the change of dielectric function spectra caused by doping engineering is different from that induced by scale effects.[52] The doping mainly modulates the exciton transition strength of 2D $MoS_2$ due to the appearance of doping levels and the change of dielectric environment, while the layer number (thickness) of 2D $MoS_2$ will influence the strength of dielectric functions as a whole. For clarity, the special spectral regions are enlarged and plotted in Figure 3a,b. It is obvious that the A and B exciton transition peaks gradually combine to be a broad peak A+B with the increasing Nb mole fraction. To figure out the specific combination processes of peaks A and B, the critical point (CP) analysis method is adopted to determine the attribute parameters of CPs A and B (corresponding to the exciton transition peaks A and B). By fitting the second-order differential spectra of $\varepsilon(E)$, the amplitude (*Amp*), the center energy $E_0$, the damping coefficient ($\Gamma$, refer to the full width at half maximum, FWHM), and the phase ($\phi$) of CPs A and B are precisely identified. Further procedures related to the CP analysis are available in the Supporting Information, and the fitting results are demonstrated in Figure S8 and Table S3.

The combining phenomenon of peaks A and B can be elaborated by analyzing the impurity-dependent evolutions of CP attribute parameters. By observing the CP parameters of #1 intrinsic sample, we find that the $Amp(\mathrm{B})/Amp(\mathrm{A}) \approx 4$ and $\Gamma(\mathrm{B})/\Gamma(\mathrm{A}) \approx 2.5$ (Figure 3c,d). In comparison, the ratios of CP parameters of #2–#4 samples are $Amp(\mathrm{B})/Amp(\mathrm{A}) > 10$ and $\Gamma(\mathrm{B})/\Gamma(\mathrm{A}) \approx 4$, suggesting the Nb doping is able to modulate the shape of peaks A and B. Further, with the increase of Nb mole fraction, the weakened exciton transition peak A gradually submerges and disappears in the enhanced and broadened exciton peak B, which can also explain the disappearance of A peak in the absorbance spectra (Figure 2b) to some extent. It is notable to mention that the absolute magnitude of *Amp* does not have a decisive reference value due to the inevitable slight influence from the quantum scale effect. Interestingly, the $E_0(\mathrm{A})$ gradually moves away from $E_0(\mathrm{B})$ with the increasing Nb mole fraction as shown in Figure 3e and Table S3. The splitting energy of CPs A and B ($E_0(\mathrm{B–A})$) expands from 100 meV to 140 meV as the mole fraction of Nb increases from 0 to 5.13%. When the percentage of dopant Nb increases to 7.59%, the value of $E_0(\mathrm{B–A})$ can no longer be given by CP analysis due to that the exciton transition peaks A and B have completely merged into an absorption peak A+B at 1.89 eV. On the basis of the above CP analysis, we speculate that the decreasing exciton peak A and the broadened exciton peak B dominate their combination and offset the negative effect from their diverging center energies. The aforementioned analysis can also be applied to interpret the impurity-induced shifts of PL (absorption) peaks shown in Figure 2. The impurity-induced blueshift of PL (absorption) peak B has not been reproduced in the exciton peak B of dielectric function spectra (Figure 3e). It can be partly interpreted as the influence of multiple transitions, which broadens and shifts the exciton peak B to some extent. However, the accompanied recombination radiations related to the non-degenerate subbands have been effectively screened during the process of collecting PL response of 2D $Mo_{1-x}Nb_xS_2$, which is conducive to observing the blueshift clearly.

Here, we want to briefly discuss the modulation efficacy of electrostatic doping and chemical doping on the basic optical constants of 2D TMDCs according to the doping-tunable optical constants of 2D $MoS_2$[59] and 2D $WS_2$[60]. As shown in Table S4, it seems that the electrostatic doping could tune the complex refractive index of 2D $MoS_2$ more effectively. In our work, the modulation efficacy of chemical doping on optical constants of 2D $MoS_2$ is evaluated based on the complex refractive indices and extrinsic carrier concentrations of #1 and #5 2D $Mo_{1-x}Nb_xS_2$. The hole concentration of #5 2D $Mo_{1-x}Nb_xS_2$ measured by Hall experiment is about $5.54\times10^{13}$ $cm^{-2}$. The lower modulation efficacy of chemical doping can be mainly attributed to the incomplete ionization of dopant Nb in 2D $MoS_2$ during the ellipsometric measurement, which may reduce the change ranges of optical constants to some extent.

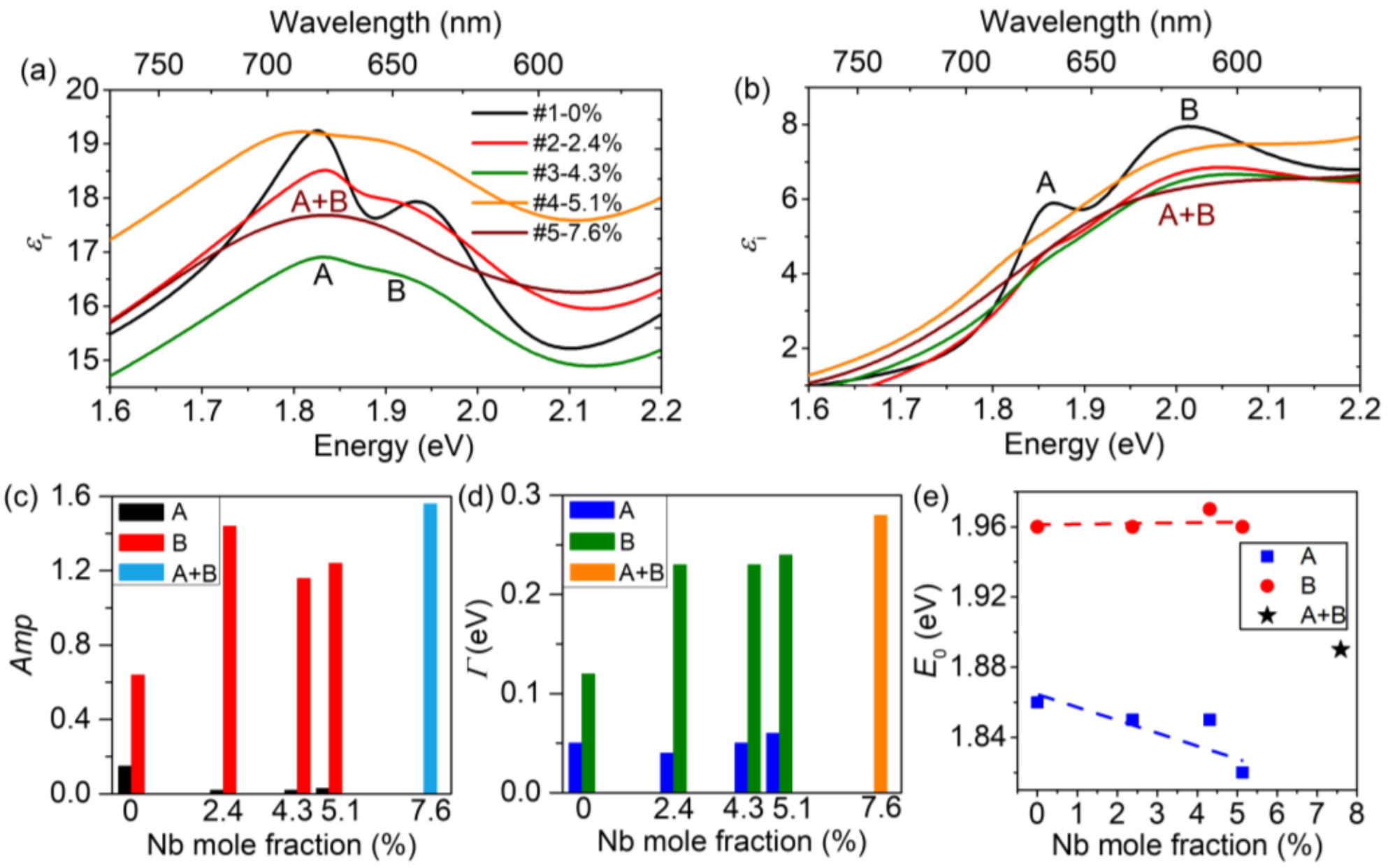

**Figure 3.** (a,b) Dielectric function spectra of #1–#5 2D $Mo_{1-x}Nb_xS_2$ samples over the energy region of 1.6–2.2 eV. (a) Real part $\varepsilon_r$ and (b) imaginary part $\varepsilon_i$. (c–e) Evolutions in the attribute parameters of CPs A, B with the Nb mole fraction. (c) Amplitude *Amp*, (d) damping coefficient $\Gamma$, and (e) center energy $E_0$.

To reveal the underlying physical origin of the distinctive combination phenomenon of exciton transition peaks A and B, the band structures and density of states (DOS) of 3L pristine and Nb-doped $MoS_2$ ($Mo_{26}NbS_{54}$, the Nb mole fraction is 3.7%) are calculated via first principles (Figure 4). The inset in Figure 4d demonstrates the crystal structure of $Mo_{26}NbS_{54}$, where one Mo atom in the $MoS_2$ supercell ($Mo_{27}S_{54}$) is substituted by an Nb atom. The best substitutional position of Nb atom can be optimally determined by making the total energy of $Mo_{26}NbS_{54}$ the lowest. As shown in Figure 4a,d, the exciton peaks A and B arise from direct transitions at K point of BZ from the spin-orbit-splitting VBs to the first CB.[61] Compared with the relatively clear band structure of intrinsic $MoS_2$, some additional VBs, especially a pair of newly emerging splitting VBs near the Fermi level ($E_F$), can be seen in the band structure of $Mo_{26}NbS_{54}$. It can be interpreted as the prominent impurity-induced VB reconfiguration. Some direct exciton transitions occurring between these two additional VBs and the first CB play important roles in broadening exciton peak B (Figure 3d). Interestingly, the effect of Nb dopant on CB is not so significant, which is different from the significant impurity-modulated VB renormalization. The DOS of $Mo_{26}NbS_{54}$ related to the exciton peak B (Figure 4e–g) is larger than that of intrinsic $MoS_2$ (Figure 4b,c), while the DOS of $Mo_{26}NbS_{54}$ associated with exciton peak A is smaller. It can partly explain the impurity-induced opposite changes of *Amp*(A) and *Amp*(B) (Figure 3c). The slight redshift of exciton transition peak A is related to the impurity-adjustable band shrinkage, mainly due to the prominent upward movement of VBs (Figure 3e). As shown in Figure 4d, the first VB of $Mo_{26}NbS_{54}$ has already crossed the $E_F$ as a result of combined contribution from Nb 4d state and Mo 4d state (Figure 4e,f),[62] indicating that $Mo_{26}NbS_{54}$ has turned into a p-type semiconductor.

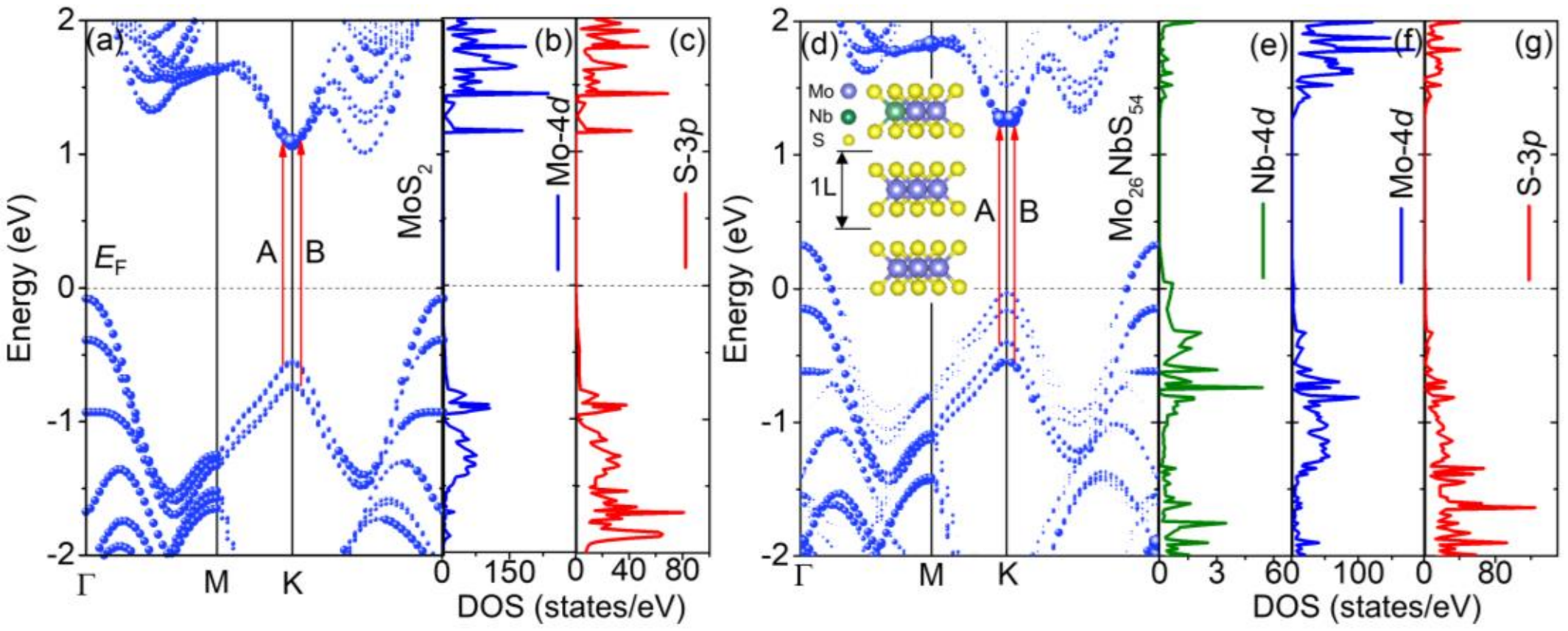

**Figure 4.** (a–c) Band structure and DOS of 3L intrinsic $MoS_2$. (d–g) Band structure and DOS of $Mo_{26}NbS_{54}$. The inset in (d) is the crystal structure of $Mo_{26}NbS_{54}$, and the size of solid spheres in (a,d) represents the strength of DOS.

Figure 5a intuitively shows the impurity-tunable combination and shift mechanisms occurring at exciton peaks A and B. The attenuation of peak A and the broadening of peak B promote the form of combined peak A+B. The doping-dependent shifts of peaks A and B in the dielectric function spectra, PL spectra, and absorbance are uniformly associated with the competition between the impurity-tunable band shrinkage and decreasing $E_b$.[53] Furthermore, the bandgap of 2D $Mo_{1-x}Nb_xS_2$ also exhibits an obvious decrease with the increasing Nb mole fraction as illustrated in Figure S9. It is directly related to the redshifted exciton transition peak A, and the physical cause is that the Nb dopants refactor the VB of 2D $Mo_{1-x}Nb_xS_2$ (Figure 4d). Compared to the electronic structure of pristine 2D $MoS_2$, the $E_F$ of 2D Nb-doped $MoS_2$ is closer to the VB (Figure 5b),[63] meaning that the 2D Nb-doped $MoS_2$ are conducive to fabricating p-type FETs. In our experiment, a 10–12L $Mo_{1-x}Nb_xS_2$ is used to prepare a p-type FET, the mole fraction of Nb dopant is about 8%. As shown in Figure 5c, the source (S) and drain (D) contact electrodes are Pb (40 nm)/Au (80 nm). The gate and gate dielectric are Al (80 nm) and $Al_2O_3$ (12 nm), respectively. The transfer characteristic of FET (Figure 5d) suggests it does attain the p-type conduction. The hole mobility and $R_c$ of the p-type FET are ~0.813 $cm^2/V{\cdot}s$ and ~$10^4$ $\Omega{\cdot}\mu m$. Theoretically, the SBH (Figure 5b) of FET can be reduced by increasing the Nb mole fraction of 2D $Mo_{1-x}Nb_xS_2$, and even an ohmic contact can be achieved. Two more FETs based on 2D $Mo_{1-x}Nb_xS_2$ are prepared to further verify the conduction type of 2D $Mo_{1-x}Nb_xS_2$. The transfer characteristic curves (Figure S10) suggest that the p-channel depletion mode FETs have been fabricated successfully, thereby the 2D $Mo_{1-x}Nb_xS_2$ synthesized here is a p-type semiconductor.

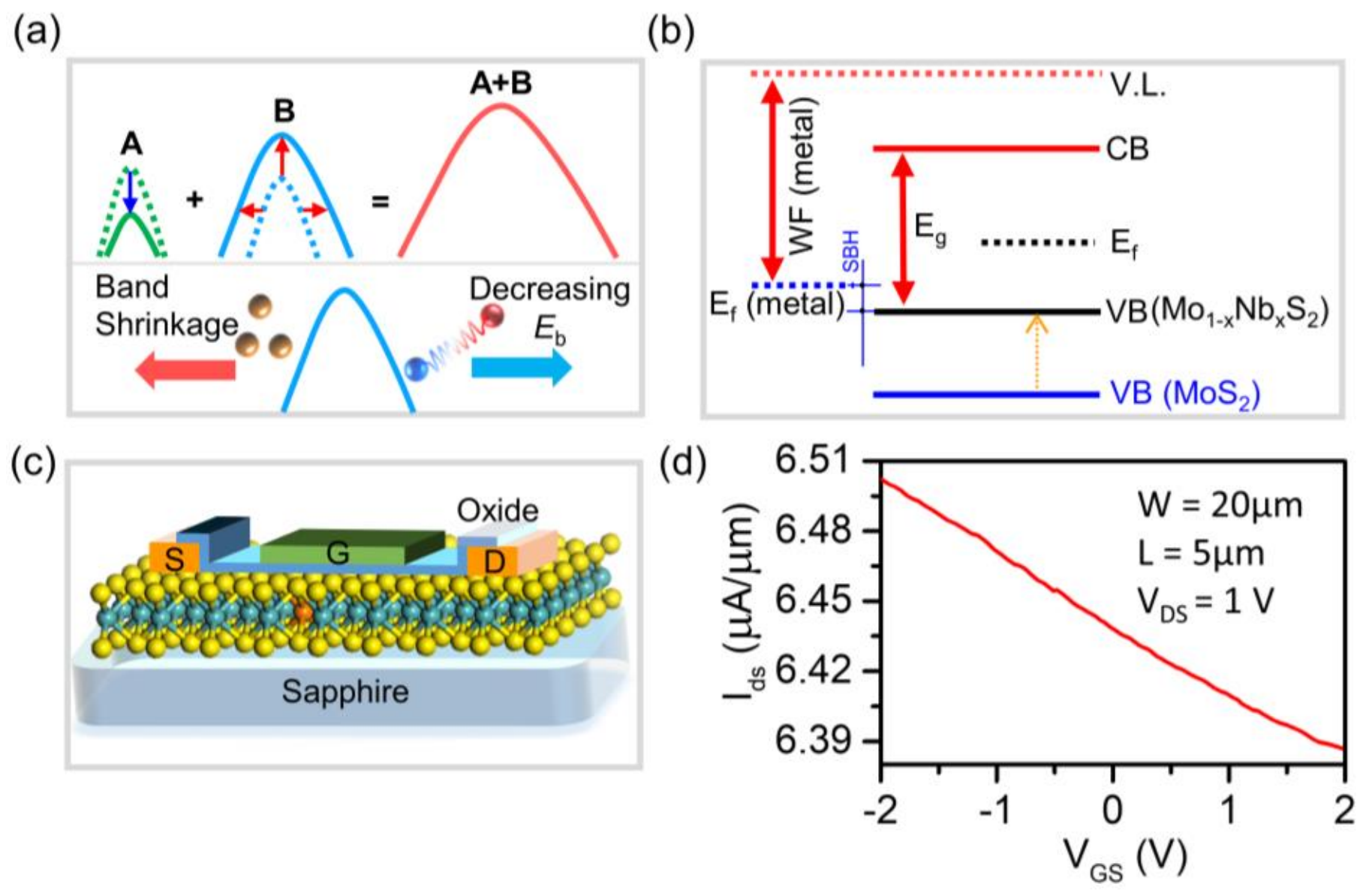

**Figure 5.** (a) The upper part shows the schematic diagram of the combination of exciton transition peaks A and B, and the bottom part presents the integrated influence of the impurity-tunable band shrinkage and the decreasing $E_b$ on the center energies of peaks A and B. (b) Schematic diagram of band alignments for the intrinsic $MoS_2$ and $Mo_{1-x}Nb_xS_2$ film with the contacted metal source or drain. (c) Schematic diagram and (d) transfer characteristic of FET based on 2D $Mo_{1-x}Nb_xS_2$.

## CONCLUSIONS

In summary, the impurity-tunable exciton transitions in 2D $Mo_{1-x}Nb_xS_2$ are investigated by combining SE, CP analysis, and first principles. We observe that the increasing Nb mole fraction can promote the first two exciton transition peaks A and B in the dielectric function spectra to combine into one broad peak A+B. In virtue of the CP analysis, we find that the direct causes of the impurity-tunable combination are the attenuation of exciton peak A and the broadening of exciton peak B. We interpret the impurity-induced broadening of exciton peak B as the multiple transitions from Nb impurity-induced VBs to the CB minimum. Additionally, the center energy difference of exciton peak A and B $E_0$(B–A) gradually increases with the Nb mole fraction. The diverging shifts of exciton transition peaks A and B are also discerned from the PL and absorbance spectra. We explain the distinctive shifts as the competition between the impurity-modulated band shrinkage and the decrease of exciton binding energy $E_b$. We have prepared a p-type FET with the 2D Nb-doped $MoS_2$ and measured its hole mobility and contact resistance, experimentally confirming the intriguing application prospects of 2D $MoS_2$ with p-type conductivity. Our study provides the "smoking gun" experimental evidence for the doping-tunable exciton transitions in 2D TMDCs, and is also instructive for the innovative design and performance optimization of p-type devices based on 2D extrinsic TMDCs.

## EXPERIMENTAL METHODS

### Sample Preparation

2D $Mo_{1-x}Nb_xS_2$ wafers were achieved by using an innovative two-step method. First, ultrathin $Mo_{1-x}Nb_xO_y$ films are deposited on 2" *c*-face sapphire substrates in a two-targets (Mo and Nb) co-sputtering system. The Nb mole fraction can be precisely controlled by the gun power of Nb as well as the shutter opening time (Table S1). Then the second step was to sulfurize the $Mo_{1-x}Nb_xO_y$ films by using $H_2S$ (10% with Ar) gaseous source in a furnace at 800℃/30 min into 2D $Mo_{1-x}Nb_xS_2$ under the pressure of 200 Torr.

### Sample Characterization

To check the quality of 2D $Mo_{1-x}Nb_xS_2$ films, an Excalab 250Xi X-ray Photoelectron Spectrometer with Al-Kα radiation source (1486.6 eV) was used to analyze the elemental composition of the 2D $Mo_{1-x}Nb_xS_2$ films. The thicknesses of the 2D $Mo_{1-x}Nb_xS_2$ films were measured by a TEM system (JEOL ARM200F). The Raman spectra of the 2D $Mo_{1-x}Nb_xS_2$ specimens were measured by using Argon ion laser Raman spectrometer (LabRAM HR800, Horiba JobinYvon) with a 532 nm line. A commercial spectroscopic ellipsometer (ME-L

Mueller matrix ellipsometer, Wuhan Eoptics Technology Co., Wuhan, China) was used to measure the ellipsometric spectra of the 2D $Mo_{1-x}Nb_xS_2$ specimens, whose energy region is 0.73–6.42 eV,[51–56] and the multi-angle measurement mode (60°, 66°, and 72°) was optimally selected in the experiment.

**First-principle Calculations**

Standard *ab* initio simulations within the density function theory (DFT) were performed by using the Vienna Ab initio Simulation Package (VASP v5.4.1),[64] which has been proved to be workable to understand the experimental phenomenon in various 2D materials.[24] These calculations were performed using projector augmented wave (PAW) potentials based on the Perdew-Burke-Ernzerhof (PBE)[65] exchange-correlation functional. The vacuum slabs of 15 Å were adopted in the geometric model to avoid interactions between adjacent atom layers, and the Van der Waals correction vdW-D2 is taken into consideration. In the geometrical optimization, the kinetic energy cut-off is set at 550 eV and Brillouin zone is sampled with a $3 \times 3 \times 1$ Γ-centered k-point mesh. The criterion for the convergence of forces and total energy are set to 0.01eV/Å and $10^{-5}$eV. In the band structure calculation, the BandUP code is used to deal with the band folding phenomenon shown in the supercell.[66,67] The calculations consider the influence of the SOC effect.

**ASSOCIATED CONTENT**

**Supporting Information**

The Supporting Information is available free of charge on the ACS Publications website at DOI: xx.xxxx/xxxxxxx.xxxxxxx.

TEM images and thicknesses of #3–#5 samples; XPS measurement and analysis results of #3–#5 samples; optical model of 2D $Mo_{1-x}Nb_xS_2$ film on the *c*-sapphire substrate;

ellipsometric analysis procedures and completed dielectric functions and complex refractive indices of #1–#5 samples over the energy range of 0.73–6.42 eV; critical point analysis results, absorption coefficients and bandgaps of #1–#5 samples; specific process parameters of 2D $Mo_{1-x}Nb_xS_2$ preparation.

## AUTHOR INFORMATION

### Corresponding Author

*E-mail: hongganggu@hust.edu.cn

*E-mail: chia500@nctu.edu.tw

*E-mail: shyliu@hust.edu.cn

### Notes

The authors declare no competing financial interest.

## ACKNOWLEDGMENTS

This work was funded by the National Natural Science Foundation of China (Grant Nos. 51727809, 51805193, and 51525502), the China Postdoctoral Science Foundation (Grant Nos. 2017T100546), and the National Science and Technology Major Project of China (Grant No. 2017ZX02101006-004). B.S. acknowledges the support from China Scholarship Council (CSC). Y.-T.H. acknowledges the financial support from the "Center for the Semiconductor Technology Research" (MOE) and the partial support from the Ministry of Science and Technology, Taiwan (Grant Nos. MOST 109-2634-F-009-029). The authors also thank the technical support from the Experiment Center for Advanced Manufacturing and Technology in School of Mechanical Science and Engineering of HUST.

# Supporting Information

## 2D Nb-doped $MoS_2$: Tuning the Exciton Transitions and Application to p-type FETs

Baokun Song,[†,#] Honggang Gu,[†,#,*] Mingsheng Fang,[†] Zhengfeng Guo,[†] Yen-Teng Ho,[*,‡] Xiuguo Chen,[†] Hao Jiang,[†] Shiyuan Liu[*,†]

[†]State Key Laboratory of Digital Manufacturing Equipment and Technology, Huazhong University of Science and Technology, Wuhan 430074, China.

[‡]International College of Semiconductor Technology, National Chiao Tung University, Hsinchu 30010, Taiwan.

[#] These authors contributed equally to this work.

[*]E-mail: hongganggu@hust.edu.cn
[*]E-mail: chia500@nctu.edu.tw
[*]E-mail: shyliu@hust.edu.cn

## HRTEM

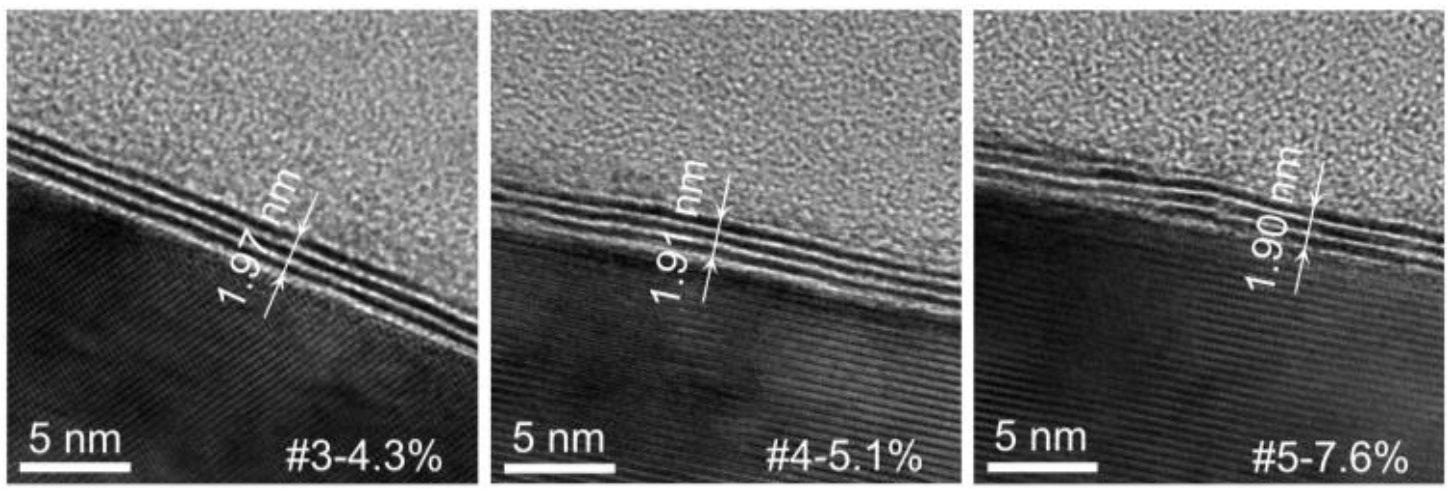

**Figure S1.** Cross-sectional crystal morphology of #3–#5 2D $Mo_{1-x}Nb_xS_2$ samples measured by high-resolution transmission electron microscopy (HRTEM).

## XPS

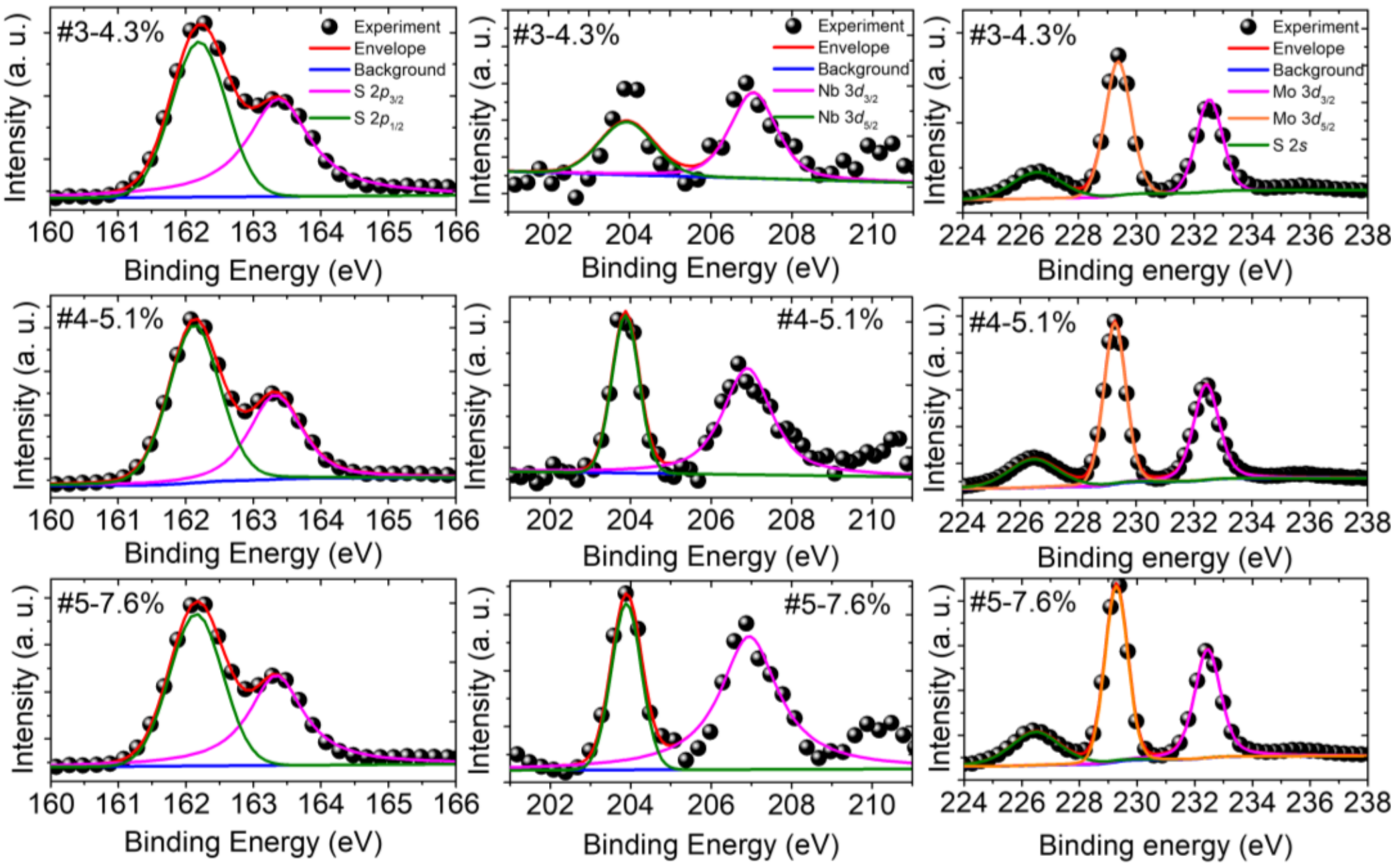

**Figure S2.** Measured XPS spectra and corresponding peak-differentiation-imitating analysis results of #3–#5 2D $Mo_{1-x}Nb_xS_2$ samples.

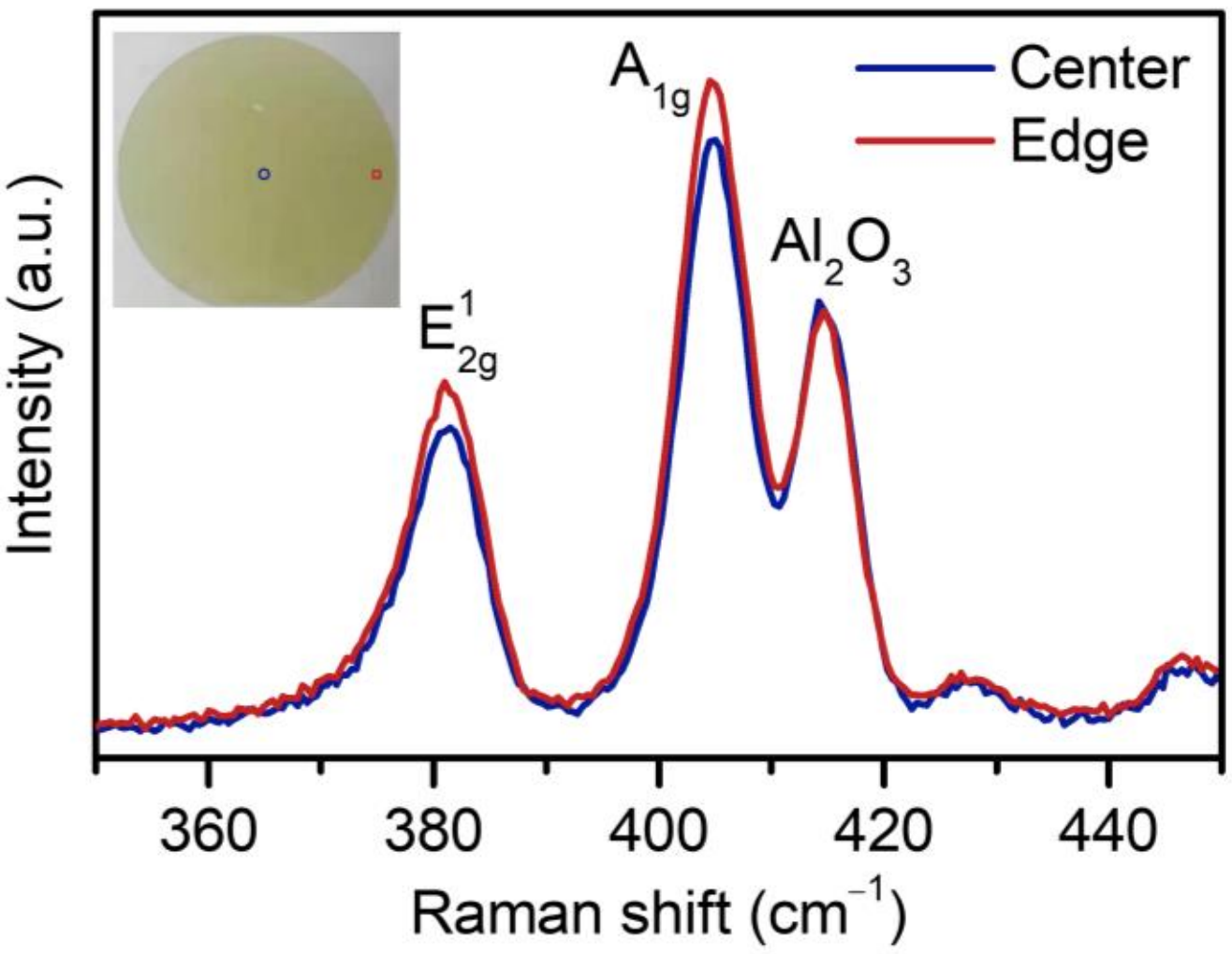

**Figure S3.** Raman spectra of 2D $Mo_{1-x}Nb_xS_2$ wafer. The detection regions are marked by the blue circle and open brown square.

**Ellipsometric analysis**

Ellipsometry is a model-based measurement technique.[1–4] Thus, a vertical stacking optical model (Figure S4) is established to approximately describe the geometric structure of the 2D $Mo_{1-x}Nb_xS_2$ film on the sapphire substrate. The optical model contains four layers, including the ambient air layer, 2D $Mo_{1-x}Nb_xS_2$ film, interlayer, and the sapphire substrate. To physically embody the dielectric properties of 2D $Mo_{1-x}Nb_xS_2$, the dielectric function is parameterized by a combined generalized oscillator model.[5,6] After constructing the optical model and dielectric function model of 2D $Mo_{1-x}Nb_xS_2$, the optical interference theory in multilayer films is used to calculate the theoretical ellipsometric spectra [$\Psi(E)$, $\Delta(E)$] of 2D $Mo_{1-x}Nb_xS_2$ samples. The dielectric functions and thicknesses of the 2D $Mo_{1-x}Nb_xS_2$ films can be simultaneously extracted by fitting the measured ellipsometric spectra with theoretically calculated ones. As shown in Figure S5, analysis results exhibit excellent goodness of fit. Further information on the ellipsometric analysis of low-dimensional films are available in the previous publications.[2,3,7,8]

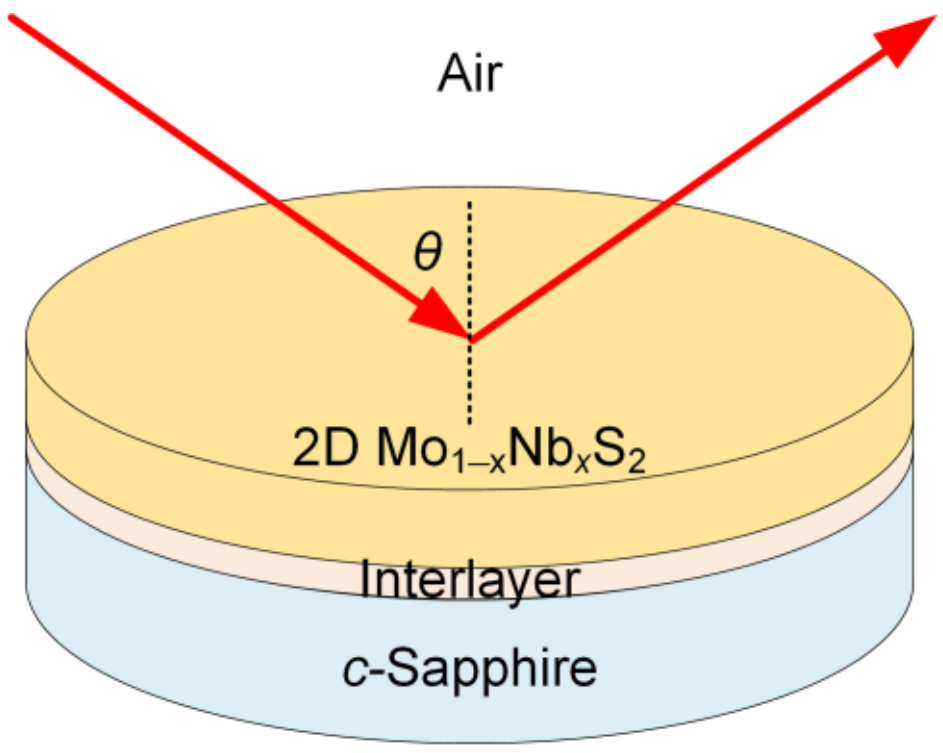

**Figure S4.** Optical model of 2D $Mo_{1-x}Nb_xS_2$ sample.

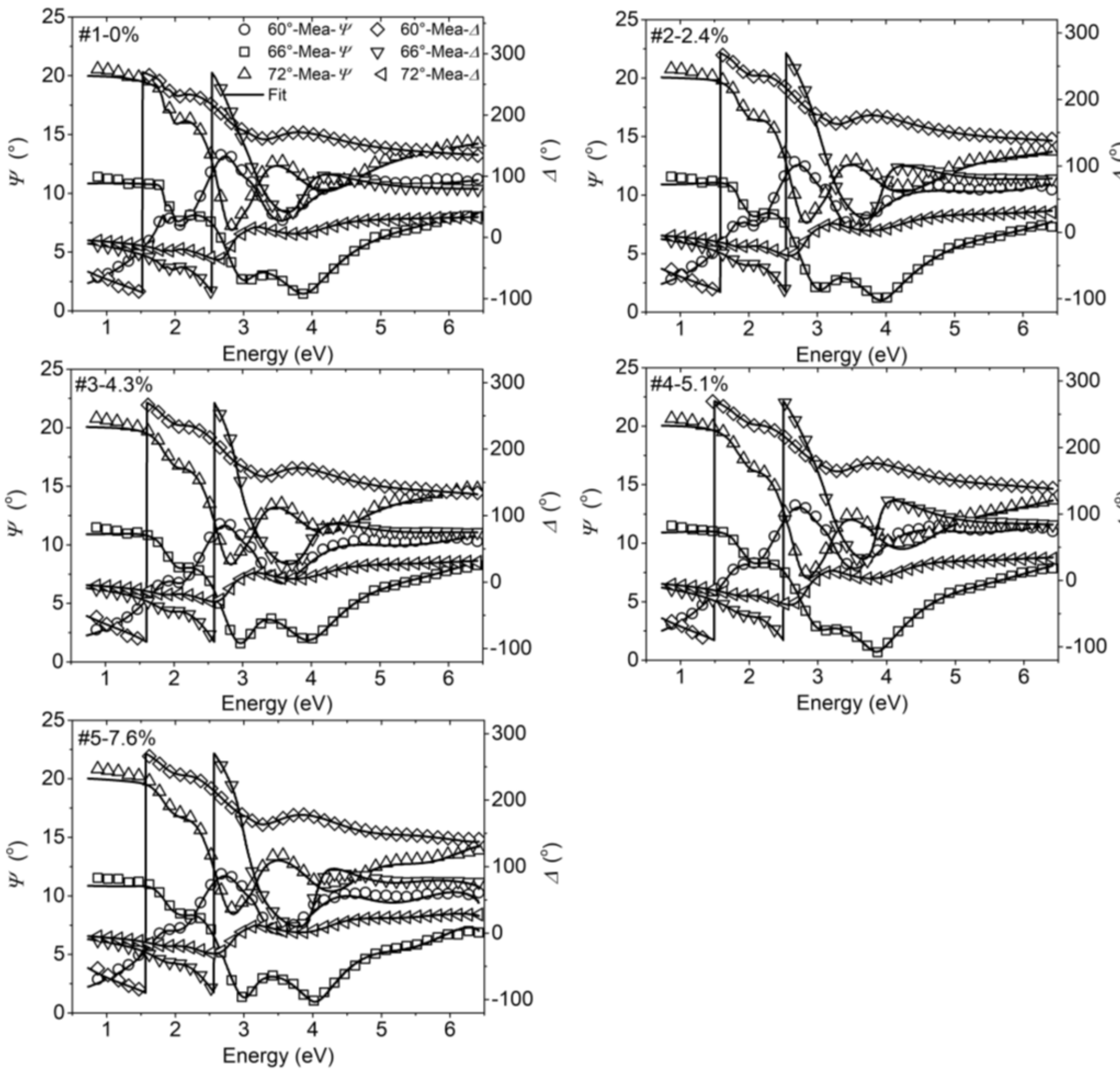

**Figure S5.** Ellipsometric analysis results of #1–#5 2D $Mo_{1-x}Nb_xS_2$ samples, the incident angles are 60°, 66°, and 72°.

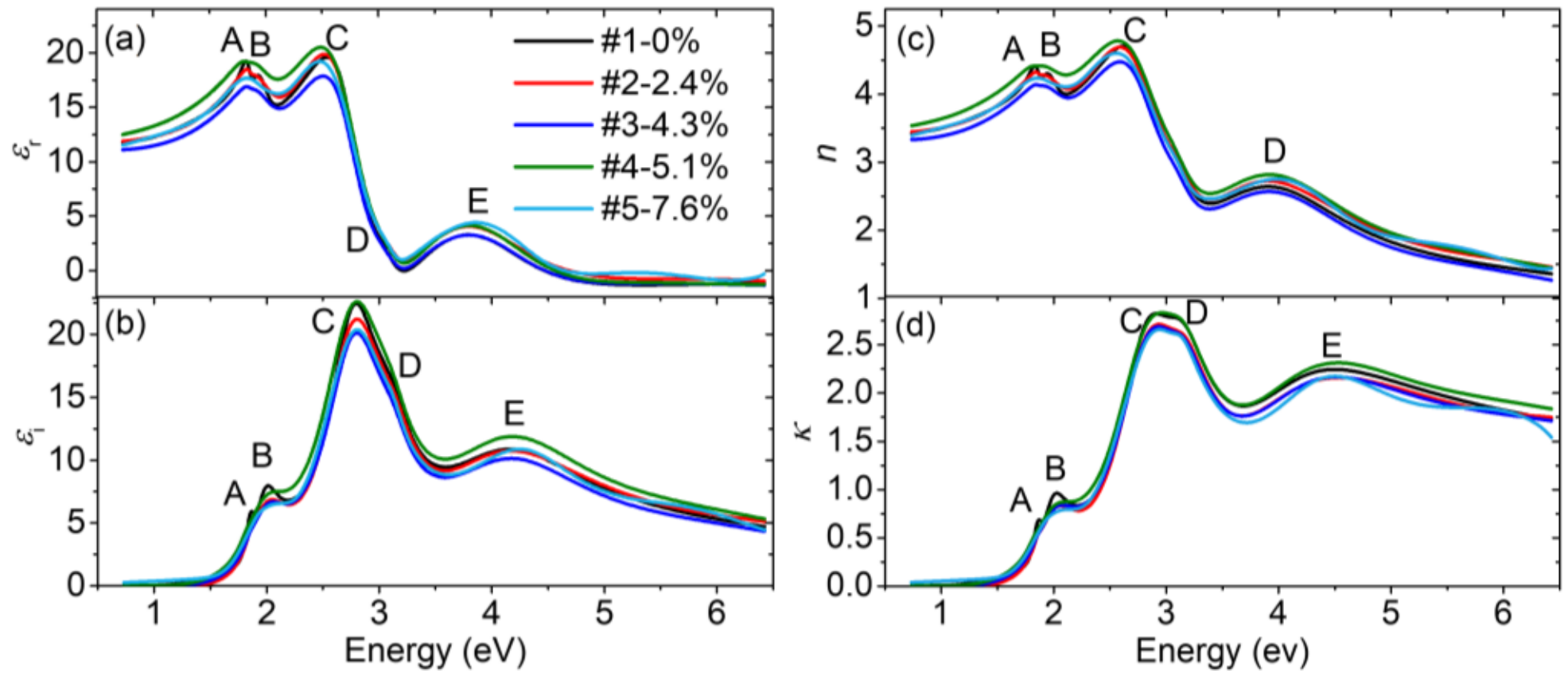

**Figure S6.** Dielectric functions and complex refractive indices of #1–#5 2D $Mo_{1-x}Nb_xS_2$ samples over the energy range of 0.73–6.42 eV.

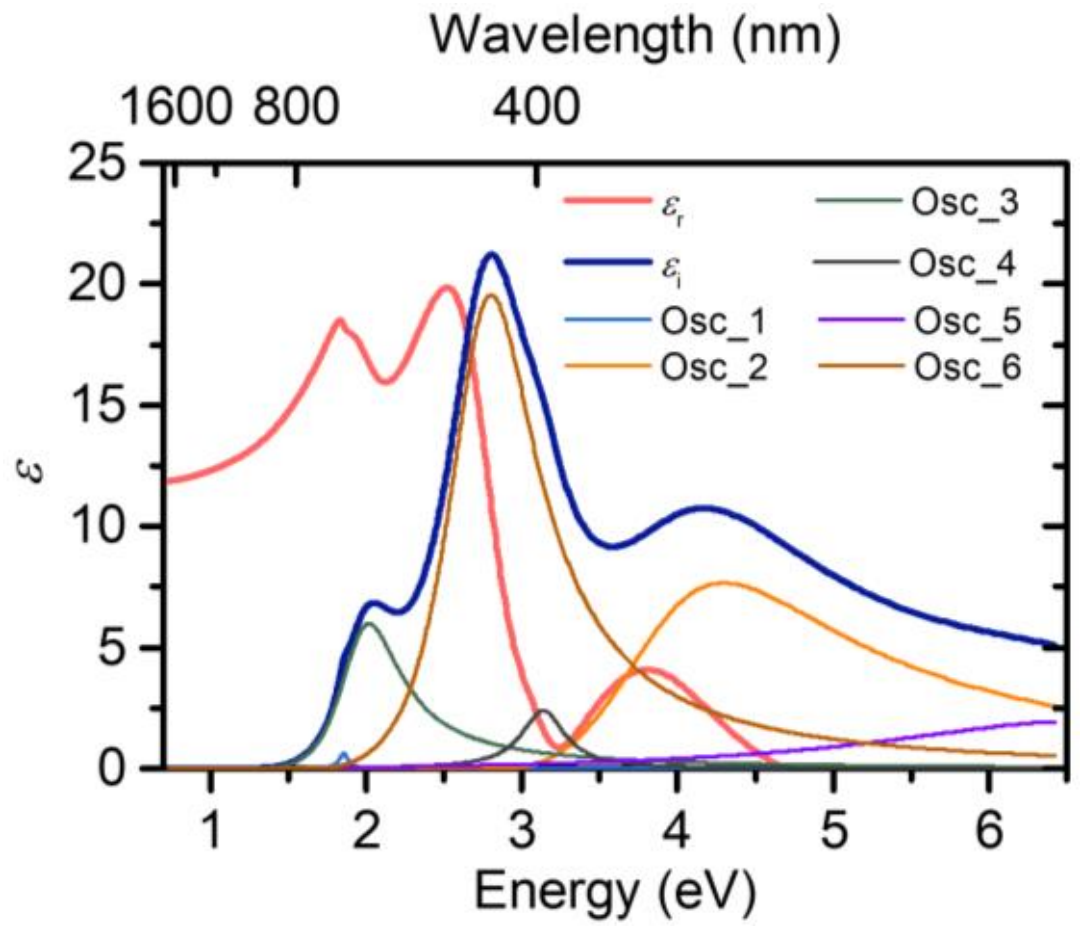

**Figure S7.** The peak-differentiation-imitating fitting results of dielectric function spectra of #2 2D $Mo_{1-x}Nb_xS_2$ film.

**Critical point analysis**

To analyze the physical origins of the combination phenomenon occurring at the exciton peaks A and B in the dielectric function spectra of 2D $Mo_{1-x}Nb_xS_2$, critical point (CP) analysis method is used to precisely determine the characteristic parameters of the CPs A and B.[1] In the CP analysis, the second derivative of the dielectric function spectrum is fitted with the following theoretical formulas

$$d^2\varepsilon / dE^2 = \begin{cases} m(m-1) Amp \exp(i\phi)(E - E_0 + i\Gamma)^{m-2} & (m \neq 0) \\ Amp \exp(i\phi)(E - E_0 + i\Gamma)^{-2} & (m = 0) \end{cases} \quad \text{(S1)}$$

here, *Amp*, $\phi$, $E_0$, and $\Gamma$ refer to the amplitude, the phase, the center energy, and the damping coefficient of the CP, and *m* stands for the dimension of the wave vector. According to the prior report,[9] we set $m = -1$ to describe the excitonic behaviour of the optical transitions occurring at CPs A and B. The second derivative spectra of the dielectric functions and the best fitting curves (Figure S8) are in well agreement with each other. The specific parameters in Equation S1 have been listed in Table S3. The CPs A and B of #5 sample cannot be distinguished by the CP analysis method due to that these two peaks have combined into one peak A+B at about 1.89 eV.

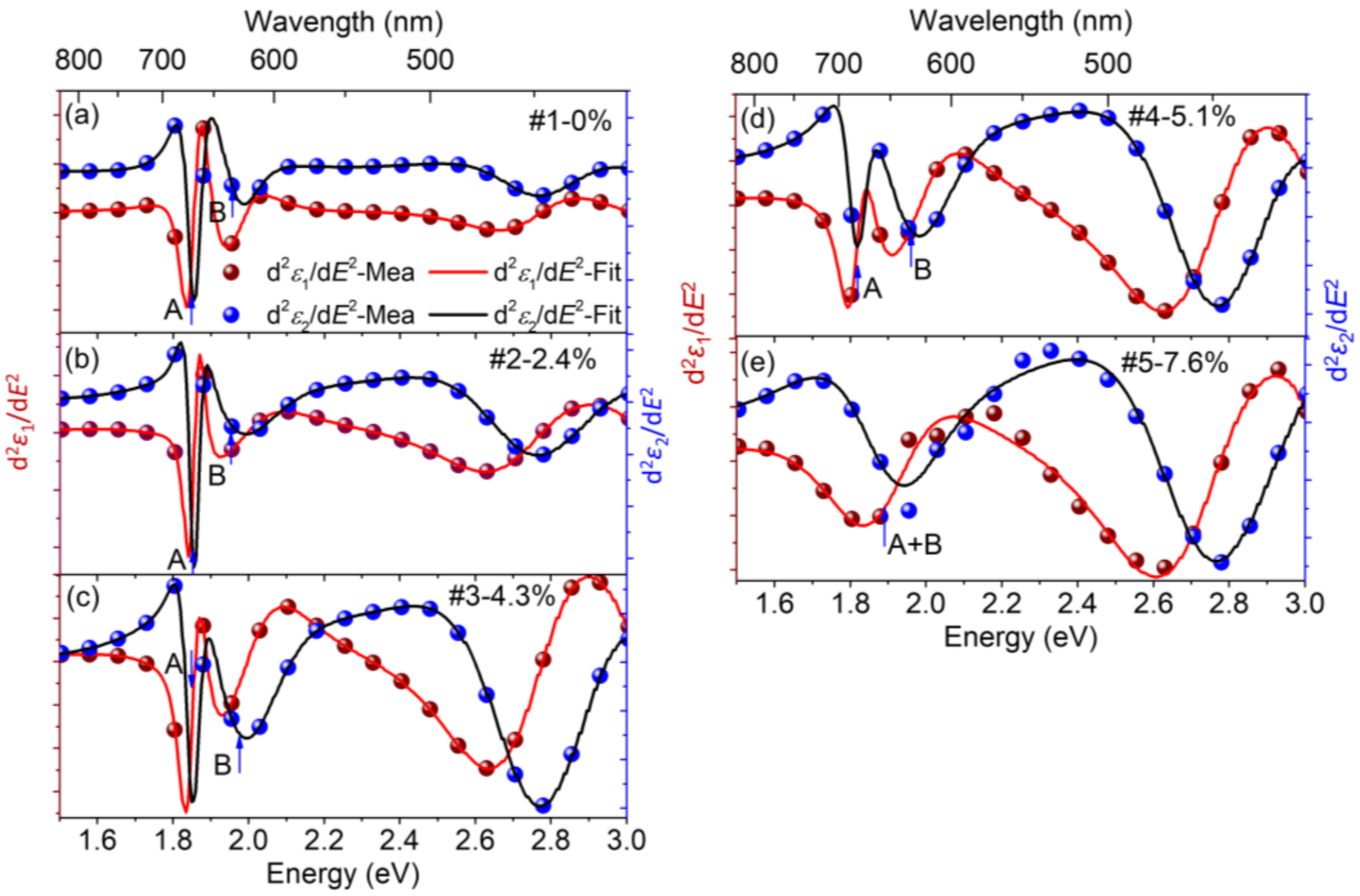

**Figure S8.** CP analysis results of #1 – #5 2D $Mo_{1-x}Nb_xS_2$ samples.

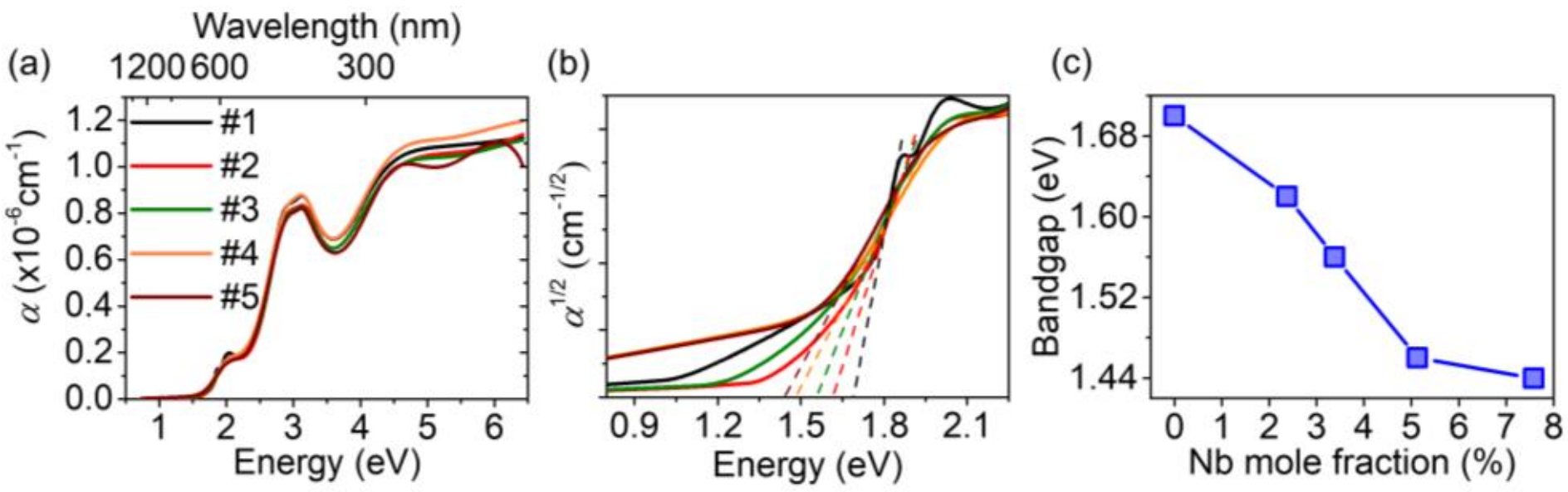

**Figure S9.** (a) Absorption coefficients of #1–#5 2D $Mo_{1-x}Nb_xS_2$ samples. (b) Bandgap analysis of #1–#5 2D $Mo_{1-x}Nb_xS_2$ samples by Tauc-plot method. (c) Evolution of bandgaps of 2D $Mo_{1-x}Nb_xS_2$ with the concentration of dopant Nb.

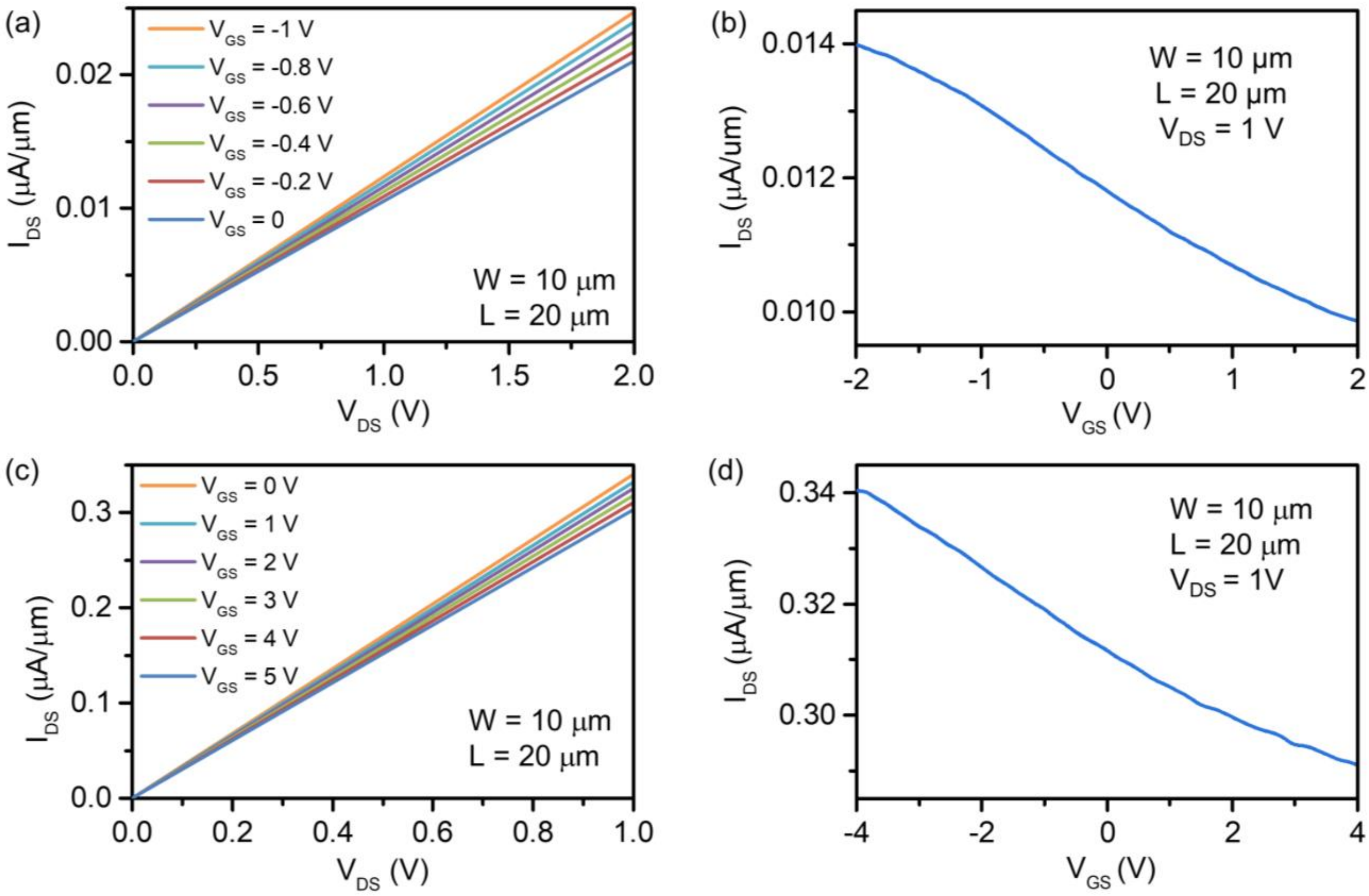

**Figure S10.** Transfer characteristic curves of FETs based on 2D $Mo_{1-x}Nb_xS_2$.

**Table S1.** Process parameters of two-step method

| No. | #1 | #2 | #3 | #4 | #5 |
|---|---|---|---|---|---|
| Nb power (RF)/W | 0 | 30 | 30 | 40 | 50 |
| Mo Power (DC)/W | 100 | 100 | 100 | 100 | 100 |
| Ar/sccm | 15 | 15 | 15 | 15 | 15 |

| | | | | | | |
|---|---|---|---|---|---|---|
| $O_2$/sccm | | 15 | 15 | 15 | 15 | 15 |
| Sputtering Time/min | Mo | 1 | 1 | 1 | 1 | 1 |
| | Nb | 0 | 0.5 | 1 | 1 | 1 |
| Sulfuration process | Pressure: 200 Torr; Volume ratio of $H_2S$ and Ar is 1:9; Temperature program: Raise the temperature to 650 ℃ at a rate of +20 ℃ and hold for 30 minutes, then raise to 750℃ and hold for 60 minutes. Finally, cool the sample to room temperature naturally. | | | | | |

**Table S2.** Thicknesses of 2D $Mo_{1-x}Nb_xS_2$

| No. | #1 | #2 | #3 | #4 | #5 |
|---|---|---|---|---|---|
| TEM (nm) | 2.11 | 1.91 | 1.97 | 1.91 | 1.90 |
| SE (nm) | 1.95 | 1.90 | 1.92 | 1.91 | 1.88 |
| Interlayer (nm) | 0.62 | 0.57 | 0.55 | 0.61 | 0.60 |

**Table S3.** CP parameters of 2D $Mo_{1-x}Nb_xS_2$

| Parameters | CPs | #1 | #2 | #3 | #4 | #5 |
|---|---|---|---|---|---|---|
| *Amp* | A | 0.15 | 0.02 | 0.02 | 0.03 | 1.56 |
| | B | 0.64 | 1.44 | 1.16 | 1.24 | |
| $E_0$ (eV) | A | 1.86 | 1.85 | 1.85 | 1.82 | 1.89 |
| | B | 1.96 | 1.96 | 1.97 | 1.96 | |
| $E_0$(B–A) (meV) | | 100 | 110 | 120 | 140 | |
| $\Gamma$ (eV) | A | 0.05 | 0.04 | 0.05 | 0.06 | 0.28 |
| | B | 0.12 | 0.23 | 0.23 | 0.24 | |
| $\phi$ (°) | A | 160.21 | 160.18 | 160.17 | 160.20 | 159.4 |
| | B | 165.68 | 266.54 | 266.66 | 266.68 | 0 |

**Table S4.** Change rates of complex refractive indices of 2D TMDCs with the carrier concentrations

| Works | Kravets et al.[10] | | Yu et al.[11] | | Ours | |
|---|---|---|---|---|---|---|
| 2D materials | 1L $MoS_2$ | | 1L $WS_2$ | | 3L $MoS_2$ | |
| Doping processes | Electrostatic doping | | Electrostatic doping | | Chemical doping | |
| Energy positions | A | B | @1.92 eV | @1.95 eV | A | B |
| $\Delta n/\Delta N$ ($\times10^{-14}$cm$^2$) | ≈ –2.18 | ≈ –2.18 | | | | |
| $\Delta\kappa/\Delta N$ ($\times 10^{-14}$cm$^2$) | ≈ –2.59 | ≈ –2.45 | | | | |
| $\Delta n/\Delta N_e$ ($\times 10^{-14}$cm$^2$) | | | ≈ –9.30 | | | |
| $\Delta\kappa/\Delta N_e$ ($\times 10^{-14}$cm$^2$) | | | | ≈ –7.62 | | |
| $\Delta n/\Delta N_h$ ($\times 10^{-14}$cm$^2$) | | | ≈ –5.00 | | ≈ –0.50 | ≈ –0.32 |
| $\Delta\kappa/\Delta N_h$ ($\times 10^{-14}$cm$^2$) | | | | ≈ –7.31 | ≈ –0.24 | ≈ –0.35 |

*$\Delta N$, $\Delta N_e$, and $\Delta N_h$ refer to the changes of the total carrier, electron, and hole concentrations; A and B denote the exiciton transition peaks in dielectric function spectra of 2D $MoS_2$.

**Table S5.** Complex refractive indices of 2D $Mo_{1-x}Nb_xS_2$

| **Wavelength (nm)** | **#1 ($x$ = 0)** | | **#2 ($x$ = 0.24)** | | **#3 ($x$ = 0.43)** | | **#4 ($x$ = 0.51)** | | **#5 ($x$ = 0.76)** | |
|---|---|---|---|---|---|---|---|---|---|---|
| | ***n*** | ***κ*** | ***n*** | ***κ*** | ***n*** | ***κ*** | ***n*** | ***κ*** | ***n*** | ***κ*** |
| 193 | 1.36 | 1.73 | 1.46 | 1.75 | 1.27 | 1.71 | 1.45 | 1.84 | 1.44 | 1.54 |
| 194 | 1.37 | 1.73 | 1.47 | 1.75 | 1.28 | 1.72 | 1.46 | 1.84 | 1.43 | 1.59 |
| 195 | 1.37 | 1.74 | 1.48 | 1.76 | 1.29 | 1.72 | 1.47 | 1.85 | 1.43 | 1.62 |
| 196 | 1.38 | 1.75 | 1.49 | 1.76 | 1.30 | 1.73 | 1.48 | 1.86 | 1.44 | 1.65 |
| 197 | 1.39 | 1.75 | 1.50 | 1.76 | 1.31 | 1.73 | 1.49 | 1.86 | 1.44 | 1.68 |
| 198 | 1.40 | 1.76 | 1.50 | 1.76 | 1.32 | 1.74 | 1.50 | 1.87 | 1.45 | 1.70 |
| 199 | 1.40 | 1.77 | 1.51 | 1.77 | 1.33 | 1.74 | 1.51 | 1.88 | 1.46 | 1.72 |
| 200 | 1.41 | 1.78 | 1.52 | 1.77 | 1.34 | 1.75 | 1.52 | 1.88 | 1.48 | 1.74 |
| 201 | 1.42 | 1.78 | 1.53 | 1.77 | 1.35 | 1.75 | 1.53 | 1.89 | 1.49 | 1.76 |
| 202 | 1.42 | 1.79 | 1.53 | 1.78 | 1.36 | 1.76 | 1.54 | 1.89 | 1.50 | 1.77 |
| 203 | 1.43 | 1.80 | 1.54 | 1.78 | 1.37 | 1.76 | 1.55 | 1.90 | 1.52 | 1.78 |
| 204 | 1.44 | 1.80 | 1.55 | 1.79 | 1.37 | 1.77 | 1.55 | 1.91 | 1.53 | 1.79 |
| 205 | 1.44 | 1.81 | 1.56 | 1.79 | 1.38 | 1.77 | 1.56 | 1.91 | 1.55 | 1.80 |
| 206 | 1.45 | 1.82 | 1.56 | 1.79 | 1.39 | 1.78 | 1.57 | 1.92 | 1.56 | 1.81 |
| 207 | 1.46 | 1.83 | 1.57 | 1.80 | 1.40 | 1.78 | 1.58 | 1.92 | 1.58 | 1.82 |
| 208 | 1.47 | 1.83 | 1.58 | 1.80 | 1.41 | 1.79 | 1.59 | 1.93 | 1.59 | 1.82 |
| 209 | 1.47 | 1.84 | 1.58 | 1.81 | 1.42 | 1.79 | 1.60 | 1.93 | 1.61 | 1.83 |
| 210 | 1.48 | 1.85 | 1.59 | 1.81 | 1.42 | 1.80 | 1.61 | 1.94 | 1.62 | 1.83 |
| 211 | 1.49 | 1.85 | 1.60 | 1.82 | 1.43 | 1.80 | 1.62 | 1.95 | 1.63 | 1.84 |
| 212 | 1.50 | 1.86 | 1.60 | 1.82 | 1.44 | 1.81 | 1.63 | 1.95 | 1.65 | 1.84 |
| 213 | 1.50 | 1.87 | 1.61 | 1.83 | 1.45 | 1.81 | 1.64 | 1.96 | 1.66 | 1.84 |
| 214 | 1.51 | 1.88 | 1.62 | 1.83 | 1.46 | 1.82 | 1.64 | 1.96 | 1.67 | 1.84 |
| 215 | 1.52 | 1.88 | 1.62 | 1.84 | 1.46 | 1.82 | 1.65 | 1.97 | 1.68 | 1.84 |
| 216 | 1.53 | 1.89 | 1.63 | 1.85 | 1.47 | 1.83 | 1.66 | 1.97 | 1.70 | 1.85 |
| 217 | 1.53 | 1.90 | 1.64 | 1.85 | 1.48 | 1.83 | 1.67 | 1.98 | 1.71 | 1.85 |
| 218 | 1.54 | 1.91 | 1.65 | 1.86 | 1.49 | 1.84 | 1.68 | 1.99 | 1.72 | 1.85 |
| 219 | 1.55 | 1.91 | 1.65 | 1.86 | 1.49 | 1.84 | 1.69 | 1.99 | 1.73 | 1.85 |
| 220 | 1.56 | 1.92 | 1.66 | 1.87 | 1.50 | 1.85 | 1.70 | 2.00 | 1.74 | 1.85 |
| 221 | 1.57 | 1.93 | 1.67 | 1.88 | 1.51 | 1.86 | 1.70 | 2.00 | 1.75 | 1.85 |
| 222 | 1.57 | 1.94 | 1.67 | 1.88 | 1.52 | 1.86 | 1.71 | 2.01 | 1.75 | 1.85 |
| 223 | 1.58 | 1.94 | 1.68 | 1.89 | 1.52 | 1.87 | 1.72 | 2.02 | 1.76 | 1.85 |
| 224 | 1.59 | 1.95 | 1.69 | 1.90 | 1.53 | 1.87 | 1.73 | 2.02 | 1.77 | 1.85 |
| 225 | 1.60 | 1.96 | 1.70 | 1.91 | 1.54 | 1.88 | 1.74 | 2.03 | 1.78 | 1.85 |
| 226 | 1.61 | 1.97 | 1.71 | 1.91 | 1.55 | 1.89 | 1.75 | 2.04 | 1.78 | 1.85 |
| 227 | 1.62 | 1.97 | 1.71 | 1.92 | 1.55 | 1.89 | 1.75 | 2.04 | 1.79 | 1.85 |
| 228 | 1.62 | 1.98 | 1.72 | 1.93 | 1.56 | 1.90 | 1.76 | 2.05 | 1.80 | 1.86 |
| 229 | 1.63 | 1.99 | 1.73 | 1.93 | 1.57 | 1.91 | 1.77 | 2.06 | 1.80 | 1.86 |
| 230 | 1.64 | 2.00 | 1.74 | 1.94 | 1.58 | 1.91 | 1.78 | 2.06 | 1.81 | 1.86 |
| 231 | 1.65 | 2.00 | 1.75 | 1.95 | 1.59 | 1.92 | 1.79 | 2.07 | 1.81 | 1.86 |
| 232 | 1.66 | 2.01 | 1.76 | 1.96 | 1.59 | 1.93 | 1.80 | 2.08 | 1.82 | 1.87 |
| 233 | 1.67 | 2.02 | 1.77 | 1.96 | 1.60 | 1.93 | 1.81 | 2.08 | 1.82 | 1.87 |
| 234 | 1.68 | 2.03 | 1.78 | 1.97 | 1.61 | 1.94 | 1.82 | 2.09 | 1.83 | 1.87 |
| 235 | 1.69 | 2.04 | 1.79 | 1.98 | 1.62 | 1.95 | 1.82 | 2.10 | 1.83 | 1.88 |
| 236 | 1.70 | 2.04 | 1.80 | 1.99 | 1.63 | 1.96 | 1.83 | 2.11 | 1.84 | 1.88 |
| 237 | 1.71 | 2.05 | 1.80 | 1.99 | 1.64 | 1.96 | 1.84 | 2.11 | 1.84 | 1.89 |
| 238 | 1.72 | 2.06 | 1.82 | 2.00 | 1.65 | 1.97 | 1.85 | 2.12 | 1.85 | 1.89 |
| 239 | 1.73 | 2.07 | 1.83 | 2.01 | 1.66 | 1.98 | 1.86 | 2.13 | 1.85 | 1.90 |
| 240 | 1.74 | 2.07 | 1.84 | 2.01 | 1.67 | 1.99 | 1.87 | 2.14 | 1.86 | 1.91 |
| 241 | 1.75 | 2.08 | 1.85 | 2.02 | 1.68 | 1.99 | 1.88 | 2.14 | 1.86 | 1.91 |
| 242 | 1.76 | 2.09 | 1.86 | 2.03 | 1.69 | 2.00 | 1.89 | 2.15 | 1.87 | 1.92 |
| 243 | 1.77 | 2.09 | 1.87 | 2.04 | 1.70 | 2.01 | 1.91 | 2.16 | 1.87 | 1.93 |
| 244 | 1.78 | 2.10 | 1.88 | 2.04 | 1.71 | 2.02 | 1.92 | 2.17 | 1.88 | 1.94 |
| 245 | 1.79 | 2.11 | 1.89 | 2.05 | 1.72 | 2.02 | 1.93 | 2.17 | 1.88 | 1.95 |
| 246 | 1.81 | 2.12 | 1.90 | 2.06 | 1.73 | 2.03 | 1.94 | 2.18 | 1.89 | 1.96 |
| 247 | 1.82 | 2.12 | 1.92 | 2.06 | 1.74 | 2.04 | 1.95 | 2.19 | 1.90 | 1.96 |
| 248 | 1.83 | 2.13 | 1.93 | 2.07 | 1.75 | 2.05 | 1.96 | 2.20 | 1.91 | 1.97 |
| 249 | 1.84 | 2.14 | 1.94 | 2.07 | 1.76 | 2.05 | 1.98 | 2.20 | 1.91 | 1.98 |
| 250 | 1.85 | 2.14 | 1.95 | 2.08 | 1.77 | 2.06 | 1.99 | 2.21 | 1.92 | 1.99 |
| 251 | 1.87 | 2.15 | 1.96 | 2.09 | 1.79 | 2.07 | 2.00 | 2.22 | 1.93 | 2.00 |
| 252 | 1.88 | 2.16 | 1.98 | 2.09 | 1.80 | 2.08 | 2.01 | 2.23 | 1.94 | 2.01 |
| 253 | 1.89 | 2.16 | 1.99 | 2.10 | 1.81 | 2.08 | 2.03 | 2.23 | 1.95 | 2.02 |
| 254 | 1.90 | 2.17 | 2.00 | 2.10 | 1.82 | 2.09 | 2.04 | 2.24 | 1.96 | 2.03 |
| 255 | 1.92 | 2.18 | 2.02 | 2.11 | 1.84 | 2.10 | 2.06 | 2.25 | 1.97 | 2.04 |
| 256 | 1.93 | 2.18 | 2.03 | 2.11 | 1.85 | 2.10 | 2.07 | 2.25 | 1.99 | 2.05 |
| 257 | 1.94 | 2.19 | 2.04 | 2.12 | 1.86 | 2.11 | 2.08 | 2.26 | 2.00 | 2.06 |
| 258 | 1.96 | 2.19 | 2.06 | 2.12 | 1.88 | 2.11 | 2.10 | 2.26 | 2.01 | 2.07 |
| 259 | 1.97 | 2.20 | 2.07 | 2.13 | 1.89 | 2.12 | 2.11 | 2.27 | 2.03 | 2.08 |
| 260 | 1.98 | 2.20 | 2.09 | 2.13 | 1.91 | 2.12 | 2.13 | 2.28 | 2.04 | 2.09 |
| 261 | 2.00 | 2.21 | 2.10 | 2.14 | 1.92 | 2.13 | 2.14 | 2.28 | 2.06 | 2.10 |
| 262 | 2.01 | 2.21 | 2.12 | 2.14 | 1.94 | 2.13 | 2.16 | 2.29 | 2.07 | 2.11 |
| 263 | 2.03 | 2.22 | 2.13 | 2.14 | 1.95 | 2.14 | 2.18 | 2.29 | 2.09 | 2.12 |
| 264 | 2.04 | 2.22 | 2.14 | 2.15 | 1.97 | 2.14 | 2.19 | 2.29 | 2.10 | 2.13 |
| 265 | 2.06 | 2.22 | 2.16 | 2.15 | 1.98 | 2.15 | 2.21 | 2.30 | 2.12 | 2.13 |
| 266 | 2.07 | 2.23 | 2.17 | 2.15 | 2.00 | 2.15 | 2.22 | 2.30 | 2.14 | 2.14 |
| 267 | 2.09 | 2.23 | 2.19 | 2.15 | 2.01 | 2.16 | 2.24 | 2.30 | 2.16 | 2.15 |
| 268 | 2.10 | 2.23 | 2.20 | 2.16 | 2.03 | 2.16 | 2.26 | 2.31 | 2.17 | 2.15 |
| 269 | 2.12 | 2.24 | 2.22 | 2.16 | 2.04 | 2.16 | 2.27 | 2.31 | 2.19 | 2.16 |
| 270 | 2.13 | 2.24 | 2.23 | 2.16 | 2.06 | 2.16 | 2.29 | 2.31 | 2.21 | 2.16 |
| 271 | 2.15 | 2.24 | 2.25 | 2.16 | 2.08 | 2.16 | 2.31 | 2.31 | 2.23 | 2.17 |
| 272 | 2.16 | 2.24 | 2.26 | 2.16 | 2.09 | 2.17 | 2.32 | 2.31 | 2.25 | 2.17 |
| 273 | 2.18 | 2.24 | 2.28 | 2.16 | 2.11 | 2.17 | 2.34 | 2.31 | 2.27 | 2.17 |
| 274 | 2.20 | 2.24 | 2.29 | 2.16 | 2.12 | 2.17 | 2.36 | 2.31 | 2.29 | 2.17 |
| 275 | 2.21 | 2.25 | 2.31 | 2.16 | 2.14 | 2.17 | 2.38 | 2.31 | 2.31 | 2.17 |
| 276 | 2.23 | 2.25 | 2.33 | 2.16 | 2.16 | 2.17 | 2.39 | 2.31 | 2.33 | 2.17 |
| 277 | 2.24 | 2.24 | 2.34 | 2.16 | 2.17 | 2.17 | 2.41 | 2.31 | 2.35 | 2.17 |
| 278 | 2.26 | 2.24 | 2.36 | 2.16 | 2.19 | 2.17 | 2.43 | 2.31 | 2.37 | 2.17 |
| 279 | 2.28 | 2.24 | 2.37 | 2.15 | 2.21 | 2.16 | 2.44 | 2.31 | 2.39 | 2.17 |
| 280 | 2.29 | 2.24 | 2.39 | 2.15 | 2.22 | 2.16 | 2.46 | 2.30 | 2.41 | 2.17 |
| 281 | 2.31 | 2.24 | 2.40 | 2.15 | 2.24 | 2.16 | 2.48 | 2.30 | 2.43 | 2.16 |
| 282 | 2.32 | 2.24 | 2.42 | 2.15 | 2.25 | 2.16 | 2.49 | 2.30 | 2.45 | 2.16 |
| 283 | 2.34 | 2.23 | 2.43 | 2.14 | 2.27 | 2.15 | 2.51 | 2.29 | 2.47 | 2.15 |
| 284 | 2.35 | 2.23 | 2.44 | 2.14 | 2.29 | 2.15 | 2.53 | 2.29 | 2.49 | 2.15 |
| 285 | 2.37 | 2.23 | 2.46 | 2.14 | 2.30 | 2.14 | 2.54 | 2.28 | 2.51 | 2.14 |
| 286 | 2.38 | 2.22 | 2.47 | 2.13 | 2.32 | 2.14 | 2.56 | 2.28 | 2.53 | 2.13 |
| 287 | 2.40 | 2.22 | 2.49 | 2.13 | 2.33 | 2.13 | 2.57 | 2.27 | 2.54 | 2.12 |
| 288 | 2.41 | 2.21 | 2.50 | 2.12 | 2.35 | 2.13 | 2.59 | 2.27 | 2.56 | 2.11 |
| 289 | 2.43 | 2.21 | 2.51 | 2.12 | 2.36 | 2.12 | 2.60 | 2.26 | 2.58 | 2.10 |
| 290 | 2.44 | 2.20 | 2.53 | 2.11 | 2.37 | 2.12 | 2.62 | 2.25 | 2.59 | 2.09 |
| 291 | 2.45 | 2.20 | 2.54 | 2.10 | 2.39 | 2.11 | 2.63 | 2.25 | 2.61 | 2.08 |
| 292 | 2.47 | 2.19 | 2.55 | 2.10 | 2.40 | 2.10 | 2.64 | 2.24 | 2.62 | 2.07 |
| 293 | 2.48 | 2.18 | 2.56 | 2.09 | 2.41 | 2.09 | 2.66 | 2.23 | 2.64 | 2.06 |
| 294 | 2.49 | 2.17 | 2.58 | 2.08 | 2.43 | 2.09 | 2.67 | 2.22 | 2.65 | 2.05 |
| 295 | 2.50 | 2.17 | 2.59 | 2.08 | 2.44 | 2.08 | 2.68 | 2.21 | 2.66 | 2.04 |
| 296 | 2.52 | 2.16 | 2.60 | 2.07 | 2.45 | 2.07 | 2.69 | 2.20 | 2.67 | 2.03 |
| 297 | 2.53 | 2.15 | 2.61 | 2.06 | 2.46 | 2.06 | 2.70 | 2.20 | 2.68 | 2.01 |
| 298 | 2.54 | 2.14 | 2.62 | 2.05 | 2.47 | 2.05 | 2.72 | 2.19 | 2.69 | 2.00 |
| 299 | 2.55 | 2.13 | 2.63 | 2.04 | 2.48 | 2.04 | 2.73 | 2.18 | 2.70 | 1.99 |
| 300 | 2.56 | 2.12 | 2.64 | 2.03 | 2.49 | 2.03 | 2.74 | 2.17 | 2.71 | 1.97 |
| 301 | 2.57 | 2.12 | 2.65 | 2.03 | 2.50 | 2.02 | 2.75 | 2.16 | 2.72 | 1.96 |
| 302 | 2.58 | 2.11 | 2.66 | 2.02 | 2.51 | 2.01 | 2.75 | 2.15 | 2.72 | 1.95 |
| 303 | 2.58 | 2.10 | 2.66 | 2.01 | 2.52 | 2.00 | 2.76 | 2.13 | 2.73 | 1.93 |
| 304 | 2.59 | 2.09 | 2.67 | 2.00 | 2.52 | 1.99 | 2.77 | 2.12 | 2.74 | 1.92 |
| 305 | 2.60 | 2.08 | 2.68 | 1.99 | 2.53 | 1.98 | 2.78 | 2.11 | 2.74 | 1.91 |
| 306 | 2.61 | 2.07 | 2.69 | 1.98 | 2.54 | 1.97 | 2.78 | 2.10 | 2.74 | 1.90 |
| 307 | 2.61 | 2.06 | 2.69 | 1.97 | 2.54 | 1.96 | 2.79 | 2.09 | 2.75 | 1.88 |
| 308 | 2.62 | 2.05 | 2.70 | 1.96 | 2.55 | 1.95 | 2.80 | 2.08 | 2.75 | 1.87 |
| 309 | 2.62 | 2.04 | 2.70 | 1.95 | 2.55 | 1.94 | 2.80 | 2.07 | 2.75 | 1.86 |
| 310 | 2.63 | 2.03 | 2.71 | 1.94 | 2.56 | 1.93 | 2.80 | 2.06 | 2.75 | 1.85 |
| 311 | 2.63 | 2.02 | 2.71 | 1.93 | 2.56 | 1.92 | 2.81 | 2.05 | 2.75 | 1.84 |
| 312 | 2.63 | 2.01 | 2.71 | 1.92 | 2.56 | 1.91 | 2.81 | 2.04 | 2.75 | 1.82 |
| 313 | 2.64 | 2.00 | 2.72 | 1.91 | 2.56 | 1.90 | 2.81 | 2.03 | 2.75 | 1.81 |
| 314 | 2.64 | 1.99 | 2.72 | 1.90 | 2.57 | 1.89 | 2.82 | 2.02 | 2.75 | 1.80 |
| 315 | 2.64 | 1.98 | 2.72 | 1.89 | 2.57 | 1.88 | 2.82 | 2.01 | 2.75 | 1.79 |
| 316 | 2.64 | 1.97 | 2.72 | 1.88 | 2.57 | 1.87 | 2.82 | 2.00 | 2.75 | 1.78 |
| 317 | 2.64 | 1.96 | 2.72 | 1.87 | 2.57 | 1.86 | 2.82 | 1.99 | 2.74 | 1.77 |
| 318 | 2.64 | 1.95 | 2.72 | 1.86 | 2.57 | 1.85 | 2.82 | 1.98 | 2.74 | 1.77 |
| 319 | 2.64 | 1.95 | 2.72 | 1.85 | 2.57 | 1.85 | 2.82 | 1.97 | 2.74 | 1.76 |
| 320 | 2.64 | 1.94 | 2.72 | 1.85 | 2.57 | 1.84 | 2.81 | 1.96 | 2.73 | 1.75 |
| 321 | 2.64 | 1.93 | 2.72 | 1.84 | 2.57 | 1.83 | 2.81 | 1.95 | 2.73 | 1.74 |
| 322 | 2.64 | 1.92 | 2.72 | 1.83 | 2.56 | 1.82 | 2.81 | 1.94 | 2.72 | 1.73 |
| 323 | 2.63 | 1.92 | 2.72 | 1.82 | 2.56 | 1.82 | 2.81 | 1.93 | 2.72 | 1.73 |
| 324 | 2.63 | 1.91 | 2.71 | 1.81 | 2.56 | 1.81 | 2.80 | 1.93 | 2.71 | 1.72 |
| 325 | 2.63 | 1.90 | 2.71 | 1.81 | 2.55 | 1.80 | 2.80 | 1.92 | 2.71 | 1.72 |
| 326 | 2.62 | 1.90 | 2.71 | 1.80 | 2.55 | 1.80 | 2.80 | 1.91 | 2.70 | 1.71 |
| 327 | 2.62 | 1.89 | 2.70 | 1.79 | 2.55 | 1.79 | 2.79 | 1.91 | 2.69 | 1.71 |
| 328 | 2.62 | 1.89 | 2.70 | 1.79 | 2.54 | 1.79 | 2.79 | 1.90 | 2.69 | 1.70 |
| 329 | 2.61 | 1.88 | 2.69 | 1.78 | 2.54 | 1.78 | 2.78 | 1.90 | 2.68 | 1.70 |
| 330 | 2.61 | 1.88 | 2.69 | 1.78 | 2.53 | 1.78 | 2.77 | 1.89 | 2.67 | 1.70 |
| 331 | 2.60 | 1.88 | 2.68 | 1.78 | 2.52 | 1.77 | 2.77 | 1.89 | 2.67 | 1.70 |
| 332 | 2.59 | 1.87 | 2.67 | 1.77 | 2.52 | 1.77 | 2.76 | 1.89 | 2.66 | 1.70 |
| 333 | 2.59 | 1.87 | 2.67 | 1.77 | 2.51 | 1.77 | 2.75 | 1.88 | 2.65 | 1.69 |
| 334 | 2.58 | 1.87 | 2.66 | 1.77 | 2.51 | 1.76 | 2.75 | 1.88 | 2.64 | 1.69 |
| 335 | 2.57 | 1.87 | 2.65 | 1.76 | 2.50 | 1.76 | 2.74 | 1.88 | 2.64 | 1.69 |
| 336 | 2.57 | 1.87 | 2.65 | 1.76 | 2.49 | 1.76 | 2.73 | 1.88 | 2.63 | 1.70 |
| 337 | 2.56 | 1.87 | 2.64 | 1.76 | 2.49 | 1.76 | 2.72 | 1.88 | 2.62 | 1.70 |
| 338 | 2.55 | 1.87 | 2.63 | 1.76 | 2.48 | 1.76 | 2.72 | 1.88 | 2.61 | 1.70 |
| 339 | 2.55 | 1.87 | 2.62 | 1.76 | 2.47 | 1.76 | 2.71 | 1.88 | 2.60 | 1.70 |
| 340 | 2.54 | 1.87 | 2.61 | 1.77 | 2.46 | 1.76 | 2.70 | 1.88 | 2.60 | 1.71 |
| 341 | 2.53 | 1.87 | 2.61 | 1.77 | 2.46 | 1.77 | 2.69 | 1.88 | 2.59 | 1.71 |
| 342 | 2.52 | 1.88 | 2.60 | 1.77 | 2.45 | 1.77 | 2.68 | 1.88 | 2.58 | 1.71 |
| 343 | 2.52 | 1.88 | 2.59 | 1.77 | 2.44 | 1.77 | 2.67 | 1.89 | 2.57 | 1.72 |
| 344 | 2.51 | 1.89 | 2.58 | 1.78 | 2.43 | 1.78 | 2.67 | 1.89 | 2.57 | 1.72 |
| 345 | 2.50 | 1.89 | 2.57 | 1.78 | 2.42 | 1.78 | 2.66 | 1.90 | 2.56 | 1.73 |
| 346 | 2.49 | 1.90 | 2.56 | 1.79 | 2.42 | 1.79 | 2.65 | 1.90 | 2.55 | 1.74 |
| 347 | 2.49 | 1.90 | 2.55 | 1.79 | 2.41 | 1.79 | 2.64 | 1.91 | 2.54 | 1.74 |
| 348 | 2.48 | 1.91 | 2.55 | 1.80 | 2.40 | 1.80 | 2.63 | 1.92 | 2.54 | 1.75 |
| 349 | 2.47 | 1.92 | 2.54 | 1.81 | 2.39 | 1.81 | 2.62 | 1.93 | 2.53 | 1.76 |
| 350 | 2.46 | 1.93 | 2.53 | 1.82 | 2.39 | 1.82 | 2.62 | 1.93 | 2.52 | 1.77 |
| 351 | 2.46 | 1.94 | 2.52 | 1.83 | 2.38 | 1.82 | 2.61 | 1.94 | 2.52 | 1.78 |
| 352 | 2.45 | 1.95 | 2.51 | 1.84 | 2.37 | 1.83 | 2.60 | 1.95 | 2.51 | 1.79 |
| 353 | 2.44 | 1.96 | 2.51 | 1.85 | 2.37 | 1.84 | 2.59 | 1.97 | 2.50 | 1.80 |
| 354 | 2.44 | 1.97 | 2.50 | 1.86 | 2.36 | 1.85 | 2.59 | 1.98 | 2.50 | 1.81 |
| 355 | 2.43 | 1.98 | 2.49 | 1.87 | 2.35 | 1.87 | 2.58 | 1.99 | 2.49 | 1.82 |
| 356 | 2.42 | 2.00 | 2.49 | 1.89 | 2.35 | 1.88 | 2.57 | 2.00 | 2.49 | 1.84 |
| 357 | 2.42 | 2.01 | 2.48 | 1.90 | 2.34 | 1.89 | 2.57 | 2.02 | 2.48 | 1.85 |
| 358 | 2.41 | 2.02 | 2.48 | 1.92 | 2.34 | 1.90 | 2.56 | 2.03 | 2.48 | 1.86 |
| 359 | 2.41 | 2.04 | 2.47 | 1.93 | 2.33 | 1.92 | 2.56 | 2.05 | 2.47 | 1.88 |
| 360 | 2.40 | 2.05 | 2.47 | 1.95 | 2.33 | 1.93 | 2.55 | 2.06 | 2.47 | 1.89 |
| 361 | 2.40 | 2.07 | 2.46 | 1.96 | 2.32 | 1.95 | 2.55 | 2.08 | 2.47 | 1.91 |
| 362 | 2.40 | 2.09 | 2.46 | 1.98 | 2.32 | 1.97 | 2.55 | 2.10 | 2.46 | 1.92 |
| 363 | 2.40 | 2.11 | 2.46 | 2.00 | 2.32 | 1.98 | 2.54 | 2.12 | 2.46 | 1.94 |
| 364 | 2.39 | 2.13 | 2.45 | 2.02 | 2.32 | 2.00 | 2.54 | 2.14 | 2.46 | 1.96 |
| 365 | 2.39 | 2.15 | 2.45 | 2.04 | 2.32 | 2.02 | 2.54 | 2.15 | 2.46 | 1.98 |
| 366 | 2.39 | 2.17 | 2.45 | 2.06 | 2.31 | 2.04 | 2.54 | 2.17 | 2.46 | 1.99 |
| 367 | 2.39 | 2.19 | 2.45 | 2.08 | 2.31 | 2.06 | 2.54 | 2.20 | 2.46 | 2.01 |
| 368 | 2.40 | 2.21 | 2.46 | 2.10 | 2.32 | 2.08 | 2.54 | 2.22 | 2.46 | 2.03 |
| 369 | 2.40 | 2.23 | 2.46 | 2.12 | 2.32 | 2.10 | 2.55 | 2.24 | 2.46 | 2.05 |
| 370 | 2.40 | 2.25 | 2.46 | 2.14 | 2.32 | 2.12 | 2.55 | 2.26 | 2.46 | 2.07 |
| 371 | 2.41 | 2.27 | 2.47 | 2.17 | 2.32 | 2.14 | 2.55 | 2.28 | 2.47 | 2.09 |
| 372 | 2.41 | 2.30 | 2.47 | 2.19 | 2.33 | 2.16 | 2.56 | 2.31 | 2.47 | 2.11 |
| 373 | 2.42 | 2.32 | 2.48 | 2.21 | 2.33 | 2.18 | 2.56 | 2.33 | 2.48 | 2.13 |
| 374 | 2.43 | 2.34 | 2.49 | 2.24 | 2.34 | 2.20 | 2.57 | 2.35 | 2.48 | 2.15 |
| 375 | 2.44 | 2.37 | 2.50 | 2.26 | 2.35 | 2.23 | 2.58 | 2.38 | 2.49 | 2.17 |
| 376 | 2.45 | 2.39 | 2.51 | 2.28 | 2.36 | 2.25 | 2.59 | 2.40 | 2.49 | 2.20 |
| 377 | 2.46 | 2.41 | 2.52 | 2.31 | 2.37 | 2.27 | 2.60 | 2.42 | 2.50 | 2.22 |
| 378 | 2.47 | 2.44 | 2.53 | 2.33 | 2.38 | 2.30 | 2.61 | 2.45 | 2.51 | 2.24 |
| 379 | 2.49 | 2.46 | 2.54 | 2.35 | 2.39 | 2.32 | 2.63 | 2.47 | 2.52 | 2.27 |
| 380 | 2.50 | 2.49 | 2.56 | 2.38 | 2.41 | 2.34 | 2.64 | 2.49 | 2.53 | 2.29 |
| 381 | 2.52 | 2.51 | 2.58 | 2.40 | 2.42 | 2.36 | 2.66 | 2.52 | 2.55 | 2.31 |
| 382 | 2.54 | 2.53 | 2.59 | 2.42 | 2.44 | 2.38 | 2.68 | 2.54 | 2.56 | 2.34 |
| 383 | 2.56 | 2.55 | 2.61 | 2.44 | 2.45 | 2.41 | 2.69 | 2.56 | 2.57 | 2.36 |
| 384 | 2.58 | 2.58 | 2.63 | 2.47 | 2.47 | 2.43 | 2.71 | 2.58 | 2.59 | 2.38 |
| 385 | 2.60 | 2.60 | 2.66 | 2.49 | 2.49 | 2.45 | 2.74 | 2.60 | 2.61 | 2.40 |
| 386 | 2.62 | 2.62 | 2.68 | 2.50 | 2.51 | 2.47 | 2.76 | 2.62 | 2.63 | 2.43 |
| 387 | 2.64 | 2.64 | 2.70 | 2.52 | 2.54 | 2.49 | 2.78 | 2.64 | 2.65 | 2.45 |
| 388 | 2.67 | 2.65 | 2.73 | 2.54 | 2.56 | 2.50 | 2.80 | 2.66 | 2.67 | 2.47 |
| 389 | 2.70 | 2.67 | 2.75 | 2.55 | 2.58 | 2.52 | 2.83 | 2.68 | 2.70 | 2.49 |
| 390 | 2.72 | 2.69 | 2.78 | 2.57 | 2.61 | 2.53 | 2.85 | 2.69 | 2.72 | 2.51 |
| 391 | 2.75 | 2.70 | 2.81 | 2.58 | 2.63 | 2.55 | 2.88 | 2.71 | 2.75 | 2.53 |
| 392 | 2.78 | 2.71 | 2.83 | 2.59 | 2.66 | 2.56 | 2.91 | 2.72 | 2.78 | 2.54 |
| 393 | 2.81 | 2.73 | 2.86 | 2.60 | 2.69 | 2.57 | 2.94 | 2.74 | 2.80 | 2.56 |
| 394 | 2.83 | 2.74 | 2.89 | 2.61 | 2.71 | 2.58 | 2.96 | 2.75 | 2.83 | 2.57 |
| 395 | 2.86 | 2.75 | 2.91 | 2.62 | 2.74 | 2.59 | 2.99 | 2.76 | 2.86 | 2.58 |
| 396 | 2.89 | 2.76 | 2.94 | 2.63 | 2.76 | 2.60 | 3.02 | 2.77 | 2.89 | 2.59 |
| 397 | 2.92 | 2.76 | 2.97 | 2.63 | 2.79 | 2.61 | 3.05 | 2.77 | 2.92 | 2.60 |
| 398 | 2.95 | 2.77 | 2.99 | 2.64 | 2.82 | 2.61 | 3.07 | 2.78 | 2.95 | 2.60 |
| 399 | 2.98 | 2.77 | 3.02 | 2.64 | 2.84 | 2.62 | 3.10 | 2.79 | 2.98 | 2.61 |
| 400 | 3.00 | 2.78 | 3.04 | 2.65 | 2.87 | 2.62 | 3.13 | 2.79 | 3.00 | 2.61 |
| 401 | 3.03 | 2.78 | 3.06 | 2.65 | 2.89 | 2.63 | 3.15 | 2.80 | 3.03 | 2.61 |
| 402 | 3.06 | 2.78 | 3.09 | 2.65 | 2.91 | 2.63 | 3.18 | 2.80 | 3.05 | 2.61 |
| 403 | 3.08 | 2.78 | 3.11 | 2.66 | 2.94 | 2.64 | 3.21 | 2.81 | 3.08 | 2.62 |
| 404 | 3.11 | 2.78 | 3.13 | 2.66 | 2.96 | 2.64 | 3.23 | 2.81 | 3.10 | 2.62 |
| 405 | 3.13 | 2.79 | 3.15 | 2.66 | 2.98 | 2.64 | 3.26 | 2.81 | 3.12 | 2.62 |
| 406 | 3.16 | 2.79 | 3.18 | 2.67 | 3.00 | 2.65 | 3.28 | 2.81 | 3.14 | 2.62 |
| 407 | 3.18 | 2.79 | 3.20 | 2.67 | 3.03 | 2.65 | 3.31 | 2.82 | 3.17 | 2.62 |
| 408 | 3.20 | 2.79 | 3.22 | 2.67 | 3.05 | 2.65 | 3.33 | 2.82 | 3.19 | 2.63 |
| 409 | 3.22 | 2.79 | 3.24 | 2.68 | 3.07 | 2.66 | 3.36 | 2.82 | 3.21 | 2.63 |
| 410 | 3.25 | 2.79 | 3.27 | 2.68 | 3.09 | 2.66 | 3.38 | 2.82 | 3.23 | 2.63 |
| 411 | 3.27 | 2.79 | 3.29 | 2.69 | 3.12 | 2.66 | 3.41 | 2.82 | 3.25 | 2.64 |
| 412 | 3.29 | 2.79 | 3.31 | 2.69 | 3.14 | 2.67 | 3.43 | 2.83 | 3.28 | 2.64 |
| 413 | 3.31 | 2.80 | 3.34 | 2.69 | 3.16 | 2.67 | 3.45 | 2.83 | 3.30 | 2.64 |
| 414 | 3.34 | 2.80 | 3.36 | 2.70 | 3.18 | 2.67 | 3.48 | 2.83 | 3.32 | 2.65 |
| 415 | 3.36 | 2.80 | 3.38 | 2.70 | 3.21 | 2.67 | 3.50 | 2.83 | 3.35 | 2.65 |
| 416 | 3.38 | 2.80 | 3.41 | 2.70 | 3.23 | 2.68 | 3.53 | 2.83 | 3.37 | 2.65 |
| 417 | 3.40 | 2.81 | 3.43 | 2.71 | 3.26 | 2.68 | 3.55 | 2.83 | 3.40 | 2.65 |
| 418 | 3.43 | 2.81 | 3.46 | 2.71 | 3.28 | 2.68 | 3.58 | 2.83 | 3.42 | 2.66 |
| 419 | 3.45 | 2.81 | 3.49 | 2.71 | 3.31 | 2.69 | 3.61 | 2.83 | 3.45 | 2.66 |
| 420 | 3.48 | 2.81 | 3.51 | 2.71 | 3.33 | 2.69 | 3.63 | 2.83 | 3.47 | 2.66 |
| 421 | 3.50 | 2.82 | 3.54 | 2.71 | 3.36 | 2.69 | 3.66 | 2.83 | 3.50 | 2.66 |
| 422 | 3.53 | 2.82 | 3.57 | 2.71 | 3.39 | 2.69 | 3.69 | 2.83 | 3.52 | 2.66 |
| 423 | 3.56 | 2.82 | 3.60 | 2.71 | 3.41 | 2.69 | 3.71 | 2.83 | 3.55 | 2.66 |
| 424 | 3.59 | 2.82 | 3.62 | 2.71 | 3.44 | 2.69 | 3.74 | 2.83 | 3.58 | 2.66 |
| 425 | 3.62 | 2.82 | 3.65 | 2.71 | 3.47 | 2.69 | 3.77 | 2.82 | 3.61 | 2.65 |
| 426 | 3.64 | 2.82 | 3.68 | 2.71 | 3.50 | 2.69 | 3.80 | 2.82 | 3.63 | 2.65 |
| 427 | 3.68 | 2.82 | 3.71 | 2.70 | 3.53 | 2.68 | 3.83 | 2.82 | 3.66 | 2.65 |
| 428 | 3.71 | 2.82 | 3.74 | 2.70 | 3.56 | 2.68 | 3.86 | 2.81 | 3.69 | 2.64 |
| 429 | 3.74 | 2.82 | 3.77 | 2.70 | 3.59 | 2.68 | 3.88 | 2.80 | 3.72 | 2.64 |
| 430 | 3.77 | 2.82 | 3.80 | 2.69 | 3.62 | 2.67 | 3.91 | 2.80 | 3.75 | 2.63 |
| 431 | 3.80 | 2.81 | 3.83 | 2.68 | 3.65 | 2.66 | 3.94 | 2.79 | 3.77 | 2.63 |
| 432 | 3.83 | 2.81 | 3.86 | 2.68 | 3.68 | 2.66 | 3.97 | 2.78 | 3.80 | 2.62 |
| 433 | 3.87 | 2.80 | 3.89 | 2.67 | 3.71 | 2.65 | 4.00 | 2.77 | 3.83 | 2.61 |
| 434 | 3.90 | 2.80 | 3.92 | 2.66 | 3.74 | 2.64 | 4.03 | 2.77 | 3.86 | 2.60 |
| 435 | 3.94 | 2.79 | 3.95 | 2.65 | 3.77 | 2.63 | 4.06 | 2.75 | 3.89 | 2.59 |
| 436 | 3.97 | 2.78 | 3.98 | 2.64 | 3.80 | 2.62 | 4.09 | 2.74 | 3.91 | 2.58 |
| 437 | 4.00 | 2.77 | 4.01 | 2.63 | 3.82 | 2.61 | 4.11 | 2.73 | 3.94 | 2.57 |
| 438 | 4.04 | 2.75 | 4.04 | 2.61 | 3.85 | 2.60 | 4.14 | 2.72 | 3.97 | 2.56 |
| 439 | 4.07 | 2.74 | 4.07 | 2.60 | 3.88 | 2.58 | 4.17 | 2.70 | 4.00 | 2.54 |
| 440 | 4.11 | 2.72 | 4.10 | 2.58 | 3.91 | 2.57 | 4.20 | 2.69 | 4.02 | 2.53 |
| 441 | 4.14 | 2.71 | 4.13 | 2.57 | 3.94 | 2.55 | 4.23 | 2.68 | 4.05 | 2.52 |
| 442 | 4.17 | 2.69 | 4.16 | 2.55 | 3.97 | 2.54 | 4.25 | 2.66 | 4.07 | 2.50 |
| 443 | 4.20 | 2.67 | 4.19 | 2.53 | 4.00 | 2.52 | 4.28 | 2.64 | 4.10 | 2.49 |
| 444 | 4.24 | 2.65 | 4.21 | 2.52 | 4.02 | 2.50 | 4.31 | 2.62 | 4.12 | 2.47 |
| 445 | 4.27 | 2.63 | 4.24 | 2.50 | 4.05 | 2.48 | 4.33 | 2.61 | 4.15 | 2.45 |
| 446 | 4.30 | 2.61 | 4.27 | 2.48 | 4.07 | 2.46 | 4.36 | 2.59 | 4.17 | 2.43 |
| 447 | 4.33 | 2.58 | 4.29 | 2.46 | 4.10 | 2.44 | 4.38 | 2.57 | 4.20 | 2.42 |
| 448 | 4.36 | 2.56 | 4.32 | 2.44 | 4.12 | 2.42 | 4.40 | 2.55 | 4.22 | 2.40 |
| 449 | 4.38 | 2.53 | 4.34 | 2.41 | 4.15 | 2.40 | 4.43 | 2.53 | 4.24 | 2.38 |
| 450 | 4.41 | 2.51 | 4.36 | 2.39 | 4.17 | 2.38 | 4.45 | 2.50 | 4.26 | 2.36 |
| 451 | 4.44 | 2.48 | 4.39 | 2.37 | 4.19 | 2.35 | 4.47 | 2.48 | 4.28 | 2.34 |
| 452 | 4.46 | 2.45 | 4.41 | 2.34 | 4.21 | 2.33 | 4.49 | 2.46 | 4.30 | 2.32 |
| 453 | 4.48 | 2.42 | 4.43 | 2.32 | 4.23 | 2.31 | 4.51 | 2.44 | 4.32 | 2.29 |
| 454 | 4.51 | 2.39 | 4.45 | 2.29 | 4.25 | 2.28 | 4.53 | 2.41 | 4.34 | 2.27 |
| 455 | 4.53 | 2.36 | 4.47 | 2.27 | 4.27 | 2.26 | 4.55 | 2.39 | 4.36 | 2.25 |
| 456 | 4.55 | 2.33 | 4.49 | 2.24 | 4.29 | 2.23 | 4.57 | 2.37 | 4.38 | 2.23 |
| 457 | 4.57 | 2.30 | 4.51 | 2.22 | 4.31 | 2.21 | 4.59 | 2.34 | 4.39 | 2.21 |
| 458 | 4.58 | 2.27 | 4.52 | 2.19 | 4.32 | 2.18 | 4.60 | 2.32 | 4.41 | 2.18 |
| 459 | 4.60 | 2.23 | 4.54 | 2.16 | 4.34 | 2.15 | 4.62 | 2.29 | 4.42 | 2.16 |
| 460 | 4.61 | 2.20 | 4.56 | 2.14 | 4.35 | 2.13 | 4.63 | 2.27 | 4.44 | 2.14 |
| 461 | 4.63 | 2.17 | 4.57 | 2.11 | 4.37 | 2.10 | 4.65 | 2.24 | 4.45 | 2.11 |
| 462 | 4.64 | 2.14 | 4.58 | 2.08 | 4.38 | 2.07 | 4.66 | 2.21 | 4.47 | 2.09 |
| 463 | 4.65 | 2.11 | 4.60 | 2.05 | 4.39 | 2.05 | 4.67 | 2.19 | 4.48 | 2.07 |
| 464 | 4.66 | 2.07 | 4.61 | 2.03 | 4.40 | 2.02 | 4.69 | 2.16 | 4.49 | 2.04 |
| 465 | 4.67 | 2.04 | 4.62 | 2.00 | 4.41 | 2.00 | 4.70 | 2.14 | 4.50 | 2.02 |
| 466 | 4.68 | 2.01 | 4.63 | 1.97 | 4.42 | 1.97 | 4.71 | 2.11 | 4.51 | 1.99 |
| 467 | 4.68 | 1.98 | 4.64 | 1.94 | 4.43 | 1.94 | 4.72 | 2.09 | 4.52 | 1.97 |
| 468 | 4.69 | 1.95 | 4.65 | 1.92 | 4.44 | 1.92 | 4.73 | 2.06 | 4.53 | 1.95 |
| 469 | 4.69 | 1.92 | 4.65 | 1.89 | 4.44 | 1.89 | 4.73 | 2.03 | 4.54 | 1.92 |
| 470 | 4.70 | 1.89 | 4.66 | 1.86 | 4.45 | 1.86 | 4.74 | 2.01 | 4.55 | 1.90 |
| 471 | 4.70 | 1.86 | 4.67 | 1.83 | 4.45 | 1.84 | 4.75 | 1.98 | 4.56 | 1.88 |
| 472 | 4.70 | 1.83 | 4.67 | 1.81 | 4.46 | 1.81 | 4.76 | 1.96 | 4.56 | 1.85 |
| 473 | 4.70 | 1.80 | 4.68 | 1.78 | 4.46 | 1.79 | 4.76 | 1.93 | 4.57 | 1.83 |
| 474 | 4.71 | 1.77 | 4.68 | 1.75 | 4.47 | 1.76 | 4.77 | 1.91 | 4.57 | 1.81 |
| 475 | 4.70 | 1.74 | 4.68 | 1.73 | 4.47 | 1.74 | 4.77 | 1.88 | 4.58 | 1.78 |
| 476 | 4.70 | 1.71 | 4.69 | 1.70 | 4.47 | 1.71 | 4.77 | 1.86 | 4.58 | 1.76 |
| 477 | 4.70 | 1.69 | 4.69 | 1.68 | 4.47 | 1.69 | 4.78 | 1.84 | 4.59 | 1.74 |
| 478 | 4.70 | 1.66 | 4.69 | 1.65 | 4.47 | 1.66 | 4.78 | 1.81 | 4.59 | 1.71 |
| 479 | 4.70 | 1.63 | 4.69 | 1.62 | 4.48 | 1.64 | 4.78 | 1.79 | 4.59 | 1.69 |
| 480 | 4.69 | 1.61 | 4.69 | 1.60 | 4.48 | 1.62 | 4.78 | 1.76 | 4.60 | 1.67 |
| 481 | 4.69 | 1.58 | 4.69 | 1.57 | 4.47 | 1.59 | 4.78 | 1.74 | 4.60 | 1.65 |
| 482 | 4.69 | 1.56 | 4.69 | 1.55 | 4.47 | 1.57 | 4.78 | 1.72 | 4.60 | 1.63 |
| 483 | 4.68 | 1.53 | 4.69 | 1.53 | 4.47 | 1.55 | 4.78 | 1.70 | 4.60 | 1.61 |
| 484 | 4.68 | 1.51 | 4.68 | 1.50 | 4.47 | 1.53 | 4.78 | 1.67 | 4.60 | 1.58 |
| 485 | 4.67 | 1.49 | 4.68 | 1.48 | 4.47 | 1.51 | 4.78 | 1.65 | 4.60 | 1.56 |
| 486 | 4.66 | 1.47 | 4.68 | 1.46 | 4.47 | 1.49 | 4.78 | 1.63 | 4.60 | 1.54 |
| 487 | 4.66 | 1.44 | 4.68 | 1.44 | 4.46 | 1.46 | 4.78 | 1.61 | 4.60 | 1.52 |
| 488 | 4.65 | 1.42 | 4.67 | 1.41 | 4.46 | 1.44 | 4.78 | 1.59 | 4.60 | 1.50 |
| 489 | 4.64 | 1.40 | 4.67 | 1.39 | 4.46 | 1.42 | 4.78 | 1.57 | 4.60 | 1.48 |
| 490 | 4.64 | 1.38 | 4.66 | 1.37 | 4.45 | 1.40 | 4.77 | 1.55 | 4.60 | 1.46 |
| 491 | 4.63 | 1.36 | 4.66 | 1.35 | 4.45 | 1.39 | 4.77 | 1.53 | 4.60 | 1.44 |
| 492 | 4.62 | 1.34 | 4.65 | 1.33 | 4.44 | 1.37 | 4.77 | 1.51 | 4.59 | 1.42 |
| 493 | 4.61 | 1.33 | 4.65 | 1.31 | 4.44 | 1.35 | 4.76 | 1.49 | 4.59 | 1.40 |
| 494 | 4.61 | 1.31 | 4.64 | 1.29 | 4.43 | 1.33 | 4.76 | 1.47 | 4.59 | 1.39 |
| 495 | 4.60 | 1.29 | 4.63 | 1.27 | 4.43 | 1.31 | 4.75 | 1.45 | 4.59 | 1.37 |
| 496 | 4.59 | 1.27 | 4.63 | 1.25 | 4.42 | 1.30 | 4.75 | 1.44 | 4.58 | 1.35 |
| 497 | 4.58 | 1.26 | 4.62 | 1.24 | 4.42 | 1.28 | 4.75 | 1.42 | 4.58 | 1.33 |
| 498 | 4.57 | 1.24 | 4.61 | 1.22 | 4.41 | 1.26 | 4.74 | 1.40 | 4.58 | 1.31 |
| 499 | 4.56 | 1.22 | 4.61 | 1.20 | 4.40 | 1.25 | 4.74 | 1.38 | 4.57 | 1.30 |
| 500 | 4.55 | 1.21 | 4.60 | 1.19 | 4.40 | 1.23 | 4.73 | 1.37 | 4.57 | 1.28 |
| 501 | 4.55 | 1.19 | 4.59 | 1.17 | 4.39 | 1.22 | 4.72 | 1.35 | 4.56 | 1.26 |
| 502 | 4.54 | 1.18 | 4.58 | 1.15 | 4.38 | 1.20 | 4.72 | 1.34 | 4.56 | 1.25 |
| 503 | 4.53 | 1.17 | 4.57 | 1.14 | 4.38 | 1.19 | 4.71 | 1.32 | 4.55 | 1.23 |
| 504 | 4.52 | 1.15 | 4.57 | 1.12 | 4.37 | 1.17 | 4.71 | 1.30 | 4.55 | 1.22 |
| 505 | 4.51 | 1.14 | 4.56 | 1.11 | 4.36 | 1.16 | 4.70 | 1.29 | 4.54 | 1.20 |
| 506 | 4.50 | 1.13 | 4.55 | 1.09 | 4.36 | 1.15 | 4.69 | 1.28 | 4.54 | 1.19 |
| 507 | 4.49 | 1.11 | 4.54 | 1.08 | 4.35 | 1.13 | 4.69 | 1.26 | 4.53 | 1.17 |
| 508 | 4.48 | 1.10 | 4.53 | 1.07 | 4.34 | 1.12 | 4.68 | 1.25 | 4.52 | 1.16 |
| 509 | 4.47 | 1.09 | 4.52 | 1.05 | 4.33 | 1.11 | 4.68 | 1.23 | 4.52 | 1.14 |
| 510 | 4.46 | 1.08 | 4.52 | 1.04 | 4.33 | 1.10 | 4.67 | 1.22 | 4.51 | 1.13 |
| 511 | 4.46 | 1.07 | 4.51 | 1.03 | 4.32 | 1.08 | 4.66 | 1.21 | 4.51 | 1.12 |
| 512 | 4.45 | 1.06 | 4.50 | 1.02 | 4.31 | 1.07 | 4.66 | 1.20 | 4.50 | 1.10 |
| 513 | 4.44 | 1.05 | 4.49 | 1.00 | 4.30 | 1.06 | 4.65 | 1.18 | 4.49 | 1.09 |
| 514 | 4.43 | 1.04 | 4.48 | 0.99 | 4.30 | 1.05 | 4.64 | 1.17 | 4.49 | 1.08 |
| 515 | 4.42 | 1.03 | 4.47 | 0.98 | 4.29 | 1.04 | 4.63 | 1.16 | 4.48 | 1.07 |
| 516 | 4.41 | 1.02 | 4.46 | 0.97 | 4.28 | 1.03 | 4.63 | 1.15 | 4.47 | 1.05 |
| 517 | 4.40 | 1.01 | 4.45 | 0.96 | 4.27 | 1.02 | 4.62 | 1.14 | 4.46 | 1.04 |
| 518 | 4.39 | 1.00 | 4.44 | 0.95 | 4.27 | 1.01 | 4.61 | 1.13 | 4.46 | 1.03 |
| 519 | 4.38 | 0.99 | 4.43 | 0.94 | 4.26 | 1.00 | 4.61 | 1.12 | 4.45 | 1.02 |
| 520 | 4.38 | 0.98 | 4.43 | 0.93 | 4.25 | 0.99 | 4.60 | 1.11 | 4.44 | 1.01 |
| 521 | 4.37 | 0.98 | 4.42 | 0.92 | 4.24 | 0.98 | 4.59 | 1.10 | 4.44 | 1.00 |
| 522 | 4.36 | 0.97 | 4.41 | 0.92 | 4.24 | 0.98 | 4.59 | 1.09 | 4.43 | 0.99 |
| 523 | 4.35 | 0.96 | 4.40 | 0.91 | 4.23 | 0.97 | 4.58 | 1.08 | 4.42 | 0.98 |
| 524 | 4.34 | 0.95 | 4.39 | 0.90 | 4.22 | 0.96 | 4.57 | 1.07 | 4.41 | 0.97 |
| 525 | 4.33 | 0.95 | 4.38 | 0.89 | 4.21 | 0.95 | 4.56 | 1.06 | 4.40 | 0.96 |
| 526 | 4.32 | 0.94 | 4.37 | 0.89 | 4.21 | 0.95 | 4.56 | 1.05 | 4.40 | 0.95 |
| 527 | 4.32 | 0.93 | 4.36 | 0.88 | 4.20 | 0.94 | 4.55 | 1.04 | 4.39 | 0.94 |
| 528 | 4.31 | 0.93 | 4.35 | 0.87 | 4.19 | 0.93 | 4.54 | 1.04 | 4.38 | 0.94 |
| 529 | 4.30 | 0.92 | 4.35 | 0.87 | 4.18 | 0.92 | 4.54 | 1.03 | 4.37 | 0.93 |
| 530 | 4.29 | 0.92 | 4.34 | 0.86 | 4.18 | 0.92 | 4.53 | 1.02 | 4.37 | 0.92 |
| 531 | 4.28 | 0.91 | 4.33 | 0.85 | 4.17 | 0.91 | 4.52 | 1.01 | 4.36 | 0.91 |
| 532 | 4.27 | 0.91 | 4.32 | 0.85 | 4.16 | 0.91 | 4.51 | 1.01 | 4.35 | 0.91 |
| 533 | 4.27 | 0.90 | 4.31 | 0.84 | 4.15 | 0.90 | 4.51 | 1.00 | 4.34 | 0.90 |
| 534 | 4.26 | 0.90 | 4.30 | 0.84 | 4.15 | 0.90 | 4.50 | 0.99 | 4.34 | 0.89 |
| 535 | 4.25 | 0.89 | 4.30 | 0.83 | 4.14 | 0.89 | 4.49 | 0.99 | 4.33 | 0.89 |
| 536 | 4.24 | 0.89 | 4.29 | 0.83 | 4.13 | 0.89 | 4.49 | 0.98 | 4.32 | 0.88 |
| 537 | 4.24 | 0.88 | 4.28 | 0.82 | 4.13 | 0.88 | 4.48 | 0.97 | 4.31 | 0.87 |
| 538 | 4.23 | 0.88 | 4.27 | 0.82 | 4.12 | 0.88 | 4.47 | 0.97 | 4.31 | 0.87 |
| 539 | 4.22 | 0.87 | 4.26 | 0.82 | 4.11 | 0.87 | 4.47 | 0.96 | 4.30 | 0.86 |
| 540 | 4.21 | 0.87 | 4.26 | 0.81 | 4.11 | 0.87 | 4.46 | 0.96 | 4.29 | 0.86 |
| 541 | 4.21 | 0.87 | 4.25 | 0.81 | 4.10 | 0.86 | 4.45 | 0.95 | 4.28 | 0.85 |
| 542 | 4.20 | 0.86 | 4.24 | 0.81 | 4.09 | 0.86 | 4.45 | 0.95 | 4.28 | 0.85 |
| 543 | 4.19 | 0.86 | 4.23 | 0.80 | 4.09 | 0.86 | 4.44 | 0.94 | 4.27 | 0.84 |
| 544 | 4.19 | 0.86 | 4.23 | 0.80 | 4.08 | 0.85 | 4.44 | 0.94 | 4.26 | 0.84 |
| 545 | 4.18 | 0.86 | 4.22 | 0.80 | 4.08 | 0.85 | 4.43 | 0.93 | 4.26 | 0.84 |
| 546 | 4.17 | 0.85 | 4.21 | 0.80 | 4.07 | 0.85 | 4.42 | 0.93 | 4.25 | 0.83 |
| 547 | 4.16 | 0.85 | 4.20 | 0.79 | 4.06 | 0.84 | 4.42 | 0.93 | 4.24 | 0.83 |
| 548 | 4.16 | 0.85 | 4.20 | 0.79 | 4.06 | 0.84 | 4.41 | 0.92 | 4.24 | 0.83 |
| 549 | 4.15 | 0.85 | 4.19 | 0.79 | 4.05 | 0.84 | 4.41 | 0.92 | 4.23 | 0.82 |
| 550 | 4.15 | 0.84 | 4.19 | 0.79 | 4.05 | 0.84 | 4.40 | 0.92 | 4.22 | 0.82 |
| 551 | 4.14 | 0.84 | 4.18 | 0.79 | 4.04 | 0.84 | 4.39 | 0.91 | 4.22 | 0.82 |
| 552 | 4.13 | 0.84 | 4.17 | 0.79 | 4.04 | 0.83 | 4.39 | 0.91 | 4.21 | 0.82 |
| 553 | 4.13 | 0.84 | 4.17 | 0.79 | 4.03 | 0.83 | 4.38 | 0.91 | 4.21 | 0.81 |
| 554 | 4.12 | 0.84 | 4.16 | 0.78 | 4.03 | 0.83 | 4.38 | 0.90 | 4.20 | 0.81 |
| 555 | 4.11 | 0.84 | 4.16 | 0.78 | 4.02 | 0.83 | 4.37 | 0.90 | 4.20 | 0.81 |
| 556 | 4.11 | 0.84 | 4.15 | 0.78 | 4.02 | 0.83 | 4.37 | 0.90 | 4.19 | 0.81 |
| 557 | 4.10 | 0.84 | 4.14 | 0.78 | 4.01 | 0.83 | 4.36 | 0.90 | 4.19 | 0.81 |
| 558 | 4.10 | 0.84 | 4.14 | 0.78 | 4.01 | 0.83 | 4.36 | 0.89 | 4.18 | 0.81 |
| 559 | 4.09 | 0.84 | 4.13 | 0.78 | 4.00 | 0.82 | 4.35 | 0.89 | 4.18 | 0.80 |
| 560 | 4.09 | 0.83 | 4.13 | 0.78 | 4.00 | 0.82 | 4.35 | 0.89 | 4.17 | 0.80 |
| 561 | 4.08 | 0.83 | 4.12 | 0.78 | 3.99 | 0.82 | 4.35 | 0.89 | 4.17 | 0.80 |
| 562 | 4.08 | 0.84 | 4.12 | 0.78 | 3.99 | 0.82 | 4.34 | 0.89 | 4.16 | 0.80 |
| 563 | 4.07 | 0.84 | 4.12 | 0.78 | 3.99 | 0.82 | 4.34 | 0.89 | 4.16 | 0.80 |
| 564 | 4.07 | 0.84 | 4.11 | 0.79 | 3.98 | 0.82 | 4.33 | 0.88 | 4.15 | 0.80 |
| 565 | 4.06 | 0.84 | 4.11 | 0.79 | 3.98 | 0.82 | 4.33 | 0.88 | 4.15 | 0.80 |
| 566 | 4.06 | 0.84 | 4.10 | 0.79 | 3.98 | 0.82 | 4.33 | 0.88 | 4.15 | 0.80 |
| 567 | 4.05 | 0.84 | 4.10 | 0.79 | 3.97 | 0.82 | 4.32 | 0.88 | 4.14 | 0.80 |
| 568 | 4.05 | 0.84 | 4.10 | 0.79 | 3.97 | 0.82 | 4.32 | 0.88 | 4.14 | 0.80 |
| 569 | 4.04 | 0.84 | 4.10 | 0.79 | 3.97 | 0.82 | 4.32 | 0.88 | 4.14 | 0.80 |
| 570 | 4.04 | 0.84 | 4.09 | 0.79 | 3.96 | 0.82 | 4.31 | 0.88 | 4.13 | 0.80 |
| 571 | 4.04 | 0.84 | 4.09 | 0.79 | 3.96 | 0.82 | 4.31 | 0.88 | 4.13 | 0.80 |
| 572 | 4.03 | 0.85 | 4.09 | 0.79 | 3.96 | 0.82 | 4.31 | 0.88 | 4.13 | 0.80 |
| 573 | 4.03 | 0.85 | 4.09 | 0.80 | 3.96 | 0.82 | 4.30 | 0.88 | 4.13 | 0.80 |
| 574 | 4.03 | 0.85 | 4.08 | 0.80 | 3.96 | 0.83 | 4.30 | 0.88 | 4.12 | 0.80 |
| 575 | 4.02 | 0.85 | 4.08 | 0.80 | 3.95 | 0.83 | 4.30 | 0.88 | 4.12 | 0.80 |
| 576 | 4.02 | 0.85 | 4.08 | 0.80 | 3.95 | 0.83 | 4.30 | 0.87 | 4.12 | 0.80 |
| 577 | 4.02 | 0.86 | 4.08 | 0.80 | 3.95 | 0.83 | 4.29 | 0.87 | 4.12 | 0.80 |
| 578 | 4.01 | 0.86 | 4.08 | 0.80 | 3.95 | 0.83 | 4.29 | 0.87 | 4.12 | 0.80 |
| 579 | 4.01 | 0.86 | 4.08 | 0.81 | 3.95 | 0.83 | 4.29 | 0.87 | 4.12 | 0.80 |
| 580 | 4.01 | 0.86 | 4.08 | 0.81 | 3.95 | 0.83 | 4.29 | 0.87 | 4.11 | 0.80 |
| 581 | 4.01 | 0.87 | 4.08 | 0.81 | 3.95 | 0.83 | 4.29 | 0.87 | 4.11 | 0.80 |
| 582 | 4.01 | 0.87 | 4.08 | 0.81 | 3.95 | 0.83 | 4.29 | 0.87 | 4.11 | 0.80 |
| 583 | 4.01 | 0.87 | 4.08 | 0.81 | 3.95 | 0.83 | 4.29 | 0.87 | 4.11 | 0.80 |
| 584 | 4.00 | 0.88 | 4.08 | 0.82 | 3.95 | 0.83 | 4.29 | 0.87 | 4.11 | 0.80 |
| 585 | 4.00 | 0.88 | 4.08 | 0.82 | 3.95 | 0.83 | 4.28 | 0.87 | 4.11 | 0.80 |
| 586 | 4.00 | 0.88 | 4.08 | 0.82 | 3.95 | 0.83 | 4.28 | 0.87 | 4.11 | 0.80 |
| 587 | 4.00 | 0.89 | 4.08 | 0.82 | 3.95 | 0.84 | 4.28 | 0.87 | 4.11 | 0.80 |
| 588 | 4.00 | 0.89 | 4.08 | 0.82 | 3.95 | 0.84 | 4.28 | 0.87 | 4.11 | 0.79 |
| 589 | 4.00 | 0.90 | 4.08 | 0.82 | 3.95 | 0.84 | 4.28 | 0.87 | 4.11 | 0.79 |
| 590 | 4.00 | 0.90 | 4.08 | 0.82 | 3.95 | 0.84 | 4.28 | 0.87 | 4.11 | 0.79 |
| 591 | 4.00 | 0.90 | 4.08 | 0.83 | 3.95 | 0.84 | 4.28 | 0.87 | 4.11 | 0.79 |
| 592 | 4.01 | 0.91 | 4.09 | 0.83 | 3.96 | 0.84 | 4.28 | 0.87 | 4.11 | 0.79 |
| 593 | 4.01 | 0.91 | 4.09 | 0.83 | 3.96 | 0.84 | 4.29 | 0.87 | 4.11 | 0.79 |
| 594 | 4.01 | 0.92 | 4.09 | 0.83 | 3.96 | 0.84 | 4.29 | 0.87 | 4.11 | 0.79 |
| 595 | 4.01 | 0.92 | 4.09 | 0.83 | 3.96 | 0.84 | 4.29 | 0.87 | 4.11 | 0.79 |
| 596 | 4.02 | 0.92 | 4.10 | 0.83 | 3.96 | 0.84 | 4.29 | 0.87 | 4.11 | 0.79 |
| 597 | 4.02 | 0.93 | 4.10 | 0.83 | 3.97 | 0.84 | 4.29 | 0.87 | 4.11 | 0.79 |
| 598 | 4.02 | 0.93 | 4.10 | 0.83 | 3.97 | 0.84 | 4.29 | 0.87 | 4.11 | 0.79 |
| 599 | 4.03 | 0.93 | 4.11 | 0.83 | 3.97 | 0.84 | 4.29 | 0.87 | 4.12 | 0.79 |
| 600 | 4.03 | 0.94 | 4.11 | 0.83 | 3.97 | 0.84 | 4.29 | 0.87 | 4.12 | 0.79 |
| 601 | 4.04 | 0.94 | 4.11 | 0.83 | 3.98 | 0.84 | 4.30 | 0.87 | 4.12 | 0.79 |
| 602 | 4.04 | 0.94 | 4.12 | 0.83 | 3.98 | 0.84 | 4.30 | 0.87 | 4.12 | 0.79 |
| 603 | 4.05 | 0.95 | 4.12 | 0.83 | 3.98 | 0.84 | 4.30 | 0.87 | 4.12 | 0.78 |
| 604 | 4.05 | 0.95 | 4.12 | 0.83 | 3.99 | 0.84 | 4.30 | 0.87 | 4.12 | 0.78 |
| 605 | 4.06 | 0.95 | 4.13 | 0.83 | 3.99 | 0.83 | 4.30 | 0.86 | 4.12 | 0.78 |
| 606 | 4.06 | 0.96 | 4.13 | 0.83 | 3.99 | 0.83 | 4.31 | 0.86 | 4.12 | 0.78 |
| 607 | 4.07 | 0.96 | 4.14 | 0.83 | 4.00 | 0.83 | 4.31 | 0.86 | 4.13 | 0.78 |
| 608 | 4.08 | 0.96 | 4.14 | 0.83 | 4.00 | 0.83 | 4.31 | 0.86 | 4.13 | 0.78 |
| 609 | 4.09 | 0.96 | 4.15 | 0.83 | 4.01 | 0.83 | 4.31 | 0.86 | 4.13 | 0.78 |
| 610 | 4.10 | 0.96 | 4.15 | 0.82 | 4.01 | 0.83 | 4.32 | 0.86 | 4.13 | 0.77 |
| 611 | 4.10 | 0.96 | 4.15 | 0.82 | 4.01 | 0.82 | 4.32 | 0.86 | 4.13 | 0.77 |
| 612 | 4.11 | 0.96 | 4.16 | 0.82 | 4.02 | 0.82 | 4.32 | 0.85 | 4.13 | 0.77 |
| 613 | 4.12 | 0.96 | 4.16 | 0.82 | 4.02 | 0.82 | 4.33 | 0.85 | 4.14 | 0.77 |
| 614 | 4.13 | 0.96 | 4.17 | 0.81 | 4.03 | 0.82 | 4.33 | 0.85 | 4.14 | 0.77 |
| 615 | 4.14 | 0.96 | 4.17 | 0.81 | 4.03 | 0.81 | 4.33 | 0.85 | 4.14 | 0.77 |
| 616 | 4.15 | 0.96 | 4.18 | 0.81 | 4.03 | 0.81 | 4.33 | 0.84 | 4.14 | 0.76 |
| 617 | 4.16 | 0.96 | 4.18 | 0.81 | 4.04 | 0.81 | 4.34 | 0.84 | 4.14 | 0.76 |
| 618 | 4.17 | 0.95 | 4.19 | 0.80 | 4.04 | 0.80 | 4.34 | 0.84 | 4.14 | 0.76 |
| 619 | 4.18 | 0.95 | 4.19 | 0.80 | 4.05 | 0.80 | 4.34 | 0.83 | 4.15 | 0.76 |
| 620 | 4.19 | 0.94 | 4.20 | 0.80 | 4.05 | 0.80 | 4.35 | 0.83 | 4.15 | 0.75 |
| 621 | 4.20 | 0.94 | 4.20 | 0.79 | 4.05 | 0.79 | 4.35 | 0.83 | 4.15 | 0.75 |
| 622 | 4.21 | 0.93 | 4.21 | 0.79 | 4.06 | 0.79 | 4.35 | 0.82 | 4.15 | 0.75 |
| 623 | 4.22 | 0.93 | 4.21 | 0.78 | 4.06 | 0.78 | 4.36 | 0.82 | 4.15 | 0.75 |
| 624 | 4.23 | 0.92 | 4.22 | 0.78 | 4.07 | 0.78 | 4.36 | 0.82 | 4.16 | 0.74 |
| 625 | 4.24 | 0.91 | 4.22 | 0.77 | 4.07 | 0.77 | 4.36 | 0.81 | 4.16 | 0.74 |
| 626 | 4.25 | 0.90 | 4.22 | 0.77 | 4.07 | 0.77 | 4.37 | 0.81 | 4.16 | 0.74 |
| 627 | 4.25 | 0.89 | 4.23 | 0.76 | 4.08 | 0.76 | 4.37 | 0.80 | 4.16 | 0.74 |
| 628 | 4.26 | 0.88 | 4.23 | 0.76 | 4.08 | 0.76 | 4.37 | 0.80 | 4.17 | 0.73 |
| 629 | 4.27 | 0.87 | 4.24 | 0.75 | 4.08 | 0.75 | 4.37 | 0.79 | 4.17 | 0.73 |
| 630 | 4.27 | 0.86 | 4.24 | 0.75 | 4.09 | 0.75 | 4.38 | 0.79 | 4.17 | 0.73 |
| 631 | 4.28 | 0.85 | 4.24 | 0.74 | 4.09 | 0.74 | 4.38 | 0.79 | 4.17 | 0.73 |
| 632 | 4.28 | 0.84 | 4.25 | 0.74 | 4.09 | 0.74 | 4.38 | 0.78 | 4.18 | 0.72 |
| 633 | 4.29 | 0.83 | 4.25 | 0.73 | 4.10 | 0.73 | 4.38 | 0.78 | 4.18 | 0.72 |
| 634 | 4.29 | 0.82 | 4.25 | 0.72 | 4.10 | 0.73 | 4.39 | 0.77 | 4.18 | 0.72 |
| 635 | 4.30 | 0.81 | 4.26 | 0.72 | 4.10 | 0.72 | 4.39 | 0.76 | 4.18 | 0.71 |
| 636 | 4.30 | 0.80 | 4.26 | 0.71 | 4.10 | 0.71 | 4.39 | 0.76 | 4.19 | 0.71 |
| 637 | 4.30 | 0.79 | 4.26 | 0.70 | 4.11 | 0.71 | 4.39 | 0.75 | 4.19 | 0.70 |
| 638 | 4.30 | 0.78 | 4.26 | 0.70 | 4.11 | 0.70 | 4.40 | 0.75 | 4.19 | 0.70 |
| 639 | 4.30 | 0.76 | 4.27 | 0.69 | 4.11 | 0.70 | 4.40 | 0.74 | 4.19 | 0.70 |
| 640 | 4.30 | 0.75 | 4.27 | 0.69 | 4.11 | 0.69 | 4.40 | 0.74 | 4.20 | 0.69 |
| 641 | 4.30 | 0.74 | 4.27 | 0.68 | 4.11 | 0.68 | 4.40 | 0.73 | 4.20 | 0.69 |
| 642 | 4.30 | 0.73 | 4.27 | 0.67 | 4.12 | 0.68 | 4.40 | 0.73 | 4.20 | 0.68 |
| 643 | 4.30 | 0.72 | 4.27 | 0.67 | 4.12 | 0.67 | 4.40 | 0.72 | 4.20 | 0.68 |
| 644 | 4.29 | 0.72 | 4.28 | 0.66 | 4.12 | 0.66 | 4.41 | 0.71 | 4.21 | 0.67 |
| 645 | 4.29 | 0.71 | 4.28 | 0.65 | 4.12 | 0.66 | 4.41 | 0.71 | 4.21 | 0.67 |
| 646 | 4.29 | 0.70 | 4.28 | 0.65 | 4.12 | 0.65 | 4.41 | 0.70 | 4.21 | 0.66 |
| 647 | 4.28 | 0.69 | 4.28 | 0.64 | 4.12 | 0.65 | 4.41 | 0.70 | 4.21 | 0.66 |
| 648 | 4.28 | 0.69 | 4.28 | 0.63 | 4.12 | 0.64 | 4.41 | 0.69 | 4.22 | 0.65 |
| 649 | 4.27 | 0.68 | 4.28 | 0.63 | 4.12 | 0.63 | 4.41 | 0.69 | 4.22 | 0.65 |
| 650 | 4.27 | 0.68 | 4.28 | 0.62 | 4.12 | 0.63 | 4.41 | 0.68 | 4.22 | 0.64 |
| 651 | 4.26 | 0.68 | 4.28 | 0.62 | 4.12 | 0.62 | 4.41 | 0.67 | 4.22 | 0.64 |
| 652 | 4.26 | 0.67 | 4.28 | 0.61 | 4.12 | 0.62 | 4.41 | 0.67 | 4.22 | 0.63 |
| 653 | 4.26 | 0.67 | 4.28 | 0.60 | 4.13 | 0.61 | 4.41 | 0.66 | 4.22 | 0.63 |
| 654 | 4.26 | 0.67 | 4.28 | 0.60 | 4.13 | 0.60 | 4.41 | 0.66 | 4.23 | 0.62 |
| 655 | 4.25 | 0.67 | 4.28 | 0.59 | 4.13 | 0.60 | 4.41 | 0.65 | 4.23 | 0.62 |
| 656 | 4.25 | 0.67 | 4.28 | 0.59 | 4.13 | 0.59 | 4.41 | 0.64 | 4.23 | 0.61 |
| 657 | 4.25 | 0.68 | 4.28 | 0.58 | 4.13 | 0.59 | 4.41 | 0.64 | 4.23 | 0.61 |
| 658 | 4.25 | 0.68 | 4.28 | 0.58 | 4.13 | 0.58 | 4.41 | 0.63 | 4.23 | 0.60 |
| 659 | 4.26 | 0.68 | 4.29 | 0.57 | 4.13 | 0.58 | 4.41 | 0.63 | 4.23 | 0.59 |
| 660 | 4.26 | 0.68 | 4.29 | 0.57 | 4.13 | 0.57 | 4.41 | 0.62 | 4.23 | 0.59 |
| 661 | 4.27 | 0.69 | 4.29 | 0.57 | 4.13 | 0.57 | 4.41 | 0.62 | 4.23 | 0.58 |
| 662 | 4.28 | 0.69 | 4.29 | 0.56 | 4.13 | 0.56 | 4.41 | 0.61 | 4.23 | 0.58 |
| 663 | 4.28 | 0.69 | 4.29 | 0.56 | 4.13 | 0.55 | 4.41 | 0.61 | 4.23 | 0.57 |
| 664 | 4.30 | 0.69 | 4.30 | 0.55 | 4.13 | 0.55 | 4.41 | 0.60 | 4.24 | 0.56 |
| 665 | 4.31 | 0.68 | 4.30 | 0.55 | 4.13 | 0.54 | 4.41 | 0.59 | 4.24 | 0.56 |
| 666 | 4.32 | 0.68 | 4.30 | 0.54 | 4.13 | 0.54 | 4.41 | 0.59 | 4.24 | 0.55 |
| 667 | 4.33 | 0.67 | 4.31 | 0.53 | 4.13 | 0.53 | 4.41 | 0.58 | 4.24 | 0.55 |
| 668 | 4.35 | 0.67 | 4.31 | 0.53 | 4.14 | 0.52 | 4.41 | 0.58 | 4.24 | 0.54 |
| 669 | 4.36 | 0.66 | 4.32 | 0.52 | 4.14 | 0.52 | 4.41 | 0.57 | 4.24 | 0.53 |
| 670 | 4.37 | 0.64 | 4.32 | 0.51 | 4.14 | 0.51 | 4.41 | 0.57 | 4.24 | 0.53 |
| 671 | 4.38 | 0.63 | 4.32 | 0.50 | 4.14 | 0.50 | 4.41 | 0.56 | 4.24 | 0.52 |
| 672 | 4.39 | 0.62 | 4.32 | 0.49 | 4.14 | 0.50 | 4.41 | 0.56 | 4.24 | 0.52 |
| 673 | 4.40 | 0.60 | 4.33 | 0.48 | 4.14 | 0.49 | 4.41 | 0.55 | 4.24 | 0.51 |
| 674 | 4.41 | 0.58 | 4.33 | 0.47 | 4.14 | 0.48 | 4.41 | 0.55 | 4.23 | 0.50 |
| 675 | 4.41 | 0.56 | 4.33 | 0.46 | 4.14 | 0.47 | 4.41 | 0.54 | 4.23 | 0.50 |
| 676 | 4.41 | 0.55 | 4.33 | 0.45 | 4.14 | 0.47 | 4.41 | 0.54 | 4.23 | 0.49 |
| 677 | 4.42 | 0.53 | 4.32 | 0.44 | 4.14 | 0.46 | 4.41 | 0.53 | 4.23 | 0.49 |
| 678 | 4.42 | 0.51 | 4.32 | 0.43 | 4.14 | 0.45 | 4.41 | 0.53 | 4.23 | 0.48 |
| 679 | 4.41 | 0.49 | 4.32 | 0.42 | 4.13 | 0.44 | 4.41 | 0.52 | 4.23 | 0.47 |
| 680 | 4.41 | 0.47 | 4.32 | 0.41 | 4.13 | 0.44 | 4.41 | 0.52 | 4.23 | 0.47 |
| 681 | 4.41 | 0.46 | 4.32 | 0.40 | 4.13 | 0.43 | 4.41 | 0.51 | 4.23 | 0.46 |
| 682 | 4.40 | 0.44 | 4.31 | 0.39 | 4.13 | 0.42 | 4.41 | 0.50 | 4.23 | 0.46 |
| 683 | 4.40 | 0.43 | 4.31 | 0.38 | 4.13 | 0.41 | 4.41 | 0.50 | 4.23 | 0.45 |
| 684 | 4.39 | 0.41 | 4.31 | 0.37 | 4.12 | 0.41 | 4.41 | 0.49 | 4.23 | 0.45 |
| 685 | 4.39 | 0.40 | 4.30 | 0.37 | 4.12 | 0.40 | 4.41 | 0.49 | 4.22 | 0.44 |
| 686 | 4.38 | 0.38 | 4.30 | 0.36 | 4.12 | 0.39 | 4.41 | 0.48 | 4.22 | 0.43 |
| 687 | 4.37 | 0.37 | 4.29 | 0.35 | 4.12 | 0.39 | 4.41 | 0.47 | 4.22 | 0.43 |
| 688 | 4.37 | 0.36 | 4.29 | 0.34 | 4.11 | 0.38 | 4.41 | 0.47 | 4.22 | 0.42 |
| 689 | 4.36 | 0.35 | 4.29 | 0.34 | 4.11 | 0.37 | 4.41 | 0.46 | 4.22 | 0.42 |
| 690 | 4.35 | 0.34 | 4.28 | 0.33 | 4.11 | 0.37 | 4.41 | 0.45 | 4.22 | 0.41 |
| 691 | 4.34 | 0.33 | 4.28 | 0.32 | 4.10 | 0.36 | 4.40 | 0.45 | 4.21 | 0.41 |
| 692 | 4.33 | 0.32 | 4.27 | 0.32 | 4.10 | 0.36 | 4.40 | 0.44 | 4.21 | 0.40 |
| 693 | 4.32 | 0.31 | 4.27 | 0.31 | 4.10 | 0.35 | 4.40 | 0.43 | 4.21 | 0.40 |
| 694 | 4.32 | 0.30 | 4.27 | 0.30 | 4.09 | 0.34 | 4.40 | 0.43 | 4.21 | 0.39 |
| 695 | 4.31 | 0.29 | 4.26 | 0.30 | 4.09 | 0.34 | 4.40 | 0.42 | 4.21 | 0.39 |
| 696 | 4.30 | 0.29 | 4.26 | 0.29 | 4.09 | 0.33 | 4.40 | 0.42 | 4.20 | 0.38 |
| 697 | 4.29 | 0.28 | 4.25 | 0.29 | 4.08 | 0.33 | 4.39 | 0.41 | 4.20 | 0.37 |
| 698 | 4.28 | 0.27 | 4.25 | 0.28 | 4.08 | 0.32 | 4.39 | 0.40 | 4.20 | 0.37 |
| 699 | 4.28 | 0.27 | 4.24 | 0.28 | 4.08 | 0.32 | 4.39 | 0.40 | 4.20 | 0.36 |
| 700 | 4.27 | 0.26 | 4.24 | 0.27 | 4.07 | 0.31 | 4.39 | 0.39 | 4.19 | 0.36 |
| 701 | 4.26 | 0.26 | 4.24 | 0.27 | 4.07 | 0.31 | 4.38 | 0.39 | 4.19 | 0.35 |
| 702 | 4.25 | 0.25 | 4.23 | 0.26 | 4.07 | 0.30 | 4.38 | 0.38 | 4.19 | 0.35 |
| 703 | 4.25 | 0.25 | 4.23 | 0.26 | 4.06 | 0.30 | 4.38 | 0.38 | 4.19 | 0.34 |
| 704 | 4.24 | 0.24 | 4.22 | 0.25 | 4.06 | 0.29 | 4.37 | 0.37 | 4.18 | 0.34 |
| 705 | 4.23 | 0.24 | 4.22 | 0.25 | 4.06 | 0.29 | 4.37 | 0.36 | 4.18 | 0.34 |
| 706 | 4.22 | 0.23 | 4.21 | 0.24 | 4.05 | 0.29 | 4.37 | 0.36 | 4.18 | 0.33 |
| 707 | 4.22 | 0.23 | 4.21 | 0.24 | 4.05 | 0.28 | 4.37 | 0.35 | 4.18 | 0.33 |
| 708 | 4.21 | 0.22 | 4.21 | 0.23 | 4.05 | 0.28 | 4.36 | 0.35 | 4.17 | 0.32 |
| 709 | 4.20 | 0.22 | 4.20 | 0.23 | 4.04 | 0.27 | 4.36 | 0.34 | 4.17 | 0.32 |
| 710 | 4.20 | 0.22 | 4.20 | 0.22 | 4.04 | 0.27 | 4.36 | 0.34 | 4.17 | 0.31 |
| 711 | 4.19 | 0.21 | 4.19 | 0.22 | 4.03 | 0.27 | 4.35 | 0.33 | 4.17 | 0.31 |
| 712 | 4.18 | 0.21 | 4.19 | 0.22 | 4.03 | 0.26 | 4.35 | 0.33 | 4.16 | 0.30 |
| 713 | 4.18 | 0.21 | 4.19 | 0.21 | 4.03 | 0.26 | 4.35 | 0.33 | 4.16 | 0.30 |
| 714 | 4.17 | 0.21 | 4.18 | 0.21 | 4.02 | 0.25 | 4.34 | 0.32 | 4.16 | 0.29 |
| 715 | 4.16 | 0.20 | 4.18 | 0.20 | 4.02 | 0.25 | 4.34 | 0.32 | 4.15 | 0.29 |
| 716 | 4.16 | 0.20 | 4.17 | 0.20 | 4.02 | 0.25 | 4.34 | 0.31 | 4.15 | 0.29 |
| 717 | 4.15 | 0.20 | 4.17 | 0.20 | 4.01 | 0.24 | 4.33 | 0.31 | 4.15 | 0.28 |
| 718 | 4.15 | 0.19 | 4.16 | 0.19 | 4.01 | 0.24 | 4.33 | 0.30 | 4.14 | 0.28 |
| 719 | 4.14 | 0.19 | 4.16 | 0.19 | 4.01 | 0.24 | 4.33 | 0.30 | 4.14 | 0.27 |
| 720 | 4.14 | 0.19 | 4.16 | 0.19 | 4.00 | 0.23 | 4.32 | 0.30 | 4.14 | 0.27 |
| 721 | 4.13 | 0.19 | 4.15 | 0.18 | 4.00 | 0.23 | 4.32 | 0.29 | 4.14 | 0.27 |
| 722 | 4.13 | 0.19 | 4.15 | 0.18 | 4.00 | 0.23 | 4.32 | 0.29 | 4.13 | 0.26 |
| 723 | 4.12 | 0.18 | 4.14 | 0.18 | 3.99 | 0.22 | 4.31 | 0.28 | 4.13 | 0.26 |
| 724 | 4.11 | 0.18 | 4.14 | 0.17 | 3.99 | 0.22 | 4.31 | 0.28 | 4.13 | 0.26 |
| 725 | 4.11 | 0.18 | 4.14 | 0.17 | 3.99 | 0.22 | 4.31 | 0.28 | 4.12 | 0.25 |
| 726 | 4.11 | 0.18 | 4.13 | 0.17 | 3.98 | 0.21 | 4.30 | 0.27 | 4.12 | 0.25 |
| 727 | 4.10 | 0.18 | 4.13 | 0.16 | 3.98 | 0.21 | 4.30 | 0.27 | 4.12 | 0.24 |
| 728 | 4.10 | 0.18 | 4.13 | 0.16 | 3.98 | 0.21 | 4.30 | 0.27 | 4.11 | 0.24 |
| 729 | 4.09 | 0.17 | 4.12 | 0.16 | 3.97 | 0.20 | 4.29 | 0.26 | 4.11 | 0.24 |
| 730 | 4.09 | 0.17 | 4.12 | 0.16 | 3.97 | 0.20 | 4.29 | 0.26 | 4.11 | 0.23 |
| 731 | 4.08 | 0.17 | 4.11 | 0.15 | 3.97 | 0.20 | 4.29 | 0.26 | 4.10 | 0.23 |
| 732 | 4.08 | 0.17 | 4.11 | 0.15 | 3.96 | 0.20 | 4.28 | 0.25 | 4.10 | 0.23 |
| 733 | 4.07 | 0.17 | 4.11 | 0.15 | 3.96 | 0.19 | 4.28 | 0.25 | 4.10 | 0.22 |
| 734 | 4.07 | 0.17 | 4.10 | 0.14 | 3.96 | 0.19 | 4.28 | 0.25 | 4.09 | 0.22 |
| 735 | 4.07 | 0.16 | 4.10 | 0.14 | 3.95 | 0.19 | 4.27 | 0.24 | 4.09 | 0.22 |
| 736 | 4.06 | 0.16 | 4.09 | 0.14 | 3.95 | 0.19 | 4.27 | 0.24 | 4.09 | 0.22 |
| 737 | 4.06 | 0.16 | 4.09 | 0.14 | 3.95 | 0.18 | 4.27 | 0.24 | 4.08 | 0.21 |
| 738 | 4.05 | 0.16 | 4.09 | 0.13 | 3.94 | 0.18 | 4.26 | 0.24 | 4.08 | 0.21 |
| 739 | 4.05 | 0.16 | 4.08 | 0.13 | 3.94 | 0.18 | 4.26 | 0.23 | 4.08 | 0.21 |
| 740 | 4.05 | 0.16 | 4.08 | 0.13 | 3.94 | 0.18 | 4.26 | 0.23 | 4.07 | 0.20 |
| 741 | 4.04 | 0.16 | 4.08 | 0.13 | 3.93 | 0.17 | 4.25 | 0.23 | 4.07 | 0.20 |
| 742 | 4.04 | 0.16 | 4.07 | 0.13 | 3.93 | 0.17 | 4.25 | 0.22 | 4.07 | 0.20 |
| 743 | 4.03 | 0.15 | 4.07 | 0.12 | 3.93 | 0.17 | 4.25 | 0.22 | 4.06 | 0.20 |
| 744 | 4.03 | 0.15 | 4.07 | 0.12 | 3.92 | 0.17 | 4.24 | 0.22 | 4.06 | 0.19 |
| 745 | 4.03 | 0.15 | 4.06 | 0.12 | 3.92 | 0.16 | 4.24 | 0.22 | 4.06 | 0.19 |
| 746 | 4.02 | 0.15 | 4.06 | 0.12 | 3.92 | 0.16 | 4.24 | 0.21 | 4.05 | 0.19 |
| 747 | 4.02 | 0.15 | 4.05 | 0.11 | 3.92 | 0.16 | 4.23 | 0.21 | 4.05 | 0.19 |
| 748 | 4.02 | 0.15 | 4.05 | 0.11 | 3.91 | 0.16 | 4.23 | 0.21 | 4.05 | 0.18 |
| 749 | 4.01 | 0.15 | 4.05 | 0.11 | 3.91 | 0.16 | 4.23 | 0.21 | 4.04 | 0.18 |
| 750 | 4.01 | 0.15 | 4.04 | 0.11 | 3.91 | 0.15 | 4.23 | 0.20 | 4.04 | 0.18 |
| 751 | 4.01 | 0.15 | 4.04 | 0.11 | 3.90 | 0.15 | 4.22 | 0.20 | 4.04 | 0.18 |
| 752 | 4.00 | 0.14 | 4.04 | 0.11 | 3.90 | 0.15 | 4.22 | 0.20 | 4.03 | 0.17 |
| 753 | 4.00 | 0.14 | 4.03 | 0.10 | 3.90 | 0.15 | 4.22 | 0.20 | 4.03 | 0.17 |
| 754 | 4.00 | 0.14 | 4.03 | 0.10 | 3.89 | 0.15 | 4.21 | 0.19 | 4.03 | 0.17 |
| 755 | 3.99 | 0.14 | 4.03 | 0.10 | 3.89 | 0.14 | 4.21 | 0.19 | 4.02 | 0.17 |
| 756 | 3.99 | 0.14 | 4.02 | 0.10 | 3.89 | 0.14 | 4.21 | 0.19 | 4.02 | 0.17 |
| 757 | 3.99 | 0.14 | 4.02 | 0.10 | 3.89 | 0.14 | 4.20 | 0.19 | 4.02 | 0.16 |
| 758 | 3.98 | 0.14 | 4.02 | 0.09 | 3.88 | 0.14 | 4.20 | 0.19 | 4.01 | 0.16 |
| 759 | 3.98 | 0.14 | 4.01 | 0.09 | 3.88 | 0.14 | 4.20 | 0.18 | 4.01 | 0.16 |
| 760 | 3.98 | 0.14 | 4.01 | 0.09 | 3.88 | 0.13 | 4.20 | 0.18 | 4.01 | 0.16 |
| 761 | 3.97 | 0.14 | 4.01 | 0.09 | 3.87 | 0.13 | 4.19 | 0.18 | 4.01 | 0.16 |
| 762 | 3.97 | 0.13 | 4.01 | 0.09 | 3.87 | 0.13 | 4.19 | 0.18 | 4.00 | 0.15 |
| 763 | 3.97 | 0.13 | 4.00 | 0.09 | 3.87 | 0.13 | 4.19 | 0.18 | 4.00 | 0.15 |
| 764 | 3.97 | 0.13 | 4.00 | 0.09 | 3.87 | 0.13 | 4.18 | 0.17 | 4.00 | 0.15 |
| 765 | 3.96 | 0.13 | 4.00 | 0.08 | 3.86 | 0.13 | 4.18 | 0.17 | 3.99 | 0.15 |
| 766 | 3.96 | 0.13 | 3.99 | 0.08 | 3.86 | 0.12 | 4.18 | 0.17 | 3.99 | 0.15 |
| 767 | 3.96 | 0.13 | 3.99 | 0.08 | 3.86 | 0.12 | 4.18 | 0.17 | 3.99 | 0.15 |
| 768 | 3.96 | 0.13 | 3.99 | 0.08 | 3.85 | 0.12 | 4.17 | 0.17 | 3.98 | 0.14 |
| 769 | 3.95 | 0.13 | 3.98 | 0.08 | 3.85 | 0.12 | 4.17 | 0.16 | 3.98 | 0.14 |
| 770 | 3.95 | 0.13 | 3.98 | 0.08 | 3.85 | 0.12 | 4.17 | 0.16 | 3.98 | 0.14 |
| 771 | 3.95 | 0.13 | 3.98 | 0.08 | 3.85 | 0.12 | 4.16 | 0.16 | 3.97 | 0.14 |
| 772 | 3.94 | 0.13 | 3.97 | 0.07 | 3.84 | 0.11 | 4.16 | 0.16 | 3.97 | 0.14 |
| 773 | 3.94 | 0.12 | 3.97 | 0.07 | 3.84 | 0.11 | 4.16 | 0.16 | 3.97 | 0.14 |
| 774 | 3.94 | 0.12 | 3.97 | 0.07 | 3.84 | 0.11 | 4.16 | 0.15 | 3.97 | 0.13 |
| 775 | 3.94 | 0.12 | 3.97 | 0.07 | 3.84 | 0.11 | 4.15 | 0.15 | 3.96 | 0.13 |
| 776 | 3.93 | 0.12 | 3.96 | 0.07 | 3.83 | 0.11 | 4.15 | 0.15 | 3.96 | 0.13 |
| 777 | 3.93 | 0.12 | 3.96 | 0.07 | 3.83 | 0.11 | 4.15 | 0.15 | 3.96 | 0.13 |
| 778 | 3.93 | 0.12 | 3.96 | 0.07 | 3.83 | 0.11 | 4.14 | 0.15 | 3.95 | 0.13 |
| 779 | 3.93 | 0.12 | 3.95 | 0.07 | 3.83 | 0.10 | 4.14 | 0.15 | 3.95 | 0.13 |
| 780 | 3.92 | 0.12 | 3.95 | 0.06 | 3.82 | 0.10 | 4.14 | 0.14 | 3.95 | 0.13 |
| 781 | 3.92 | 0.12 | 3.95 | 0.06 | 3.82 | 0.10 | 4.14 | 0.14 | 3.95 | 0.13 |
| 782 | 3.92 | 0.12 | 3.95 | 0.06 | 3.82 | 0.10 | 4.13 | 0.14 | 3.94 | 0.12 |
| 783 | 3.92 | 0.12 | 3.94 | 0.06 | 3.82 | 0.10 | 4.13 | 0.14 | 3.94 | 0.12 |
| 784 | 3.91 | 0.12 | 3.94 | 0.06 | 3.81 | 0.10 | 4.13 | 0.14 | 3.94 | 0.12 |
| 785 | 3.91 | 0.12 | 3.94 | 0.06 | 3.81 | 0.10 | 4.13 | 0.14 | 3.93 | 0.12 |
| 786 | 3.91 | 0.11 | 3.94 | 0.06 | 3.81 | 0.10 | 4.12 | 0.14 | 3.93 | 0.12 |
| 787 | 3.91 | 0.11 | 3.93 | 0.06 | 3.81 | 0.09 | 4.12 | 0.13 | 3.93 | 0.12 |
| 788 | 3.91 | 0.11 | 3.93 | 0.06 | 3.80 | 0.09 | 4.12 | 0.13 | 3.93 | 0.12 |
| 789 | 3.90 | 0.11 | 3.93 | 0.05 | 3.80 | 0.09 | 4.12 | 0.13 | 3.92 | 0.12 |
| 790 | 3.90 | 0.11 | 3.92 | 0.05 | 3.80 | 0.09 | 4.11 | 0.13 | 3.92 | 0.12 |
| 791 | 3.90 | 0.11 | 3.92 | 0.05 | 3.80 | 0.09 | 4.11 | 0.13 | 3.92 | 0.11 |
| 792 | 3.90 | 0.11 | 3.92 | 0.05 | 3.79 | 0.09 | 4.11 | 0.13 | 3.92 | 0.11 |
| 793 | 3.89 | 0.11 | 3.92 | 0.05 | 3.79 | 0.09 | 4.11 | 0.13 | 3.91 | 0.11 |
| 794 | 3.89 | 0.11 | 3.91 | 0.05 | 3.79 | 0.09 | 4.10 | 0.12 | 3.91 | 0.11 |
| 795 | 3.89 | 0.11 | 3.91 | 0.05 | 3.79 | 0.09 | 4.10 | 0.12 | 3.91 | 0.11 |
| 796 | 3.89 | 0.11 | 3.91 | 0.05 | 3.78 | 0.08 | 4.10 | 0.12 | 3.90 | 0.11 |
| 797 | 3.89 | 0.11 | 3.91 | 0.05 | 3.78 | 0.08 | 4.10 | 0.12 | 3.90 | 0.11 |
| 798 | 3.88 | 0.11 | 3.90 | 0.05 | 3.78 | 0.08 | 4.09 | 0.12 | 3.90 | 0.11 |
| 799 | 3.88 | 0.11 | 3.90 | 0.05 | 3.78 | 0.08 | 4.09 | 0.12 | 3.90 | 0.11 |
| 800 | 3.88 | 0.10 | 3.90 | 0.04 | 3.78 | 0.08 | 4.09 | 0.12 | 3.89 | 0.11 |
| 801 | 3.88 | 0.10 | 3.90 | 0.04 | 3.77 | 0.08 | 4.09 | 0.12 | 3.89 | 0.11 |
| 802 | 3.88 | 0.10 | 3.89 | 0.04 | 3.77 | 0.08 | 4.08 | 0.11 | 3.89 | 0.11 |
| 803 | 3.87 | 0.10 | 3.89 | 0.04 | 3.77 | 0.08 | 4.08 | 0.11 | 3.89 | 0.10 |
| 804 | 3.87 | 0.10 | 3.89 | 0.04 | 3.77 | 0.08 | 4.08 | 0.11 | 3.88 | 0.10 |
| 805 | 3.87 | 0.10 | 3.89 | 0.04 | 3.76 | 0.07 | 4.08 | 0.11 | 3.88 | 0.10 |
| 806 | 3.87 | 0.10 | 3.88 | 0.04 | 3.76 | 0.07 | 4.08 | 0.11 | 3.88 | 0.10 |
| 807 | 3.87 | 0.10 | 3.88 | 0.04 | 3.76 | 0.07 | 4.07 | 0.11 | 3.88 | 0.10 |
| 808 | 3.86 | 0.10 | 3.88 | 0.04 | 3.76 | 0.07 | 4.07 | 0.11 | 3.88 | 0.10 |
| 809 | 3.86 | 0.10 | 3.88 | 0.04 | 3.76 | 0.07 | 4.07 | 0.11 | 3.87 | 0.10 |
| 810 | 3.86 | 0.10 | 3.87 | 0.04 | 3.75 | 0.07 | 4.07 | 0.11 | 3.87 | 0.10 |
| 811 | 3.86 | 0.10 | 3.87 | 0.04 | 3.75 | 0.07 | 4.06 | 0.10 | 3.87 | 0.10 |
| 812 | 3.86 | 0.10 | 3.87 | 0.04 | 3.75 | 0.07 | 4.06 | 0.10 | 3.87 | 0.10 |
| 813 | 3.85 | 0.10 | 3.87 | 0.04 | 3.75 | 0.07 | 4.06 | 0.10 | 3.86 | 0.10 |
| 814 | 3.85 | 0.10 | 3.87 | 0.03 | 3.74 | 0.07 | 4.06 | 0.10 | 3.86 | 0.10 |
| 815 | 3.85 | 0.10 | 3.86 | 0.03 | 3.74 | 0.07 | 4.06 | 0.10 | 3.86 | 0.10 |
| 816 | 3.85 | 0.09 | 3.86 | 0.03 | 3.74 | 0.07 | 4.05 | 0.10 | 3.86 | 0.10 |
| 817 | 3.85 | 0.09 | 3.86 | 0.03 | 3.74 | 0.06 | 4.05 | 0.10 | 3.85 | 0.10 |
| 818 | 3.84 | 0.09 | 3.86 | 0.03 | 3.74 | 0.06 | 4.05 | 0.10 | 3.85 | 0.10 |
| 819 | 3.84 | 0.09 | 3.85 | 0.03 | 3.73 | 0.06 | 4.05 | 0.10 | 3.85 | 0.09 |
| 820 | 3.84 | 0.09 | 3.85 | 0.03 | 3.73 | 0.06 | 4.04 | 0.09 | 3.85 | 0.09 |
| 821 | 3.84 | 0.09 | 3.85 | 0.03 | 3.73 | 0.06 | 4.04 | 0.09 | 3.85 | 0.09 |
| 822 | 3.84 | 0.09 | 3.85 | 0.03 | 3.73 | 0.06 | 4.04 | 0.09 | 3.84 | 0.09 |
| 823 | 3.84 | 0.09 | 3.85 | 0.03 | 3.73 | 0.06 | 4.04 | 0.09 | 3.84 | 0.09 |
| 824 | 3.83 | 0.09 | 3.84 | 0.03 | 3.72 | 0.06 | 4.04 | 0.09 | 3.84 | 0.09 |
| 825 | 3.83 | 0.09 | 3.84 | 0.03 | 3.72 | 0.06 | 4.03 | 0.09 | 3.84 | 0.09 |
| 826 | 3.83 | 0.09 | 3.84 | 0.03 | 3.72 | 0.06 | 4.03 | 0.09 | 3.84 | 0.09 |
| 827 | 3.83 | 0.09 | 3.84 | 0.03 | 3.72 | 0.06 | 4.03 | 0.09 | 3.83 | 0.09 |
| 828 | 3.83 | 0.09 | 3.84 | 0.03 | 3.72 | 0.06 | 4.03 | 0.09 | 3.83 | 0.09 |
| 829 | 3.82 | 0.09 | 3.83 | 0.03 | 3.71 | 0.06 | 4.03 | 0.09 | 3.83 | 0.09 |
| 830 | 3.82 | 0.09 | 3.83 | 0.03 | 3.71 | 0.05 | 4.02 | 0.09 | 3.83 | 0.09 |
| 831 | 3.82 | 0.09 | 3.83 | 0.03 | 3.71 | 0.05 | 4.02 | 0.08 | 3.83 | 0.09 |
| 832 | 3.82 | 0.09 | 3.83 | 0.02 | 3.71 | 0.05 | 4.02 | 0.08 | 3.82 | 0.09 |
| 833 | 3.82 | 0.09 | 3.83 | 0.02 | 3.71 | 0.05 | 4.02 | 0.08 | 3.82 | 0.09 |
| 834 | 3.82 | 0.08 | 3.82 | 0.02 | 3.71 | 0.05 | 4.02 | 0.08 | 3.82 | 0.09 |
| 835 | 3.81 | 0.08 | 3.82 | 0.02 | 3.70 | 0.05 | 4.01 | 0.08 | 3.82 | 0.09 |
| 836 | 3.81 | 0.08 | 3.82 | 0.02 | 3.70 | 0.05 | 4.01 | 0.08 | 3.82 | 0.09 |
| 837 | 3.81 | 0.08 | 3.82 | 0.02 | 3.70 | 0.05 | 4.01 | 0.08 | 3.81 | 0.09 |
| 838 | 3.81 | 0.08 | 3.82 | 0.02 | 3.70 | 0.05 | 4.01 | 0.08 | 3.81 | 0.09 |
| 839 | 3.81 | 0.08 | 3.81 | 0.02 | 3.70 | 0.05 | 4.01 | 0.08 | 3.81 | 0.09 |
| 840 | 3.81 | 0.08 | 3.81 | 0.02 | 3.69 | 0.05 | 4.00 | 0.08 | 3.81 | 0.09 |
| 841 | 3.81 | 0.08 | 3.81 | 0.02 | 3.69 | 0.05 | 4.00 | 0.08 | 3.81 | 0.09 |
| 842 | 3.80 | 0.08 | 3.81 | 0.02 | 3.69 | 0.05 | 4.00 | 0.08 | 3.80 | 0.09 |
| 843 | 3.80 | 0.08 | 3.81 | 0.02 | 3.69 | 0.05 | 4.00 | 0.07 | 3.80 | 0.09 |
| 844 | 3.80 | 0.08 | 3.80 | 0.02 | 3.69 | 0.05 | 4.00 | 0.07 | 3.80 | 0.08 |
| 845 | 3.80 | 0.08 | 3.80 | 0.02 | 3.69 | 0.04 | 3.99 | 0.07 | 3.80 | 0.08 |
| 846 | 3.80 | 0.08 | 3.80 | 0.02 | 3.68 | 0.04 | 3.99 | 0.07 | 3.80 | 0.08 |
| 847 | 3.80 | 0.08 | 3.80 | 0.02 | 3.68 | 0.04 | 3.99 | 0.07 | 3.80 | 0.08 |
| 848 | 3.79 | 0.08 | 3.80 | 0.02 | 3.68 | 0.04 | 3.99 | 0.07 | 3.79 | 0.08 |
| 849 | 3.79 | 0.08 | 3.80 | 0.02 | 3.68 | 0.04 | 3.99 | 0.07 | 3.79 | 0.08 |
| 850 | 3.79 | 0.08 | 3.79 | 0.02 | 3.68 | 0.04 | 3.99 | 0.07 | 3.79 | 0.08 |
| 851 | 3.79 | 0.08 | 3.79 | 0.02 | 3.67 | 0.04 | 3.98 | 0.07 | 3.79 | 0.08 |
| 852 | 3.79 | 0.08 | 3.79 | 0.02 | 3.67 | 0.04 | 3.98 | 0.07 | 3.79 | 0.08 |
| 853 | 3.79 | 0.08 | 3.79 | 0.02 | 3.67 | 0.04 | 3.98 | 0.07 | 3.79 | 0.08 |
| 854 | 3.79 | 0.08 | 3.79 | 0.02 | 3.67 | 0.04 | 3.98 | 0.07 | 3.78 | 0.08 |
| 855 | 3.78 | 0.07 | 3.78 | 0.02 | 3.67 | 0.04 | 3.98 | 0.07 | 3.78 | 0.08 |
| 856 | 3.78 | 0.07 | 3.78 | 0.02 | 3.67 | 0.04 | 3.97 | 0.07 | 3.78 | 0.08 |
| 857 | 3.78 | 0.07 | 3.78 | 0.02 | 3.66 | 0.04 | 3.97 | 0.07 | 3.78 | 0.08 |
| 858 | 3.78 | 0.07 | 3.78 | 0.01 | 3.66 | 0.04 | 3.97 | 0.06 | 3.78 | 0.08 |
| 859 | 3.78 | 0.07 | 3.78 | 0.01 | 3.66 | 0.04 | 3.97 | 0.06 | 3.78 | 0.08 |
| 860 | 3.78 | 0.07 | 3.78 | 0.01 | 3.66 | 0.04 | 3.97 | 0.06 | 3.77 | 0.08 |
| 861 | 3.78 | 0.07 | 3.77 | 0.01 | 3.66 | 0.04 | 3.97 | 0.06 | 3.77 | 0.08 |
| 862 | 3.77 | 0.07 | 3.77 | 0.01 | 3.66 | 0.04 | 3.96 | 0.06 | 3.77 | 0.08 |
| 863 | 3.77 | 0.07 | 3.77 | 0.01 | 3.66 | 0.04 | 3.96 | 0.06 | 3.77 | 0.08 |
| 864 | 3.77 | 0.07 | 3.77 | 0.01 | 3.65 | 0.04 | 3.96 | 0.06 | 3.77 | 0.08 |
| 865 | 3.77 | 0.07 | 3.77 | 0.01 | 3.65 | 0.03 | 3.96 | 0.06 | 3.77 | 0.08 |
| 866 | 3.77 | 0.07 | 3.77 | 0.01 | 3.65 | 0.03 | 3.96 | 0.06 | 3.77 | 0.08 |
| 867 | 3.77 | 0.07 | 3.76 | 0.01 | 3.65 | 0.03 | 3.96 | 0.06 | 3.76 | 0.08 |
| 868 | 3.77 | 0.07 | 3.76 | 0.01 | 3.65 | 0.03 | 3.95 | 0.06 | 3.76 | 0.08 |
| 869 | 3.76 | 0.07 | 3.76 | 0.01 | 3.65 | 0.03 | 3.95 | 0.06 | 3.76 | 0.08 |
| 870 | 3.76 | 0.07 | 3.76 | 0.01 | 3.64 | 0.03 | 3.95 | 0.06 | 3.76 | 0.08 |
| 871 | 3.76 | 0.07 | 3.76 | 0.01 | 3.64 | 0.03 | 3.95 | 0.06 | 3.76 | 0.08 |
| 872 | 3.76 | 0.07 | 3.76 | 0.01 | 3.64 | 0.03 | 3.95 | 0.06 | 3.76 | 0.08 |
| 873 | 3.76 | 0.07 | 3.75 | 0.01 | 3.64 | 0.03 | 3.95 | 0.06 | 3.75 | 0.08 |
| 874 | 3.76 | 0.07 | 3.75 | 0.01 | 3.64 | 0.03 | 3.94 | 0.05 | 3.75 | 0.08 |
| 875 | 3.76 | 0.07 | 3.75 | 0.01 | 3.64 | 0.03 | 3.94 | 0.05 | 3.75 | 0.08 |
| 876 | 3.76 | 0.07 | 3.75 | 0.01 | 3.64 | 0.03 | 3.94 | 0.05 | 3.75 | 0.08 |
| 877 | 3.75 | 0.07 | 3.75 | 0.01 | 3.63 | 0.03 | 3.94 | 0.05 | 3.75 | 0.08 |
| 878 | 3.75 | 0.06 | 3.75 | 0.01 | 3.63 | 0.03 | 3.94 | 0.05 | 3.75 | 0.08 |
| 879 | 3.75 | 0.06 | 3.75 | 0.01 | 3.63 | 0.03 | 3.94 | 0.05 | 3.75 | 0.08 |
| 880 | 3.75 | 0.06 | 3.74 | 0.01 | 3.63 | 0.03 | 3.93 | 0.05 | 3.74 | 0.08 |
| 881 | 3.75 | 0.06 | 3.74 | 0.01 | 3.63 | 0.03 | 3.93 | 0.05 | 3.74 | 0.08 |
| 882 | 3.75 | 0.06 | 3.74 | 0.01 | 3.63 | 0.03 | 3.93 | 0.05 | 3.74 | 0.08 |
| 883 | 3.75 | 0.06 | 3.74 | 0.01 | 3.62 | 0.03 | 3.93 | 0.05 | 3.74 | 0.08 |
| 884 | 3.75 | 0.06 | 3.74 | 0.01 | 3.62 | 0.03 | 3.93 | 0.05 | 3.74 | 0.08 |
| 885 | 3.74 | 0.06 | 3.74 | 0.01 | 3.62 | 0.03 | 3.93 | 0.05 | 3.74 | 0.08 |
| 886 | 3.74 | 0.06 | 3.74 | 0.01 | 3.62 | 0.03 | 3.92 | 0.05 | 3.74 | 0.08 |
| 887 | 3.74 | 0.06 | 3.73 | 0.01 | 3.62 | 0.03 | 3.92 | 0.05 | 3.74 | 0.08 |
| 888 | 3.74 | 0.06 | 3.73 | 0.01 | 3.62 | 0.03 | 3.92 | 0.05 | 3.73 | 0.08 |
| 889 | 3.74 | 0.06 | 3.73 | 0.01 | 3.62 | 0.03 | 3.92 | 0.05 | 3.73 | 0.08 |
| 890 | 3.74 | 0.06 | 3.73 | 0.01 | 3.62 | 0.02 | 3.92 | 0.05 | 3.73 | 0.08 |
| 891 | 3.74 | 0.06 | 3.73 | 0.01 | 3.61 | 0.02 | 3.92 | 0.05 | 3.73 | 0.08 |
| 892 | 3.74 | 0.06 | 3.73 | 0.01 | 3.61 | 0.02 | 3.92 | 0.05 | 3.73 | 0.08 |
| 893 | 3.73 | 0.06 | 3.73 | 0.01 | 3.61 | 0.02 | 3.91 | 0.05 | 3.73 | 0.08 |
| 894 | 3.73 | 0.06 | 3.72 | 0.01 | 3.61 | 0.02 | 3.91 | 0.04 | 3.73 | 0.08 |
| 895 | 3.73 | 0.06 | 3.72 | 0.01 | 3.61 | 0.02 | 3.91 | 0.04 | 3.73 | 0.07 |
| 896 | 3.73 | 0.06 | 3.72 | 0.01 | 3.61 | 0.02 | 3.91 | 0.04 | 3.72 | 0.07 |
| 897 | 3.73 | 0.06 | 3.72 | 0.01 | 3.61 | 0.02 | 3.91 | 0.04 | 3.72 | 0.07 |
| 898 | 3.73 | 0.06 | 3.72 | 0.01 | 3.60 | 0.02 | 3.91 | 0.04 | 3.72 | 0.07 |
| 899 | 3.73 | 0.06 | 3.72 | 0.01 | 3.60 | 0.02 | 3.91 | 0.04 | 3.72 | 0.07 |
| 900 | 3.73 | 0.06 | 3.72 | 0.01 | 3.60 | 0.02 | 3.90 | 0.04 | 3.72 | 0.07 |
| 901 | 3.72 | 0.06 | 3.71 | 0.01 | 3.60 | 0.02 | 3.90 | 0.04 | 3.72 | 0.07 |
| 902 | 3.72 | 0.06 | 3.71 | 0.01 | 3.60 | 0.02 | 3.90 | 0.04 | 3.72 | 0.07 |
| 903 | 3.72 | 0.06 | 3.71 | 0.01 | 3.60 | 0.02 | 3.90 | 0.04 | 3.72 | 0.07 |
| 904 | 3.72 | 0.05 | 3.71 | 0.01 | 3.60 | 0.02 | 3.90 | 0.04 | 3.71 | 0.07 |
| 905 | 3.72 | 0.05 | 3.71 | 0.01 | 3.60 | 0.02 | 3.90 | 0.04 | 3.71 | 0.07 |
| 906 | 3.72 | 0.05 | 3.71 | 0.01 | 3.59 | 0.02 | 3.90 | 0.04 | 3.71 | 0.07 |
| 907 | 3.72 | 0.05 | 3.71 | 0.01 | 3.59 | 0.02 | 3.89 | 0.04 | 3.71 | 0.07 |
| 908 | 3.72 | 0.05 | 3.71 | 0.01 | 3.59 | 0.02 | 3.89 | 0.04 | 3.71 | 0.07 |
| 909 | 3.72 | 0.05 | 3.70 | 0.01 | 3.59 | 0.02 | 3.89 | 0.04 | 3.71 | 0.07 |
| 910 | 3.71 | 0.05 | 3.70 | 0.01 | 3.59 | 0.02 | 3.89 | 0.04 | 3.71 | 0.07 |
| 911 | 3.71 | 0.05 | 3.70 | 0.01 | 3.59 | 0.02 | 3.89 | 0.04 | 3.71 | 0.07 |
| 912 | 3.71 | 0.05 | 3.70 | 0.00 | 3.59 | 0.02 | 3.89 | 0.04 | 3.71 | 0.07 |
| 913 | 3.71 | 0.05 | 3.70 | 0.00 | 3.59 | 0.02 | 3.89 | 0.04 | 3.70 | 0.07 |
| 914 | 3.71 | 0.05 | 3.70 | 0.00 | 3.58 | 0.02 | 3.88 | 0.04 | 3.70 | 0.07 |
| 915 | 3.71 | 0.05 | 3.70 | 0.00 | 3.58 | 0.02 | 3.88 | 0.04 | 3.70 | 0.07 |
| 916 | 3.71 | 0.05 | 3.70 | 0.00 | 3.58 | 0.02 | 3.88 | 0.04 | 3.70 | 0.07 |
| 917 | 3.71 | 0.05 | 3.69 | 0.00 | 3.58 | 0.02 | 3.88 | 0.04 | 3.70 | 0.07 |
| 918 | 3.71 | 0.05 | 3.69 | 0.00 | 3.58 | 0.02 | 3.88 | 0.04 | 3.70 | 0.07 |
| 919 | 3.70 | 0.05 | 3.69 | 0.00 | 3.58 | 0.02 | 3.88 | 0.03 | 3.70 | 0.07 |
| 920 | 3.70 | 0.05 | 3.69 | 0.00 | 3.58 | 0.02 | 3.88 | 0.03 | 3.70 | 0.07 |
| 921 | 3.70 | 0.05 | 3.69 | 0.00 | 3.58 | 0.02 | 3.88 | 0.03 | 3.70 | 0.07 |
| 922 | 3.70 | 0.05 | 3.69 | 0.00 | 3.57 | 0.02 | 3.87 | 0.03 | 3.69 | 0.07 |
| 923 | 3.70 | 0.05 | 3.69 | 0.00 | 3.57 | 0.02 | 3.87 | 0.03 | 3.69 | 0.07 |
| 924 | 3.70 | 0.05 | 3.69 | 0.00 | 3.57 | 0.02 | 3.87 | 0.03 | 3.69 | 0.07 |
| 925 | 3.70 | 0.05 | 3.68 | 0.00 | 3.57 | 0.02 | 3.87 | 0.03 | 3.69 | 0.07 |
| 926 | 3.70 | 0.05 | 3.68 | 0.00 | 3.57 | 0.02 | 3.87 | 0.03 | 3.69 | 0.07 |
| 927 | 3.70 | 0.05 | 3.68 | 0.00 | 3.57 | 0.01 | 3.87 | 0.03 | 3.69 | 0.07 |
| 928 | 3.70 | 0.05 | 3.68 | 0.00 | 3.57 | 0.01 | 3.87 | 0.03 | 3.69 | 0.07 |
| 929 | 3.69 | 0.05 | 3.68 | 0.00 | 3.57 | 0.01 | 3.86 | 0.03 | 3.69 | 0.07 |
| 930 | 3.69 | 0.05 | 3.68 | 0.00 | 3.57 | 0.01 | 3.86 | 0.03 | 3.69 | 0.07 |
| 931 | 3.69 | 0.05 | 3.68 | 0.00 | 3.56 | 0.01 | 3.86 | 0.03 | 3.69 | 0.07 |
| 932 | 3.69 | 0.05 | 3.68 | 0.00 | 3.56 | 0.01 | 3.86 | 0.03 | 3.68 | 0.07 |
| 933 | 3.69 | 0.05 | 3.68 | 0.00 | 3.56 | 0.01 | 3.86 | 0.03 | 3.68 | 0.07 |
| 934 | 3.69 | 0.04 | 3.67 | 0.00 | 3.56 | 0.01 | 3.86 | 0.03 | 3.68 | 0.07 |
| 935 | 3.69 | 0.04 | 3.67 | 0.00 | 3.56 | 0.01 | 3.86 | 0.03 | 3.68 | 0.07 |
| 936 | 3.69 | 0.04 | 3.67 | 0.00 | 3.56 | 0.01 | 3.86 | 0.03 | 3.68 | 0.07 |
| 937 | 3.69 | 0.04 | 3.67 | 0.00 | 3.56 | 0.01 | 3.85 | 0.03 | 3.68 | 0.07 |
| 938 | 3.69 | 0.04 | 3.67 | 0.00 | 3.56 | 0.01 | 3.85 | 0.03 | 3.68 | 0.07 |
| 939 | 3.68 | 0.04 | 3.67 | 0.00 | 3.56 | 0.01 | 3.85 | 0.03 | 3.68 | 0.07 |
| 940 | 3.68 | 0.04 | 3.67 | 0.00 | 3.55 | 0.01 | 3.85 | 0.03 | 3.68 | 0.07 |
| 941 | 3.68 | 0.04 | 3.67 | 0.00 | 3.55 | 0.01 | 3.85 | 0.03 | 3.68 | 0.07 |
| 942 | 3.68 | 0.04 | 3.67 | 0.00 | 3.55 | 0.01 | 3.85 | 0.03 | 3.67 | 0.07 |
| 943 | 3.68 | 0.04 | 3.67 | 0.00 | 3.55 | 0.01 | 3.85 | 0.03 | 3.67 | 0.07 |
| 944 | 3.68 | 0.04 | 3.66 | 0.00 | 3.55 | 0.01 | 3.85 | 0.03 | 3.67 | 0.07 |
| 945 | 3.68 | 0.04 | 3.66 | 0.00 | 3.55 | 0.01 | 3.85 | 0.03 | 3.67 | 0.07 |
| 946 | 3.68 | 0.04 | 3.66 | 0.00 | 3.55 | 0.01 | 3.84 | 0.03 | 3.67 | 0.07 |
| 947 | 3.68 | 0.04 | 3.66 | 0.00 | 3.55 | 0.01 | 3.84 | 0.03 | 3.67 | 0.07 |
| 948 | 3.68 | 0.04 | 3.66 | 0.00 | 3.55 | 0.01 | 3.84 | 0.03 | 3.67 | 0.07 |
| 949 | 3.67 | 0.04 | 3.66 | 0.00 | 3.54 | 0.01 | 3.84 | 0.03 | 3.67 | 0.07 |
| 950 | 3.67 | 0.04 | 3.66 | 0.00 | 3.54 | 0.01 | 3.84 | 0.03 | 3.67 | 0.07 |
| 951 | 3.67 | 0.04 | 3.66 | 0.00 | 3.54 | 0.01 | 3.84 | 0.02 | 3.67 | 0.07 |
| 952 | 3.67 | 0.04 | 3.66 | 0.00 | 3.54 | 0.01 | 3.84 | 0.02 | 3.67 | 0.07 |
| 953 | 3.67 | 0.04 | 3.66 | 0.00 | 3.54 | 0.01 | 3.84 | 0.02 | 3.66 | 0.07 |
| 954 | 3.67 | 0.04 | 3.65 | 0.00 | 3.54 | 0.01 | 3.83 | 0.02 | 3.66 | 0.07 |
| 955 | 3.67 | 0.04 | 3.65 | 0.00 | 3.54 | 0.01 | 3.83 | 0.02 | 3.66 | 0.07 |
| 956 | 3.67 | 0.04 | 3.65 | 0.00 | 3.54 | 0.01 | 3.83 | 0.02 | 3.66 | 0.07 |
| 957 | 3.67 | 0.04 | 3.65 | 0.00 | 3.54 | 0.01 | 3.83 | 0.02 | 3.66 | 0.07 |
| 958 | 3.67 | 0.04 | 3.65 | 0.00 | 3.54 | 0.01 | 3.83 | 0.02 | 3.66 | 0.07 |
| 959 | 3.67 | 0.04 | 3.65 | 0.00 | 3.53 | 0.01 | 3.83 | 0.02 | 3.66 | 0.07 |
| 960 | 3.66 | 0.04 | 3.65 | 0.00 | 3.53 | 0.01 | 3.83 | 0.02 | 3.66 | 0.07 |
| 961 | 3.66 | 0.04 | 3.65 | 0.00 | 3.53 | 0.01 | 3.83 | 0.02 | 3.66 | 0.07 |
| 962 | 3.66 | 0.04 | 3.65 | 0.00 | 3.53 | 0.01 | 3.83 | 0.02 | 3.66 | 0.07 |
| 963 | 3.66 | 0.04 | 3.65 | 0.00 | 3.53 | 0.01 | 3.82 | 0.02 | 3.66 | 0.07 |
| 964 | 3.66 | 0.04 | 3.65 | 0.00 | 3.53 | 0.01 | 3.82 | 0.02 | 3.65 | 0.07 |
| 965 | 3.66 | 0.04 | 3.64 | 0.00 | 3.53 | 0.01 | 3.82 | 0.02 | 3.65 | 0.07 |
| 966 | 3.66 | 0.04 | 3.64 | 0.00 | 3.53 | 0.01 | 3.82 | 0.02 | 3.65 | 0.07 |
| 967 | 3.66 | 0.04 | 3.64 | 0.00 | 3.53 | 0.01 | 3.82 | 0.02 | 3.65 | 0.07 |
| 968 | 3.66 | 0.04 | 3.64 | 0.00 | 3.53 | 0.01 | 3.82 | 0.02 | 3.65 | 0.07 |
| 969 | 3.66 | 0.04 | 3.64 | 0.00 | 3.52 | 0.01 | 3.82 | 0.02 | 3.65 | 0.07 |
| 970 | 3.66 | 0.04 | 3.64 | 0.00 | 3.52 | 0.01 | 3.82 | 0.02 | 3.65 | 0.07 |
| 971 | 3.65 | 0.03 | 3.64 | 0.00 | 3.52 | 0.01 | 3.82 | 0.02 | 3.65 | 0.07 |
| 972 | 3.65 | 0.03 | 3.64 | 0.00 | 3.52 | 0.01 | 3.82 | 0.02 | 3.65 | 0.07 |
| 973 | 3.65 | 0.03 | 3.64 | 0.00 | 3.52 | 0.01 | 3.81 | 0.02 | 3.65 | 0.07 |
| 974 | 3.65 | 0.03 | 3.64 | 0.00 | 3.52 | 0.01 | 3.81 | 0.02 | 3.65 | 0.07 |
| 975 | 3.65 | 0.03 | 3.64 | 0.00 | 3.52 | 0.01 | 3.81 | 0.02 | 3.65 | 0.07 |
| 976 | 3.65 | 0.03 | 3.63 | 0.00 | 3.52 | 0.01 | 3.81 | 0.02 | 3.64 | 0.07 |
| 977 | 3.65 | 0.03 | 3.63 | 0.00 | 3.52 | 0.01 | 3.81 | 0.02 | 3.64 | 0.07 |
| 978 | 3.65 | 0.03 | 3.63 | 0.00 | 3.52 | 0.01 | 3.81 | 0.02 | 3.64 | 0.07 |
| 979 | 3.65 | 0.03 | 3.63 | 0.00 | 3.52 | 0.01 | 3.81 | 0.02 | 3.64 | 0.07 |
| 980 | 3.65 | 0.03 | 3.63 | 0.00 | 3.51 | 0.01 | 3.81 | 0.02 | 3.64 | 0.07 |
| 981 | 3.65 | 0.03 | 3.63 | 0.00 | 3.51 | 0.01 | 3.81 | 0.02 | 3.64 | 0.07 |
| 982 | 3.65 | 0.03 | 3.63 | 0.00 | 3.51 | 0.01 | 3.80 | 0.02 | 3.64 | 0.07 |
| 983 | 3.64 | 0.03 | 3.63 | 0.00 | 3.51 | 0.01 | 3.80 | 0.02 | 3.64 | 0.07 |
| 984 | 3.64 | 0.03 | 3.63 | 0.00 | 3.51 | 0.01 | 3.80 | 0.02 | 3.64 | 0.07 |
| 985 | 3.64 | 0.03 | 3.63 | 0.00 | 3.51 | 0.01 | 3.80 | 0.02 | 3.64 | 0.07 |
| 986 | 3.64 | 0.03 | 3.63 | 0.00 | 3.51 | 0.01 | 3.80 | 0.02 | 3.64 | 0.07 |
| 987 | 3.64 | 0.03 | 3.63 | 0.00 | 3.51 | 0.01 | 3.80 | 0.02 | 3.64 | 0.07 |
| 988 | 3.64 | 0.03 | 3.63 | 0.00 | 3.51 | 0.01 | 3.80 | 0.02 | 3.64 | 0.07 |
| 989 | 3.64 | 0.03 | 3.62 | 0.00 | 3.51 | 0.01 | 3.80 | 0.02 | 3.63 | 0.07 |
| 990 | 3.64 | 0.03 | 3.62 | 0.00 | 3.51 | 0.01 | 3.80 | 0.02 | 3.63 | 0.07 |
| 991 | 3.64 | 0.03 | 3.62 | 0.00 | 3.51 | 0.01 | 3.80 | 0.02 | 3.63 | 0.07 |
| 992 | 3.64 | 0.03 | 3.62 | 0.00 | 3.50 | 0.01 | 3.79 | 0.02 | 3.63 | 0.07 |
| 993 | 3.64 | 0.03 | 3.62 | 0.00 | 3.50 | 0.01 | 3.79 | 0.02 | 3.63 | 0.07 |
| 994 | 3.64 | 0.03 | 3.62 | 0.00 | 3.50 | 0.01 | 3.79 | 0.02 | 3.63 | 0.07 |
| 995 | 3.63 | 0.03 | 3.62 | 0.00 | 3.50 | 0.01 | 3.79 | 0.02 | 3.63 | 0.07 |
| 996 | 3.63 | 0.03 | 3.62 | 0.00 | 3.50 | 0.01 | 3.79 | 0.02 | 3.63 | 0.07 |
| 997 | 3.63 | 0.03 | 3.62 | 0.00 | 3.50 | 0.01 | 3.79 | 0.02 | 3.63 | 0.07 |
| 998 | 3.63 | 0.03 | 3.62 | 0.00 | 3.50 | 0.01 | 3.79 | 0.02 | 3.63 | 0.07 |
| 999 | 3.63 | 0.03 | 3.62 | 0.00 | 3.50 | 0.01 | 3.79 | 0.02 | 3.63 | 0.07 |
| 1000 | 3.63 | 0.03 | 3.62 | 0.00 | 3.50 | 0.01 | 3.79 | 0.01 | 3.63 | 0.07 |
| 1002.5 | 3.63 | 0.03 | 3.61 | 0.00 | 3.50 | 0.01 | 3.78 | 0.01 | 3.63 | 0.07 |
| 1005 | 3.63 | 0.03 | 3.61 | 0.00 | 3.49 | 0.00 | 3.78 | 0.01 | 3.62 | 0.06 |
| 1007.5 | 3.62 | 0.03 | 3.61 | 0.00 | 3.49 | 0.00 | 3.78 | 0.01 | 3.62 | 0.06 |
| 1010 | 3.62 | 0.03 | 3.61 | 0.00 | 3.49 | 0.00 | 3.78 | 0.01 | 3.62 | 0.06 |
| 1012.5 | 3.62 | 0.03 | 3.61 | 0.00 | 3.49 | 0.00 | 3.78 | 0.01 | 3.62 | 0.06 |
| 1015 | 3.62 | 0.03 | 3.61 | 0.00 | 3.49 | 0.00 | 3.77 | 0.01 | 3.62 | 0.06 |
| 1017.5 | 3.62 | 0.02 | 3.60 | 0.00 | 3.48 | 0.00 | 3.77 | 0.01 | 3.61 | 0.06 |
| 1020 | 3.62 | 0.02 | 3.60 | 0.00 | 3.48 | 0.00 | 3.77 | 0.01 | 3.61 | 0.06 |
| 1022.5 | 3.61 | 0.02 | 3.60 | 0.00 | 3.48 | 0.00 | 3.77 | 0.01 | 3.61 | 0.06 |
| 1025 | 3.61 | 0.02 | 3.60 | 0.00 | 3.48 | 0.00 | 3.76 | 0.01 | 3.61 | 0.06 |
| 1027.5 | 3.61 | 0.02 | 3.60 | 0.00 | 3.48 | 0.00 | 3.76 | 0.01 | 3.61 | 0.06 |
| 1030 | 3.61 | 0.02 | 3.60 | 0.00 | 3.48 | 0.00 | 3.76 | 0.01 | 3.61 | 0.06 |
| 1032.5 | 3.61 | 0.02 | 3.59 | 0.00 | 3.47 | 0.00 | 3.76 | 0.01 | 3.60 | 0.06 |
| 1035 | 3.60 | 0.02 | 3.59 | 0.00 | 3.47 | 0.00 | 3.76 | 0.01 | 3.60 | 0.06 |
| 1037.5 | 3.60 | 0.02 | 3.59 | 0.00 | 3.47 | 0.00 | 3.75 | 0.01 | 3.60 | 0.06 |
| 1040 | 3.60 | 0.02 | 3.59 | 0.00 | 3.47 | 0.00 | 3.75 | 0.01 | 3.60 | 0.06 |
| 1042.5 | 3.60 | 0.02 | 3.59 | 0.00 | 3.47 | 0.00 | 3.75 | 0.01 | 3.60 | 0.06 |
| 1045 | 3.60 | 0.02 | 3.59 | 0.00 | 3.47 | 0.00 | 3.75 | 0.01 | 3.60 | 0.06 |
| 1047.5 | 3.60 | 0.02 | 3.59 | 0.00 | 3.46 | 0.00 | 3.75 | 0.01 | 3.60 | 0.06 |
| 1050 | 3.59 | 0.02 | 3.58 | 0.00 | 3.46 | 0.00 | 3.74 | 0.01 | 3.59 | 0.06 |
| 1052.5 | 3.59 | 0.02 | 3.58 | 0.00 | 3.46 | 0.00 | 3.74 | 0.01 | 3.59 | 0.06 |
| 1055 | 3.59 | 0.02 | 3.58 | 0.00 | 3.46 | 0.00 | 3.74 | 0.01 | 3.59 | 0.06 |
| 1057.5 | 3.59 | 0.02 | 3.58 | 0.00 | 3.46 | 0.00 | 3.74 | 0.01 | 3.59 | 0.06 |
| 1060 | 3.59 | 0.02 | 3.58 | 0.00 | 3.46 | 0.00 | 3.74 | 0.01 | 3.59 | 0.06 |
| 1062.5 | 3.59 | 0.02 | 3.58 | 0.00 | 3.46 | 0.00 | 3.73 | 0.01 | 3.59 | 0.06 |
| 1065 | 3.58 | 0.02 | 3.58 | 0.00 | 3.45 | 0.00 | 3.73 | 0.01 | 3.59 | 0.06 |
| 1067.5 | 3.58 | 0.02 | 3.57 | 0.00 | 3.45 | 0.00 | 3.73 | 0.01 | 3.58 | 0.06 |
| 1070 | 3.58 | 0.02 | 3.57 | 0.00 | 3.45 | 0.00 | 3.73 | 0.01 | 3.58 | 0.06 |
| 1072.5 | 3.58 | 0.02 | 3.57 | 0.00 | 3.45 | 0.00 | 3.73 | 0.01 | 3.58 | 0.06 |
| 1075 | 3.58 | 0.02 | 3.57 | 0.00 | 3.45 | 0.00 | 3.73 | 0.01 | 3.58 | 0.06 |
| 1077.5 | 3.58 | 0.02 | 3.57 | 0.00 | 3.45 | 0.00 | 3.72 | 0.01 | 3.58 | 0.06 |
| 1080 | 3.58 | 0.02 | 3.57 | 0.00 | 3.45 | 0.00 | 3.72 | 0.01 | 3.58 | 0.06 |
| 1082.5 | 3.57 | 0.02 | 3.57 | 0.00 | 3.44 | 0.00 | 3.72 | 0.01 | 3.58 | 0.06 |
| 1085 | 3.57 | 0.01 | 3.57 | 0.00 | 3.44 | 0.00 | 3.72 | 0.01 | 3.57 | 0.06 |
| 1087.5 | 3.57 | 0.01 | 3.56 | 0.00 | 3.44 | 0.00 | 3.72 | 0.01 | 3.57 | 0.06 |
| 1090 | 3.57 | 0.01 | 3.56 | 0.00 | 3.44 | 0.00 | 3.72 | 0.01 | 3.57 | 0.06 |
| 1092.5 | 3.57 | 0.01 | 3.56 | 0.00 | 3.44 | 0.00 | 3.71 | 0.01 | 3.57 | 0.06 |
| 1095 | 3.57 | 0.01 | 3.56 | 0.00 | 3.44 | 0.00 | 3.71 | 0.01 | 3.57 | 0.06 |
| 1097.5 | 3.57 | 0.01 | 3.56 | 0.00 | 3.44 | 0.00 | 3.71 | 0.01 | 3.57 | 0.06 |
| 1100 | 3.56 | 0.01 | 3.56 | 0.00 | 3.44 | 0.00 | 3.71 | 0.01 | 3.57 | 0.06 |
| 1102.5 | 3.56 | 0.01 | 3.56 | 0.00 | 3.43 | 0.00 | 3.71 | 0.01 | 3.57 | 0.06 |
| 1105 | 3.56 | 0.01 | 3.56 | 0.00 | 3.43 | 0.00 | 3.71 | 0.01 | 3.56 | 0.06 |
| 1107.5 | 3.56 | 0.01 | 3.55 | 0.00 | 3.43 | 0.00 | 3.70 | 0.01 | 3.56 | 0.06 |
| 1110 | 3.56 | 0.01 | 3.55 | 0.00 | 3.43 | 0.00 | 3.70 | 0.01 | 3.56 | 0.06 |
| 1112.5 | 3.56 | 0.01 | 3.55 | 0.00 | 3.43 | 0.00 | 3.70 | 0.01 | 3.56 | 0.06 |
| 1115 | 3.56 | 0.01 | 3.55 | 0.00 | 3.43 | 0.00 | 3.70 | 0.01 | 3.56 | 0.06 |
| 1117.5 | 3.55 | 0.01 | 3.55 | 0.00 | 3.43 | 0.00 | 3.70 | 0.01 | 3.56 | 0.06 |
| 1120 | 3.55 | 0.01 | 3.55 | 0.00 | 3.43 | 0.00 | 3.70 | 0.00 | 3.56 | 0.06 |
| 1122.5 | 3.55 | 0.01 | 3.55 | 0.00 | 3.43 | 0.00 | 3.70 | 0.00 | 3.56 | 0.06 |
| 1125 | 3.55 | 0.01 | 3.55 | 0.00 | 3.42 | 0.00 | 3.69 | 0.00 | 3.55 | 0.06 |
| 1127.5 | 3.55 | 0.01 | 3.55 | 0.00 | 3.42 | 0.00 | 3.69 | 0.00 | 3.55 | 0.06 |
| 1130 | 3.55 | 0.01 | 3.54 | 0.00 | 3.42 | 0.00 | 3.69 | 0.00 | 3.55 | 0.06 |
| 1132.5 | 3.55 | 0.01 | 3.54 | 0.00 | 3.42 | 0.00 | 3.69 | 0.00 | 3.55 | 0.06 |
| 1135 | 3.55 | 0.01 | 3.54 | 0.00 | 3.42 | 0.00 | 3.69 | 0.00 | 3.55 | 0.06 |
| 1137.5 | 3.54 | 0.01 | 3.54 | 0.00 | 3.42 | 0.00 | 3.69 | 0.00 | 3.55 | 0.06 |
| 1140 | 3.54 | 0.01 | 3.54 | 0.00 | 3.42 | 0.00 | 3.69 | 0.00 | 3.55 | 0.06 |
| 1142.5 | 3.54 | 0.01 | 3.54 | 0.00 | 3.42 | 0.00 | 3.68 | 0.00 | 3.55 | 0.06 |
| 1145 | 3.54 | 0.01 | 3.54 | 0.00 | 3.42 | 0.00 | 3.68 | 0.00 | 3.55 | 0.06 |
| 1147.5 | 3.54 | 0.01 | 3.54 | 0.00 | 3.41 | 0.00 | 3.68 | 0.00 | 3.54 | 0.06 |
| 1150 | 3.54 | 0.01 | 3.54 | 0.00 | 3.41 | 0.00 | 3.68 | 0.00 | 3.54 | 0.06 |
| 1152.5 | 3.54 | 0.01 | 3.54 | 0.00 | 3.41 | 0.00 | 3.68 | 0.00 | 3.54 | 0.06 |
| 1155 | 3.54 | 0.01 | 3.53 | 0.00 | 3.41 | 0.00 | 3.68 | 0.00 | 3.54 | 0.06 |
| 1157.5 | 3.53 | 0.01 | 3.53 | 0.00 | 3.41 | 0.00 | 3.68 | 0.00 | 3.54 | 0.06 |
| 1160 | 3.53 | 0.01 | 3.53 | 0.00 | 3.41 | 0.00 | 3.68 | 0.00 | 3.54 | 0.06 |
| 1162.5 | 3.53 | 0.01 | 3.53 | 0.00 | 3.41 | 0.00 | 3.67 | 0.00 | 3.54 | 0.06 |
| 1165 | 3.53 | 0.01 | 3.53 | 0.00 | 3.41 | 0.00 | 3.67 | 0.00 | 3.54 | 0.06 |
| 1167.5 | 3.53 | 0.01 | 3.53 | 0.00 | 3.41 | 0.00 | 3.67 | 0.00 | 3.54 | 0.06 |
| 1170 | 3.53 | 0.01 | 3.53 | 0.00 | 3.41 | 0.00 | 3.67 | 0.00 | 3.54 | 0.06 |
| 1172.5 | 3.53 | 0.01 | 3.53 | 0.00 | 3.41 | 0.00 | 3.67 | 0.00 | 3.53 | 0.05 |
| 1175 | 3.53 | 0.01 | 3.53 | 0.00 | 3.40 | 0.00 | 3.67 | 0.00 | 3.53 | 0.05 |
| 1177.5 | 3.53 | 0.01 | 3.53 | 0.00 | 3.40 | 0.00 | 3.67 | 0.00 | 3.53 | 0.05 |
| 1180 | 3.52 | 0.01 | 3.53 | 0.00 | 3.40 | 0.00 | 3.67 | 0.00 | 3.53 | 0.05 |
| 1182.5 | 3.52 | 0.01 | 3.53 | 0.00 | 3.40 | 0.00 | 3.66 | 0.00 | 3.53 | 0.05 |
| 1185 | 3.52 | 0.01 | 3.52 | 0.00 | 3.40 | 0.00 | 3.66 | 0.00 | 3.53 | 0.05 |
| 1187.5 | 3.52 | 0.01 | 3.52 | 0.00 | 3.40 | 0.00 | 3.66 | 0.00 | 3.53 | 0.05 |
| 1190 | 3.52 | 0.01 | 3.52 | 0.00 | 3.40 | 0.00 | 3.66 | 0.00 | 3.53 | 0.05 |
| 1192.5 | 3.52 | 0.01 | 3.52 | 0.00 | 3.40 | 0.00 | 3.66 | 0.00 | 3.53 | 0.05 |
| 1195 | 3.52 | 0.01 | 3.52 | 0.00 | 3.40 | 0.00 | 3.66 | 0.00 | 3.53 | 0.05 |
| 1197.5 | 3.52 | 0.01 | 3.52 | 0.00 | 3.40 | 0.00 | 3.66 | 0.00 | 3.52 | 0.05 |
| 1200 | 3.52 | 0.01 | 3.52 | 0.00 | 3.40 | 0.00 | 3.66 | 0.00 | 3.52 | 0.05 |
| 1202.5 | 3.51 | 0.01 | 3.52 | 0.00 | 3.40 | 0.00 | 3.65 | 0.00 | 3.52 | 0.05 |
| 1205 | 3.51 | 0.01 | 3.52 | 0.00 | 3.40 | 0.00 | 3.65 | 0.00 | 3.52 | 0.05 |
| 1207.5 | 3.51 | 0.01 | 3.52 | 0.00 | 3.39 | 0.00 | 3.65 | 0.00 | 3.52 | 0.05 |
| 1210 | 3.51 | 0.01 | 3.52 | 0.00 | 3.39 | 0.00 | 3.65 | 0.00 | 3.52 | 0.05 |
| 1212.5 | 3.51 | 0.01 | 3.52 | 0.00 | 3.39 | 0.00 | 3.65 | 0.00 | 3.52 | 0.05 |
| 1215 | 3.51 | 0.01 | 3.51 | 0.00 | 3.39 | 0.00 | 3.65 | 0.00 | 3.52 | 0.05 |
| 1217.5 | 3.51 | 0.01 | 3.51 | 0.00 | 3.39 | 0.00 | 3.65 | 0.00 | 3.52 | 0.05 |
| 1220 | 3.51 | 0.01 | 3.51 | 0.00 | 3.39 | 0.00 | 3.65 | 0.00 | 3.52 | 0.05 |
| 1222.5 | 3.51 | 0.01 | 3.51 | 0.00 | 3.39 | 0.00 | 3.65 | 0.00 | 3.52 | 0.05 |
| 1225 | 3.51 | 0.01 | 3.51 | 0.00 | 3.39 | 0.00 | 3.65 | 0.00 | 3.51 | 0.05 |
| 1227.5 | 3.51 | 0.01 | 3.51 | 0.00 | 3.39 | 0.00 | 3.64 | 0.00 | 3.51 | 0.05 |
| 1230 | 3.50 | 0.01 | 3.51 | 0.00 | 3.39 | 0.00 | 3.64 | 0.00 | 3.51 | 0.05 |
| 1232.5 | 3.50 | 0.01 | 3.51 | 0.00 | 3.39 | 0.00 | 3.64 | 0.00 | 3.51 | 0.05 |
| 1235 | 3.50 | 0.01 | 3.51 | 0.00 | 3.39 | 0.00 | 3.64 | 0.00 | 3.51 | 0.05 |
| 1237.5 | 3.50 | 0.01 | 3.51 | 0.00 | 3.39 | 0.00 | 3.64 | 0.00 | 3.51 | 0.05 |
| 1240 | 3.50 | 0.01 | 3.51 | 0.00 | 3.39 | 0.00 | 3.64 | 0.00 | 3.51 | 0.05 |
| 1242.5 | 3.50 | 0.01 | 3.51 | 0.00 | 3.38 | 0.00 | 3.64 | 0.00 | 3.51 | 0.05 |
| 1245 | 3.50 | 0.01 | 3.51 | 0.00 | 3.38 | 0.00 | 3.64 | 0.00 | 3.51 | 0.05 |
| 1247.5 | 3.50 | 0.01 | 3.51 | 0.00 | 3.38 | 0.00 | 3.64 | 0.00 | 3.51 | 0.05 |
| 1250 | 3.50 | 0.01 | 3.51 | 0.00 | 3.38 | 0.00 | 3.64 | 0.00 | 3.51 | 0.05 |
| 1252.5 | 3.50 | 0.01 | 3.50 | 0.00 | 3.38 | 0.00 | 3.63 | 0.00 | 3.50 | 0.05 |
| 1255 | 3.50 | 0.01 | 3.50 | 0.00 | 3.38 | 0.00 | 3.63 | 0.00 | 3.50 | 0.05 |
| 1257.5 | 3.50 | 0.01 | 3.50 | 0.00 | 3.38 | 0.00 | 3.63 | 0.00 | 3.50 | 0.05 |
| 1260 | 3.49 | 0.01 | 3.50 | 0.00 | 3.38 | 0.00 | 3.63 | 0.00 | 3.50 | 0.05 |
| 1262.5 | 3.49 | 0.01 | 3.50 | 0.00 | 3.38 | 0.00 | 3.63 | 0.00 | 3.50 | 0.05 |
| 1265 | 3.49 | 0.01 | 3.50 | 0.00 | 3.38 | 0.00 | 3.63 | 0.00 | 3.50 | 0.05 |
| 1267.5 | 3.49 | 0.01 | 3.50 | 0.00 | 3.38 | 0.00 | 3.63 | 0.00 | 3.50 | 0.05 |
| 1270 | 3.49 | 0.01 | 3.50 | 0.00 | 3.38 | 0.00 | 3.63 | 0.00 | 3.50 | 0.05 |
| 1272.5 | 3.49 | 0.01 | 3.50 | 0.00 | 3.38 | 0.00 | 3.63 | 0.00 | 3.50 | 0.05 |
| 1275 | 3.49 | 0.01 | 3.50 | 0.00 | 3.38 | 0.00 | 3.63 | 0.00 | 3.50 | 0.05 |
| 1277.5 | 3.49 | 0.01 | 3.50 | 0.00 | 3.38 | 0.00 | 3.63 | 0.00 | 3.50 | 0.05 |
| 1280 | 3.49 | 0.01 | 3.50 | 0.00 | 3.38 | 0.00 | 3.63 | 0.00 | 3.50 | 0.05 |
| 1282.5 | 3.49 | 0.01 | 3.50 | 0.00 | 3.38 | 0.00 | 3.62 | 0.00 | 3.49 | 0.05 |
| 1285 | 3.49 | 0.01 | 3.50 | 0.00 | 3.37 | 0.00 | 3.62 | 0.00 | 3.49 | 0.05 |
| 1287.5 | 3.49 | 0.01 | 3.50 | 0.00 | 3.37 | 0.00 | 3.62 | 0.00 | 3.49 | 0.05 |
| 1290 | 3.49 | 0.01 | 3.50 | 0.00 | 3.37 | 0.00 | 3.62 | 0.00 | 3.49 | 0.05 |
| 1292.5 | 3.48 | 0.01 | 3.49 | 0.00 | 3.37 | 0.00 | 3.62 | 0.00 | 3.49 | 0.05 |
| 1295 | 3.48 | 0.01 | 3.49 | 0.00 | 3.37 | 0.00 | 3.62 | 0.00 | 3.49 | 0.05 |
| 1297.5 | 3.48 | 0.01 | 3.49 | 0.00 | 3.37 | 0.00 | 3.62 | 0.00 | 3.49 | 0.05 |
| 1300 | 3.48 | 0.01 | 3.49 | 0.00 | 3.37 | 0.00 | 3.62 | 0.00 | 3.49 | 0.05 |
| 1302.5 | 3.48 | 0.01 | 3.49 | 0.00 | 3.37 | 0.00 | 3.62 | 0.00 | 3.49 | 0.05 |
| 1305 | 3.48 | 0.01 | 3.49 | 0.00 | 3.37 | 0.00 | 3.62 | 0.00 | 3.49 | 0.05 |
| 1307.5 | 3.48 | 0.01 | 3.49 | 0.00 | 3.37 | 0.00 | 3.62 | 0.00 | 3.49 | 0.05 |
| 1310 | 3.48 | 0.01 | 3.49 | 0.00 | 3.37 | 0.00 | 3.62 | 0.00 | 3.49 | 0.05 |
| 1312.5 | 3.48 | 0.01 | 3.49 | 0.00 | 3.37 | 0.00 | 3.61 | 0.00 | 3.49 | 0.05 |
| 1315 | 3.48 | 0.01 | 3.49 | 0.00 | 3.37 | 0.00 | 3.61 | 0.00 | 3.48 | 0.05 |
| 1317.5 | 3.48 | 0.01 | 3.49 | 0.00 | 3.37 | 0.00 | 3.61 | 0.00 | 3.48 | 0.05 |
| 1320 | 3.48 | 0.01 | 3.49 | 0.00 | 3.37 | 0.00 | 3.61 | 0.00 | 3.48 | 0.05 |
| 1322.5 | 3.48 | 0.01 | 3.49 | 0.00 | 3.37 | 0.00 | 3.61 | 0.00 | 3.48 | 0.05 |
| 1325 | 3.48 | 0.01 | 3.49 | 0.00 | 3.37 | 0.00 | 3.61 | 0.00 | 3.48 | 0.05 |
| 1327.5 | 3.48 | 0.01 | 3.49 | 0.00 | 3.37 | 0.00 | 3.61 | 0.00 | 3.48 | 0.05 |
| 1330 | 3.48 | 0.01 | 3.49 | 0.00 | 3.37 | 0.00 | 3.61 | 0.00 | 3.48 | 0.05 |
| 1332.5 | 3.47 | 0.01 | 3.49 | 0.00 | 3.37 | 0.00 | 3.61 | 0.00 | 3.48 | 0.05 |
| 1335 | 3.47 | 0.01 | 3.49 | 0.00 | 3.36 | 0.00 | 3.61 | 0.00 | 3.48 | 0.05 |
| 1337.5 | 3.47 | 0.01 | 3.49 | 0.00 | 3.36 | 0.00 | 3.61 | 0.00 | 3.48 | 0.05 |
| 1340 | 3.47 | 0.01 | 3.48 | 0.00 | 3.36 | 0.00 | 3.61 | 0.00 | 3.48 | 0.05 |
| 1342.5 | 3.47 | 0.01 | 3.48 | 0.00 | 3.36 | 0.00 | 3.61 | 0.00 | 3.48 | 0.05 |
| 1345 | 3.47 | 0.01 | 3.48 | 0.00 | 3.36 | 0.00 | 3.60 | 0.00 | 3.48 | 0.05 |
| 1347.5 | 3.47 | 0.01 | 3.48 | 0.00 | 3.36 | 0.00 | 3.60 | 0.00 | 3.48 | 0.05 |
| 1350 | 3.47 | 0.01 | 3.48 | 0.00 | 3.36 | 0.00 | 3.60 | 0.00 | 3.47 | 0.05 |
| 1352.5 | 3.47 | 0.01 | 3.48 | 0.00 | 3.36 | 0.00 | 3.60 | 0.00 | 3.47 | 0.05 |
| 1355 | 3.47 | 0.01 | 3.48 | 0.00 | 3.36 | 0.00 | 3.60 | 0.00 | 3.47 | 0.05 |
| 1357.5 | 3.47 | 0.01 | 3.48 | 0.00 | 3.36 | 0.00 | 3.60 | 0.00 | 3.47 | 0.05 |
| 1360 | 3.47 | 0.01 | 3.48 | 0.00 | 3.36 | 0.00 | 3.60 | 0.00 | 3.47 | 0.05 |
| 1362.5 | 3.47 | 0.01 | 3.48 | 0.00 | 3.36 | 0.00 | 3.60 | 0.00 | 3.47 | 0.05 |
| 1365 | 3.47 | 0.01 | 3.48 | 0.00 | 3.36 | 0.00 | 3.60 | 0.00 | 3.47 | 0.05 |
| 1367.5 | 3.47 | 0.01 | 3.48 | 0.00 | 3.36 | 0.00 | 3.60 | 0.00 | 3.47 | 0.05 |
| 1370 | 3.47 | 0.01 | 3.48 | 0.00 | 3.36 | 0.00 | 3.60 | 0.00 | 3.47 | 0.05 |
| 1372.5 | 3.47 | 0.01 | 3.48 | 0.00 | 3.36 | 0.00 | 3.60 | 0.00 | 3.47 | 0.05 |
| 1375 | 3.47 | 0.01 | 3.48 | 0.00 | 3.36 | 0.00 | 3.60 | 0.00 | 3.47 | 0.05 |
| 1377.5 | 3.46 | 0.01 | 3.48 | 0.00 | 3.36 | 0.00 | 3.60 | 0.00 | 3.47 | 0.05 |
| 1380 | 3.46 | 0.01 | 3.48 | 0.00 | 3.36 | 0.00 | 3.60 | 0.00 | 3.47 | 0.05 |
| 1382.5 | 3.46 | 0.01 | 3.48 | 0.00 | 3.36 | 0.00 | 3.59 | 0.00 | 3.47 | 0.05 |
| 1385 | 3.46 | 0.01 | 3.48 | 0.00 | 3.36 | 0.00 | 3.59 | 0.00 | 3.46 | 0.05 |
| 1387.5 | 3.46 | 0.01 | 3.48 | 0.00 | 3.36 | 0.00 | 3.59 | 0.00 | 3.46 | 0.05 |
| 1390 | 3.46 | 0.01 | 3.48 | 0.00 | 3.36 | 0.00 | 3.59 | 0.00 | 3.46 | 0.05 |
| 1392.5 | 3.46 | 0.01 | 3.48 | 0.00 | 3.36 | 0.00 | 3.59 | 0.00 | 3.46 | 0.05 |
| 1395 | 3.46 | 0.01 | 3.48 | 0.00 | 3.36 | 0.00 | 3.59 | 0.00 | 3.46 | 0.05 |
| 1397.5 | 3.46 | 0.01 | 3.47 | 0.00 | 3.35 | 0.00 | 3.59 | 0.00 | 3.46 | 0.05 |
| 1400 | 3.46 | 0.01 | 3.47 | 0.00 | 3.35 | 0.00 | 3.59 | 0.00 | 3.46 | 0.05 |
| 1402.5 | 3.46 | 0.01 | 3.47 | 0.00 | 3.35 | 0.00 | 3.59 | 0.00 | 3.46 | 0.05 |
| 1405 | 3.46 | 0.01 | 3.47 | 0.00 | 3.35 | 0.00 | 3.59 | 0.00 | 3.46 | 0.05 |
| 1407.5 | 3.46 | 0.01 | 3.47 | 0.00 | 3.35 | 0.00 | 3.59 | 0.00 | 3.46 | 0.05 |
| 1410 | 3.46 | 0.01 | 3.47 | 0.00 | 3.35 | 0.00 | 3.59 | 0.00 | 3.46 | 0.05 |
| 1412.5 | 3.46 | 0.01 | 3.47 | 0.00 | 3.35 | 0.00 | 3.59 | 0.00 | 3.46 | 0.05 |
| 1415 | 3.46 | 0.01 | 3.47 | 0.00 | 3.35 | 0.00 | 3.59 | 0.00 | 3.46 | 0.05 |
| 1417.5 | 3.46 | 0.01 | 3.47 | 0.00 | 3.35 | 0.00 | 3.59 | 0.00 | 3.46 | 0.05 |
| 1420 | 3.46 | 0.01 | 3.47 | 0.00 | 3.35 | 0.00 | 3.59 | 0.00 | 3.46 | 0.04 |
| 1422.5 | 3.46 | 0.01 | 3.47 | 0.00 | 3.35 | 0.00 | 3.58 | 0.00 | 3.45 | 0.04 |
| 1425 | 3.46 | 0.01 | 3.47 | 0.00 | 3.35 | 0.00 | 3.58 | 0.00 | 3.45 | 0.04 |
| 1427.5 | 3.46 | 0.01 | 3.47 | 0.00 | 3.35 | 0.00 | 3.58 | 0.00 | 3.45 | 0.04 |
| 1430 | 3.45 | 0.00 | 3.47 | 0.00 | 3.35 | 0.00 | 3.58 | 0.00 | 3.45 | 0.04 |
| 1432.5 | 3.45 | 0.00 | 3.47 | 0.00 | 3.35 | 0.00 | 3.58 | 0.00 | 3.45 | 0.04 |
| 1435 | 3.45 | 0.00 | 3.47 | 0.00 | 3.35 | 0.00 | 3.58 | 0.00 | 3.45 | 0.04 |
| 1437.5 | 3.45 | 0.00 | 3.47 | 0.00 | 3.35 | 0.00 | 3.58 | 0.00 | 3.45 | 0.04 |
| 1440 | 3.45 | 0.00 | 3.47 | 0.00 | 3.35 | 0.00 | 3.58 | 0.00 | 3.45 | 0.04 |
| 1442.5 | 3.45 | 0.00 | 3.47 | 0.00 | 3.35 | 0.00 | 3.58 | 0.00 | 3.45 | 0.04 |
| 1445 | 3.45 | 0.00 | 3.47 | 0.00 | 3.35 | 0.00 | 3.58 | 0.00 | 3.45 | 0.04 |
| 1447.5 | 3.45 | 0.00 | 3.47 | 0.00 | 3.35 | 0.00 | 3.58 | 0.00 | 3.45 | 0.04 |
| 1450 | 3.45 | 0.00 | 3.47 | 0.00 | 3.35 | 0.00 | 3.58 | 0.00 | 3.45 | 0.04 |
| 1452.5 | 3.45 | 0.00 | 3.47 | 0.00 | 3.35 | 0.00 | 3.58 | 0.00 | 3.45 | 0.04 |
| 1455 | 3.45 | 0.00 | 3.47 | 0.00 | 3.35 | 0.00 | 3.58 | 0.00 | 3.45 | 0.04 |
| 1457.5 | 3.45 | 0.00 | 3.47 | 0.00 | 3.35 | 0.00 | 3.58 | 0.00 | 3.45 | 0.04 |
| 1460 | 3.45 | 0.00 | 3.47 | 0.00 | 3.35 | 0.00 | 3.58 | 0.00 | 3.45 | 0.04 |
| 1462.5 | 3.45 | 0.00 | 3.47 | 0.00 | 3.35 | 0.00 | 3.58 | 0.00 | 3.44 | 0.04 |
| 1465 | 3.45 | 0.00 | 3.47 | 0.00 | 3.35 | 0.00 | 3.58 | 0.00 | 3.44 | 0.04 |
| 1467.5 | 3.45 | 0.00 | 3.46 | 0.00 | 3.35 | 0.00 | 3.57 | 0.00 | 3.44 | 0.04 |
| 1470 | 3.45 | 0.00 | 3.46 | 0.00 | 3.35 | 0.00 | 3.57 | 0.00 | 3.44 | 0.04 |
| 1472.5 | 3.45 | 0.00 | 3.46 | 0.00 | 3.35 | 0.00 | 3.57 | 0.00 | 3.44 | 0.04 |
| 1475 | 3.45 | 0.00 | 3.46 | 0.00 | 3.35 | 0.00 | 3.57 | 0.00 | 3.44 | 0.04 |
| 1477.5 | 3.45 | 0.00 | 3.46 | 0.00 | 3.35 | 0.00 | 3.57 | 0.00 | 3.44 | 0.04 |
| 1480 | 3.45 | 0.00 | 3.46 | 0.00 | 3.35 | 0.00 | 3.57 | 0.00 | 3.44 | 0.04 |
| 1482.5 | 3.45 | 0.00 | 3.46 | 0.00 | 3.35 | 0.00 | 3.57 | 0.00 | 3.44 | 0.04 |
| 1485 | 3.45 | 0.00 | 3.46 | 0.00 | 3.35 | 0.00 | 3.57 | 0.00 | 3.44 | 0.04 |
| 1487.5 | 3.45 | 0.00 | 3.46 | 0.00 | 3.34 | 0.00 | 3.57 | 0.00 | 3.44 | 0.04 |
| 1490 | 3.45 | 0.00 | 3.46 | 0.00 | 3.34 | 0.00 | 3.57 | 0.00 | 3.44 | 0.04 |
| 1492.5 | 3.45 | 0.00 | 3.46 | 0.00 | 3.34 | 0.00 | 3.57 | 0.00 | 3.44 | 0.04 |
| 1495 | 3.44 | 0.00 | 3.46 | 0.00 | 3.34 | 0.00 | 3.57 | 0.00 | 3.44 | 0.04 |
| 1497.5 | 3.44 | 0.00 | 3.46 | 0.00 | 3.34 | 0.00 | 3.57 | 0.00 | 3.44 | 0.04 |
| 1500 | 3.44 | 0.00 | 3.46 | 0.00 | 3.34 | 0.00 | 3.57 | 0.00 | 3.44 | 0.04 |
| 1502.5 | 3.44 | 0.00 | 3.46 | 0.00 | 3.34 | 0.00 | 3.57 | 0.00 | 3.44 | 0.04 |
| 1505 | 3.44 | 0.00 | 3.46 | 0.00 | 3.34 | 0.00 | 3.57 | 0.00 | 3.43 | 0.04 |
| 1507.5 | 3.44 | 0.00 | 3.46 | 0.00 | 3.34 | 0.00 | 3.57 | 0.00 | 3.43 | 0.04 |
| 1510 | 3.44 | 0.00 | 3.46 | 0.00 | 3.34 | 0.00 | 3.57 | 0.00 | 3.43 | 0.04 |
| 1512.5 | 3.44 | 0.00 | 3.46 | 0.00 | 3.34 | 0.00 | 3.57 | 0.00 | 3.43 | 0.04 |
| 1515 | 3.44 | 0.00 | 3.46 | 0.00 | 3.34 | 0.00 | 3.57 | 0.00 | 3.43 | 0.04 |
| 1517.5 | 3.44 | 0.00 | 3.46 | 0.00 | 3.34 | 0.00 | 3.56 | 0.00 | 3.43 | 0.04 |
| 1520 | 3.44 | 0.00 | 3.46 | 0.00 | 3.34 | 0.00 | 3.56 | 0.00 | 3.43 | 0.04 |
| 1522.5 | 3.44 | 0.00 | 3.46 | 0.00 | 3.34 | 0.00 | 3.56 | 0.00 | 3.43 | 0.04 |
| 1525 | 3.44 | 0.00 | 3.46 | 0.00 | 3.34 | 0.00 | 3.56 | 0.00 | 3.43 | 0.04 |
| 1527.5 | 3.44 | 0.00 | 3.46 | 0.00 | 3.34 | 0.00 | 3.56 | 0.00 | 3.43 | 0.04 |
| 1530 | 3.44 | 0.00 | 3.46 | 0.00 | 3.34 | 0.00 | 3.56 | 0.00 | 3.43 | 0.04 |
| 1532.5 | 3.44 | 0.00 | 3.46 | 0.00 | 3.34 | 0.00 | 3.56 | 0.00 | 3.43 | 0.04 |
| 1535 | 3.44 | 0.00 | 3.46 | 0.00 | 3.34 | 0.00 | 3.56 | 0.00 | 3.43 | 0.04 |
| 1537.5 | 3.44 | 0.00 | 3.46 | 0.00 | 3.34 | 0.00 | 3.56 | 0.00 | 3.43 | 0.04 |
| 1540 | 3.44 | 0.00 | 3.46 | 0.00 | 3.34 | 0.00 | 3.56 | 0.00 | 3.43 | 0.04 |
| 1542.5 | 3.44 | 0.00 | 3.46 | 0.00 | 3.34 | 0.00 | 3.56 | 0.00 | 3.43 | 0.04 |
| 1545 | 3.44 | 0.00 | 3.46 | 0.00 | 3.34 | 0.00 | 3.56 | 0.00 | 3.43 | 0.04 |
| 1547.5 | 3.44 | 0.00 | 3.46 | 0.00 | 3.34 | 0.00 | 3.56 | 0.00 | 3.43 | 0.04 |
| 1550 | 3.44 | 0.00 | 3.46 | 0.00 | 3.34 | 0.00 | 3.56 | 0.00 | 3.42 | 0.04 |
| 1552.5 | 3.44 | 0.00 | 3.46 | 0.00 | 3.34 | 0.00 | 3.56 | 0.00 | 3.42 | 0.04 |
| 1555 | 3.44 | 0.00 | 3.46 | 0.00 | 3.34 | 0.00 | 3.56 | 0.00 | 3.42 | 0.04 |
| 1557.5 | 3.44 | 0.00 | 3.46 | 0.00 | 3.34 | 0.00 | 3.56 | 0.00 | 3.42 | 0.04 |
| 1560 | 3.44 | 0.00 | 3.46 | 0.00 | 3.34 | 0.00 | 3.56 | 0.00 | 3.42 | 0.04 |
| 1562.5 | 3.44 | 0.00 | 3.45 | 0.00 | 3.34 | 0.00 | 3.56 | 0.00 | 3.42 | 0.04 |
| 1565 | 3.44 | 0.00 | 3.45 | 0.00 | 3.34 | 0.00 | 3.56 | 0.00 | 3.42 | 0.04 |
| 1567.5 | 3.44 | 0.00 | 3.45 | 0.00 | 3.34 | 0.00 | 3.56 | 0.00 | 3.42 | 0.04 |
| 1570 | 3.44 | 0.00 | 3.45 | 0.00 | 3.34 | 0.00 | 3.55 | 0.00 | 3.42 | 0.04 |
| 1572.5 | 3.44 | 0.00 | 3.45 | 0.00 | 3.34 | 0.00 | 3.55 | 0.00 | 3.42 | 0.04 |
| 1575 | 3.44 | 0.00 | 3.45 | 0.00 | 3.34 | 0.00 | 3.55 | 0.00 | 3.42 | 0.04 |
| 1577.5 | 3.43 | 0.00 | 3.45 | 0.00 | 3.34 | 0.00 | 3.55 | 0.00 | 3.42 | 0.04 |
| 1580 | 3.43 | 0.00 | 3.45 | 0.00 | 3.34 | 0.00 | 3.55 | 0.00 | 3.42 | 0.04 |
| 1582.5 | 3.43 | 0.00 | 3.45 | 0.00 | 3.34 | 0.00 | 3.55 | 0.00 | 3.42 | 0.04 |
| 1585 | 3.43 | 0.00 | 3.45 | 0.00 | 3.34 | 0.00 | 3.55 | 0.00 | 3.42 | 0.04 |
| 1587.5 | 3.43 | 0.00 | 3.45 | 0.00 | 3.34 | 0.00 | 3.55 | 0.00 | 3.42 | 0.04 |
| 1590 | 3.43 | 0.00 | 3.45 | 0.00 | 3.34 | 0.00 | 3.55 | 0.00 | 3.42 | 0.04 |
| 1592.5 | 3.43 | 0.00 | 3.45 | 0.00 | 3.34 | 0.00 | 3.55 | 0.00 | 3.42 | 0.04 |
| 1595 | 3.43 | 0.00 | 3.45 | 0.00 | 3.34 | 0.00 | 3.55 | 0.00 | 3.41 | 0.04 |
| 1597.5 | 3.43 | 0.00 | 3.45 | 0.00 | 3.34 | 0.00 | 3.55 | 0.00 | 3.41 | 0.04 |
| 1600 | 3.43 | 0.00 | 3.45 | 0.00 | 3.34 | 0.00 | 3.55 | 0.00 | 3.41 | 0.04 |
| 1602.5 | 3.43 | 0.00 | 3.45 | 0.00 | 3.34 | 0.00 | 3.55 | 0.00 | 3.41 | 0.04 |
| 1605 | 3.43 | 0.00 | 3.45 | 0.00 | 3.34 | 0.00 | 3.55 | 0.00 | 3.41 | 0.04 |
| 1607.5 | 3.43 | 0.00 | 3.45 | 0.00 | 3.34 | 0.00 | 3.55 | 0.00 | 3.41 | 0.04 |
| 1610 | 3.43 | 0.00 | 3.45 | 0.00 | 3.34 | 0.00 | 3.55 | 0.00 | 3.41 | 0.04 |
| 1612.5 | 3.43 | 0.00 | 3.45 | 0.00 | 3.34 | 0.00 | 3.55 | 0.00 | 3.41 | 0.04 |
| 1615 | 3.43 | 0.00 | 3.45 | 0.00 | 3.34 | 0.00 | 3.55 | 0.00 | 3.41 | 0.04 |
| 1617.5 | 3.43 | 0.00 | 3.45 | 0.00 | 3.34 | 0.00 | 3.55 | 0.00 | 3.41 | 0.04 |
| 1620 | 3.43 | 0.00 | 3.45 | 0.00 | 3.34 | 0.00 | 3.55 | 0.00 | 3.41 | 0.04 |
| 1622.5 | 3.43 | 0.00 | 3.45 | 0.00 | 3.34 | 0.00 | 3.55 | 0.00 | 3.41 | 0.04 |
| 1625 | 3.43 | 0.00 | 3.45 | 0.00 | 3.34 | 0.00 | 3.55 | 0.00 | 3.41 | 0.04 |
| 1627.5 | 3.43 | 0.00 | 3.45 | 0.00 | 3.34 | 0.00 | 3.55 | 0.00 | 3.41 | 0.04 |
| 1630 | 3.43 | 0.00 | 3.45 | 0.00 | 3.34 | 0.00 | 3.55 | 0.00 | 3.41 | 0.04 |
| 1632.5 | 3.43 | 0.00 | 3.45 | 0.00 | 3.34 | 0.00 | 3.54 | 0.00 | 3.41 | 0.04 |
| 1635 | 3.43 | 0.00 | 3.45 | 0.00 | 3.34 | 0.00 | 3.54 | 0.00 | 3.41 | 0.04 |
| 1637.5 | 3.43 | 0.00 | 3.45 | 0.00 | 3.34 | 0.00 | 3.54 | 0.00 | 3.41 | 0.04 |
| 1640 | 3.43 | 0.00 | 3.45 | 0.00 | 3.34 | 0.00 | 3.54 | 0.00 | 3.41 | 0.04 |
| 1642.5 | 3.43 | 0.00 | 3.45 | 0.00 | 3.34 | 0.00 | 3.54 | 0.00 | 3.40 | 0.04 |
| 1645 | 3.43 | 0.00 | 3.45 | 0.00 | 3.34 | 0.00 | 3.54 | 0.00 | 3.40 | 0.04 |
| 1647.5 | 3.43 | 0.00 | 3.45 | 0.00 | 3.34 | 0.00 | 3.54 | 0.00 | 3.40 | 0.04 |
| 1650 | 3.43 | 0.00 | 3.45 | 0.00 | 3.34 | 0.00 | 3.54 | 0.00 | 3.40 | 0.04 |
| 1652.5 | 3.43 | 0.00 | 3.45 | 0.00 | 3.34 | 0.00 | 3.54 | 0.00 | 3.40 | 0.04 |
| 1655 | 3.43 | 0.00 | 3.45 | 0.00 | 3.33 | 0.00 | 3.54 | 0.00 | 3.40 | 0.04 |
| 1657.5 | 3.43 | 0.00 | 3.45 | 0.00 | 3.33 | 0.00 | 3.54 | 0.00 | 3.40 | 0.04 |
| 1660 | 3.43 | 0.00 | 3.45 | 0.00 | 3.33 | 0.00 | 3.54 | 0.00 | 3.40 | 0.04 |
| 1662.5 | 3.43 | 0.00 | 3.45 | 0.00 | 3.33 | 0.00 | 3.54 | 0.00 | 3.40 | 0.04 |
| 1665 | 3.43 | 0.00 | 3.45 | 0.00 | 3.33 | 0.00 | 3.54 | 0.00 | 3.40 | 0.04 |
| 1667.5 | 3.43 | 0.00 | 3.45 | 0.00 | 3.33 | 0.00 | 3.54 | 0.00 | 3.40 | 0.04 |
| 1670 | 3.43 | 0.00 | 3.45 | 0.00 | 3.33 | 0.00 | 3.54 | 0.00 | 3.40 | 0.04 |
| 1672.5 | 3.43 | 0.00 | 3.45 | 0.00 | 3.33 | 0.00 | 3.54 | 0.00 | 3.40 | 0.04 |
| 1675 | 3.43 | 0.00 | 3.45 | 0.00 | 3.33 | 0.00 | 3.54 | 0.00 | 3.40 | 0.04 |
| 1677.5 | 3.43 | 0.00 | 3.45 | 0.00 | 3.33 | 0.00 | 3.54 | 0.00 | 3.40 | 0.04 |
| 1680 | 3.43 | 0.00 | 3.45 | 0.00 | 3.33 | 0.00 | 3.54 | 0.00 | 3.40 | 0.04 |
| 1682.5 | 3.43 | 0.00 | 3.45 | 0.00 | 3.33 | 0.00 | 3.54 | 0.00 | 3.40 | 0.04 |
| 1685 | 3.43 | 0.00 | 3.45 | 0.00 | 3.33 | 0.00 | 3.54 | 0.00 | 3.40 | 0.04 |
| 1687.5 | 3.43 | 0.00 | 3.45 | 0.00 | 3.33 | 0.00 | 3.54 | 0.00 | 3.40 | 0.04 |
| 1690 | 3.42 | 0.00 | 3.45 | 0.00 | 3.33 | 0.00 | 3.54 | 0.00 | 3.40 | 0.04 |